%% file: 00_main.tex
\pdfoutput=1 

\documentclass[times,twocolumn,longtitle,final]{elsarticle}

\usepackage{medima}
\usepackage{framed,multirow}

\usepackage{amssymb}
\usepackage{latexsym}
\usepackage{xspace}
\usepackage{pdfpages}
\usepackage{url}
\usepackage{xcolor}
\usepackage{hyperref}
\definecolor{newcolor}{rgb}{.8,.349,.1}

\usepackage{soul}
\usepackage{amsmath}
\usepackage[american]{babel}
\usepackage{hyphenat}
\usepackage{multirow}
\usepackage{booktabs}
\usepackage{graphicx}
\usepackage{pifont}
\usepackage{lscape}
\usepackage{rotating} 
\usepackage{longtable}	
\usepackage{color, colortbl}
\usepackage{caption}
\usepackage{subcaption} 
\newsavebox{\twosubbox} 
\usepackage{pdflscape}
\usepackage{afterpage}
\usepackage{cleveref}
\usepackage{chngcntr} 
\usepackage{placeins}
\usepackage[utf8]{inputenc}
\usepackage{newunicodechar}
\newunicodechar{−}{-}
\usepackage{adjustbox}
\usepackage{caption}
\usepackage{subcaption}
\usepackage{tabularx}

\journal{Medical Image Analysis}
\begin{document}
\verso{V. Rogowski \textit{et~al.}}

\begin{frontmatter}

\title{Generating synthetic computed tomography for radiotherapy: SynthRAD2025 challenge report}

\author[1,2]{Viktor \snm{Rogowski}\fnref{fn2}}
\address[1]{Radiation Physics, Department of Hematology, Oncology, and Radiation Physics, Skåne University Hospital, Lund, Sweden}
\address[2]{Medical Radiation Physics, Department of Clinical Sciences Lund, Lund University, Lund, Sweden}

\author[3,4]{Maarten L. \snm{Terpstra}\fnref{fn2}}
\address[3]{Radiotherapy Department, University Medical Center Utrecht, Utrecht, The Netherlands}
\address[4]{Computational Imaging Group for MR Diagnostics $\&$ Therapy, University Medical Center Utrecht, Utrecht, The Netherlands}

\author[5,6]{Niklas \snm{Wahl}\fnref{fn2}}
\address[5]{Division of Medical Physics in Radiation Oncology, Deutsches Krebsforschungszentrum (DKFZ), Heidelberg, Germany}
\address[6]{Heidelberg Institute for Radiation Oncology (HIRO) and National Center for Radiation Research in Oncology (NCRO), Heidelberg, Germany}

\author[7]{Florian \snm{Kamp}\fnref{fn2}}
\address[7]{Department of Radiation Oncology and Cyberknife Center, University Hospital of Cologne, Cologne, Germany}

\author[8]{Erik \snm{van der Bijl}\fnref{fn2}}

\author[9]{Arthur Jr. \snm{Galapon}\fnref{fn2}}

\author[10]{Christopher \snm{Kurz}\fnref{fn2}}


\author[12,13]{Bowen \snm{Xin}} 
\author[13]{Zhengxiang \snm{Sun}} 
\author[12]{Hollie \snm{Min}} 
\author[12]{Gregg \snm{Belous}} 
\author[12]{Jason \snm{Dowling}}

\author[15]{Yan \snm{Xia}} 
\author[16]{Siyuan \snm{Mei}} 
\author[17]{Fuxin \snm{Fan}}

\author[18]{Arthur \snm{Longuefosse}} 
\author[19]{Javier \snm{Sequeiro Gonzalez}} 
\author[19]{Miguel \snm{Diaz Benito}}
\author[19]{Alvaro Garcia \snm{Martin}}
\author[18]{Fabien \snm{Baldacci}}

\author[20]{Valentin \snm{Boussot}} 
\author[20]{Cédric \snm{Hémon}} 
\author[20]{Jean-Claude \snm{Nunes}} 
\author[20]{Jean-Louis \snm{Dillenseger}}

\author[21]{Zhiyuan \snm{Zhang}} 
\author[22]{Jinghua \snm{Cai}} 

\author[23]{Han \snm{Bing}} 
\author[23]{Tan \snm{Zuopeng}}

\author[24]{Ricardo \snm{Brioso}} 
\author[24]{Daniele \snm{Loiacono}} 

\author[10,11]{Guillaume \snm{Landry}\fnref{fn2}}
\author[10,25]{Adrian \snm{Thummerer}\fnref{fn2,fn3}}
\author[3,4]{Matteo \snm{Maspero}\corref{cor1}\fnref{fn2,fn3}}
\ead{m.maspero@umcutrecht.nl}
\fntext[fn2]{Challenge Organizer}
\fntext[fn3]{Equally contributing senior authors}
\cortext[cor1]{Corresponding author: Heidelberglaan 100, 3508 GA, UMC Utrecht, P.O. Box 85500 Utrecht, The Netherlands, Tel.: +31-88 75 67492;}

\address[8]{Department of Radiation Oncology, Radboud University Medical Center, Nijmegen, The Netherlands}
\address[9]{Department of Radiation Oncology, University Medical Center Groningen, University of Groningen, Groningen, The Netherlands}
\address[10]{Department of Radiation Oncology, LMU University Hospital, LMU Munich, Munich, Germany}
\address[11]{Bavarian Cancer Research Center (BZKF), Munich, Germany}
\address[12]{Australian eHealth Research Center, CSIRO, Brisbane, Australia}
\address[13]{School of Computer Science, University of Sydney, Sydney, Australia}
\address[15]{Department of Orthodontics and Orofacial Orthopedics, FAU Erlangen-Nuremberg, Germany}
\address[16]{Pattern Recognition Lab, FAU Erlangen-Nuremberg, Germany}
\address[17]{Digital Technology and Innovation, Siemens Healthineers, Shanghai, China}
\address[18]{RIKEN Center for Integrative Medical Sciences, Tokyo, Japan}
\address[19]{Erasmus Mundus Joint Master's Degree IPCVai, University of Bordeaux, France}
\address[20]{Univ Rennes 1, CLCC Eugène Marquis, INSERM, LTSI - UMR 1099, France}
\address[21]{Computer Science, Huazhong University of Science and Technology, Wuhan, China}
\address[22]{Huazhong University of Science and Technology, Wuhan, China}
\address[23]{Canon Medical Systems (China) CO., LTD., Beijing, China}
\address[24]{DEIB, Politecnico di Milano, Milan, Italy}
\address[25]{Department of Radiation Oncology, Inselspital, Bern University Hospital and University of Bern, Bern, Switzerland}

\received{n/a}
\begin{abstract}
Radiation therapy (RT) requires precise dose delivery to tumors while sparing surrounding healthy tissues over multiple fractions. Computed tomography (CT) is fundamental for treatment planning, providing the electron density necessary for accurate dose calculation. However, repeated CT acquisitions introduce additional radiation exposure and logistical burdens, particularly within adaptive radiotherapy workflows. While magnetic resonance imaging (MRI) offers superior soft-tissue contrast, it lacks electron density information. Conversely, cone-beam CT (CBCT) enables daily three-dimensional imaging at the treatment unit but requires correction for reliable dose calculation.\\
Synthetic CT (sCT) generation addresses these limitations by converting MRI or CBCT into CT-equivalent images with accurate Hounsfield Unit (HU) values, enabling MRI-only RT and CBCT-based adaptive workflows. Expanding upon the SynthRAD2023 framework, the SynthRAD2025 challenge benchmarked sCT generation methods using a significantly larger multi-center dataset. This cohort included 2,362 patients from five European centers and covered three anatomically complex regions: head and neck, thorax, and abdomen. The challenge was divided into two tasks: MRI-to-CT (890 cases) and CBCT-to-CT (1,472 cases). Evaluation included image similarity metrics (MAE, PSNR, MS-SSIM), segmentation-based measures (Dice, HD95), and dosimetric accuracy derived from photon and proton treatment plans.\\
The challenge attracted substantial participation (803 single users), resulting in 12 and 13 valid submissions for Tasks 1 and 2, respectively. In Task 1, the top-performing method achieved a MAE value of $64.8 \pm 21.3$ HU, PSNR of $\sim30$ dB, MS-SSIM of $\sim0.936$, Dice up to $0.79$, and photon $\gamma_{2\%/2\text{mm}}$ pass rates above $98\%$. In Task 2, performance improved further, with a MAE value of $48.3 \pm 13.4$ HU, PSNR up to $32.6$ dB, MS-SSIM up to $0.968$, Dice up to $0.86$, and photon $\gamma$-pass rates exceeding $99\%$. Proton $\gamma$-pass rates reached approximately $85\%$ and $89\%$ for Tasks 1 and 2, respectively. Correlation analysis showed strong relationships between image similarity and segmentation metrics ($\rho=0.78$--$0.79$ between MS-SSIM and Dice), but only moderate associations with dose metrics, confirming that image quality alone is not a sufficient surrogate for dosimetric accuracy.
Across anatomical regions, head-and-neck cases achieved the most consistent performance, whereas thoracic and abdominal cases showed greater variability due to motion and anatomical complexity. Qualitative analysis demonstrated that residual errors are concentrated at tissue interfaces, particularly air-tissue and bone boundaries, and propagate along beam paths, affecting proton dose calculations more strongly than photon dose.\\
SynthRAD2025 advances the benchmarking of sCT generation across diverse anatomical regions and demonstrates that deep learning methods can produce high-quality, clinically relevant sCTs, particularly for CBCT-to-CT synthesis. The results identify persistent challenges in MRI-to-CT translation and underscore the necessity of multi-metric evaluation, specifically dose-based assessment, for robust clinical validation. These findings support the continued development and clinical translation of CT-free radiotherapy.


\end{abstract}

\end{frontmatter}


\input{01_Introduction}

\input{02_Methods}

\input{03_Participants}

\input{04_Results}

\input{05_Discussion}

\input{06_Conclusion}

\section*{Authors contribution}
\subsection*{Organizers}
\textbf{Viktor Rogowski:} Conceptualization, Formal Analysis, Investigation, Methodology, Software, Validation, Visualization, Writing – original draft, Writing – review \& editing.
\textbf{Maarten Terpstra:} Conceptualization, Formal Analysis, Investigation, Methodology, Software, Validation, Visualization, Writing – original draft,  Writing – review \& editing.
\textbf{Niklas Wahl:} Conceptualization,  Formal Analysis, Investigation, Methodology,  Project administration, Software, Visualization, Writing – review \& editing.
\textbf{Florian Kamp:} Conceptualization, Data curation, Writing – review \& editing.
\textbf{Erik van der Bijl:} Conceptualization, Data curation, Writing – review \& editing.
\textbf{Arthur Jr. Galapon:} Conceptualization, Data curation, Formal Analysis, Investigation, Resources, Visualization, Writing – review \& editing.
\textbf{Christopher Kurz:} Conceptualization, Data curation, Software, Writing - review \& editing, Validation.
\textbf{Guillaume Landry:} Conceptualization, Data curation, Software, Writing - review \& editing, Validation.
\textbf{Adrian Thummerer:} Conceptualization, Investigation, Data curation, Software, Methodology, Validation, Funding acquisition, Resources, Writing – original draft, Writing – review \& editing.
\textbf{Matteo Maspero:} Conceptualization, Investigation, Data curation, Software, Methodology, Validation, Funding acquisition, Project administration, Resources, Supervision, Writing – original draft, Writing – review \& editing.

\subsection*{Participants}
All the participants: Methodology, Software, Writing – review \& editing.\\
\textbf{KoalAI} (Bowen Xin, Zhengxiang Sun, Hang Min, Gregg Belous, Jason Dowling),
\textbf{ImagePasNet} (Javier Sequeiro Gonzalez, Arthur
Longuefosse, Miguel Diaz, Alvaro Garcia Martin, Fabien Baldacci),
\textbf{BreizhCT} (Valentin Boussot, Cédric Hemon,Jean-Claude Nunes, Jean-Louis Dillenseger),
\textbf{MixCT} (Siyuan Mei, Yan Xia, Fuxin Fan),
\textbf{Et} (Zhiyuan Zhang, Jinghua Cai), 
\textbf{QWER} (Bing Han, Zuopeng Tan),
\textbf{DeepSyn} (Xuewen Hou, Jing Ni, Hongcheng Yang),
\textbf{Faking\_it} (Nikou Moradi, Gijs Luijten, Behrus Puladi, Jens Kleesiek, Victor Alves, Jan Egger, André
Ferreira),
\textbf{MEVIS} (Daniel Mensing,  Stefan Heldmann),
\textbf{HiLab} (Xianhao Zhou,  Ku Zhao, Shaoting Zhang, Guotai Wang),
\textbf{sk} (Satoshi Kondo, Satoshi Kasai),
\textbf{SEU \& Rennes team} (Yahui Liu, Songyue Han, Jiasong Wu)

\textbf{imi-graz} (Arnela Hadzic, Simon Johannes
Joham, Martin Urschler),
\textbf{RicardoBrioso} (Ricardo Coimbra Brioso, Damiano Dei,Nicola Lambri, Pietro Mancosu, Marta Scorsetti, Daniele Loiacono),
\textbf{SynthRADShu} (Junfeng Liu, Qi Wang, Yuhan Tang),
\textbf{abdullahshazli18 } (Abdullah Shazly, Abbas Mohamed Rezk, Abdulkhalek Al-Fakih, Daniel Kim, Kanghyun Ryu, Mohammed A. Al-masni),
\textbf{Sirus} (Wenzhou Xia, Zhe Xiong, Xiaoqun Zhang, Qiaoqiao Ding, Wenxiang Ding, Yihao Wang)

\section*{Acknowledgments}
The organizers would like to thank Cornelis (Nico) AT van den Berg via the AI Lab Imaging and Image-Guided Interventions \url{https://3ai.umcutrecht.nl/pillars/research/ai-lab-imaging-and-image-guided-interventions/} for their support of the challenge prizes.
We thank the European Society of Radiation Oncology (ESTRO) \url{https://www.estro.org/}, the European Federation of Organisations for Medical Physics  (EFOMP) \url{https://www.efomp.org/}, the Nederlandse Vereniging voor Klinische Fysica (NVKF) \url{https://www.nvkf.nl/}, the Deutsche Gesellschaft f\"ur Medizinische Physik e.V. (DGPM) \url{https://www.dgmp.de/}, the Svensk F\"orening för Radiofysik \url{https://www.radiofysik.org/}, the Swedish Society of Medical Physics (SSHF)  \url{https://sjukhusfysiker.se/} for endorsing the event.

The organizers thank SASHIMI2025 \url{https://2025.sashimiworkshop.org/} for sharing the satellite event at MICCAI2025.

\section*{Funding}
The SynthRAD2025 challenge was funded by a grant from ``Stiftungen zu Gunsten der Medizinischen Fakultät der Ludwig-Maximilians-Universität M\"unchen" awarded to Adrian Thummerer to support the computation costs. Adrian Thummerer was funded by a grant from Deutsche Krebshilfe (70114849). None of the centers received compensation for sharing the dataset.

Prizes were supported by the AI Lab Imaging and Image-Guided Interventions \url{https://3ai.umcutrecht.nl/pillars/research/ai-lab-imaging-and-image-guided-interventions/}.

\section*{Declaration of generative AI}
While preparing this work, the authors used ChatGPT 5.2 and Claude Sonnet 4.6 to enhance the writing structure and refine grammar. After using these tools, the authors reviewed and edited the content as needed and took full responsibility for the publication's content.

\section*{Appendix: Participation rules and prize policies}
To ensure fairness and transparency in SynthRAD2025, organizers, data providers, and contributors were prohibited from participating in the challenge, as data providers and organizers had access to the data, including the test-set ground-truth CTs. However, members affiliated with the organizers' institutes were allowed to participate, provided they had not co-authored any publications with the organizers in the year preceding the challenge.

Participants were required to develop fully automated methods that run in the Amazon Web Services (AWS) cloud environment using a single \texttt{g4dn.2xlarge} instance. This instance includes a GPU with 16 GB VRAM, an 8-core CPU, and 32 GB RAM. In this environment, the inference time to generate an sCT for a single case (one patient) is capped at 15 minutes.

Teams receiving a prize had to present their methodology at MICCAI 2025, sign all necessary prize acceptance documents, and submit a detailed paper in LNCS format outlining their methods. Additionally, participants committed to citing both the data challenge paper \citep{thummerer2025synthrad2025} and this challenge overview paper in subsequent publications, whether scientific or non-scientific. Although sharing codes was strongly encouraged, it was not mandatory. The challenge results and rankings were publicly announced after the test phase concluded. The top three teams for both tasks were awarded a total of €3,500, with the following distribution: €875, €550, and €325.

The complete challenge design can be found at \citep{Thummerer_2024_14051075}.

\bibliographystyle{model2-names.bst}\biboptions{authoryear}
\bibliography{refs}


\newpage
\onecolumn
\setcounter{section}{1}
\input{07_suppA}
\newpage
\clearpage
\setcounter{section}{1}
\setcounter{subsection}{0}
\renewcommand{\figurename}{Supplementary Fig.}
\setcounter{figure}{0}    
\input{08_suppB}

\clearpage
\newpage
\setcounter{section}{1}
\setcounter{subsection}{0}
\renewcommand{\figurename}{Supplementary Fig.}
\setcounter{figure}{0}    
\input{09_suppC}

\end{document}

%% file: 01_Introduction.tex
\section{Introduction}
\label{sec:introduction}

Radiotherapy treatment planning relies on accurate electron density mapping derived from computed tomography (CT)  to enable precise radiation dose calculations \citep{dobbs1983use}. CT provides  Hounsfield unit (HU) values required for tissue characterization. However, CT presents several limitations, including limited soft-tissue contrast compared to magnetic resonance imaging (MRI), additional ionizing radiation exposure, and logistical constraints that impede the implementation of adaptive radiation therapy (ART) workflows \citep{glide2021adaptive, sonke2019adaptive}. As ART emerges as a paradigm shift in modern radiation oncology \citep{Jaffray2012}, allowing adjustments of the daily treatment plan based on anatomical and physiological changes, e.g., tumor response or organ motion, reliance on CT-based registration introduces substantial workflow inefficiencies and geometric uncertainties that limit the potential benefits of real-time treatment adaptation. MRI integration has substantially improved soft-tissue visualization for target delineation. At the same time, cone-beam CT (CBCT) has enabled daily three-dimensional imaging at treatment units for precise positioning and real-time adaptation \citep{schmidt2015radiotherapy}. However, neither modality provides the accurate electron-density information required for dose calculation, necessitating the registration of planning CT images. Synthetic CT (sCT) generation addresses these limitations by transforming MRI or CBCT  into CT-equivalent images with accurate HU values, enabling MRI-only planning workflows and direct dose calculations in adaptive radiotherapy workflows based on daily images \citep{Edmund2017, spadea2021sctreview, rusanov2022cbctreview}. The field has evolved from conventional approaches, including atlas-based methods and bulk density assignment \citep{spadea2021sctreview}, to sophisticated deep learning architectures, including convolutional neural networks (CNNs) \citep{huijben2024generating}, generative adversarial networks (GANs) \citep{gan,CycleGAN2017,pix2pix2017}, U-Net variants \citep{unet}, diffusion \citep{diffusion}, and transformer-based models \citep{transformer}. These methods have demonstrated promising performance and clinically acceptable dosimetric accuracy in single-institution studies. Despite methodological advances, fundamental challenges impede widespread clinical implementation. The generalizability of algorithms across heterogeneous imaging protocols, scanner manufacturers, and institutional parameters remains inadequately characterized. Robustness to anatomical variations, pathological alterations, and imaging artifacts requires systematic evaluation. Computational efficiency becomes critical in online adaptive workflows with strict time constraints. \\
Furthermore, conventional evaluation paradigms that emphasize image similarity metrics show limited correlation with dosimetric accuracy, the clinically relevant endpoint. The field requires standardized, multi-institutional benchmark datasets and comprehensive evaluation frameworks to rigorously assess clinical readiness. The SynthRAD 2023 Grand Challenge addressed this gap by establishing the first large-scale, multi-institutional benchmark, consisting of more than 1,000 paired datasets of brain and pelvic anatomies from three Dutch medical centers \citep{huijben2024generating,Thummerer2023synth}. Results from the challenge revealed a negligible correlation between image similarity metrics and dose-distribution accuracy, highlighting the critical need for dosimetric evaluation in clinical validation. The SynthRAD 2025 Grand Challenge, held at MICCAI 2025, expanded this benchmark by increasing the dataset scale, anatomical diversity, and institutional heterogeneity. SynthRAD 2025 encompasses three new anatomical regions (head and neck, thorax, and abdomen) representing sites with increased complexity due to respiratory motion, intricate organ geometry, greater inter-patient variability, and diverse pathological presentations. The challenge focused on evaluating MRI-to-CT synthesis for MRI-only and MR-guided radiotherapy (Task 1), and CBCT-to-CT synthesis for image-guided and online adaptive radiotherapy (Task 2). Building on the SynthRAD 2023 evaluation criteria, the 2025 edition expanded the assessment protocol to include additional metrics, such as segmentation-based measures and a perceptual image-quality metric \cite{wang2003multiscale}. This study reports on challenge participation, algorithm evaluation, and ranking. It further examines trends among submitted algorithms and their relationship to overall performance, with a particular focus on dataset variability, evaluation metric choice, and ranking stability.

%% file: 02_Methods.tex
\section{Material and methods}
\subsection{Challenge setup}
\begin{figure*}[t]
\centering
\includegraphics[width=0.8\linewidth]{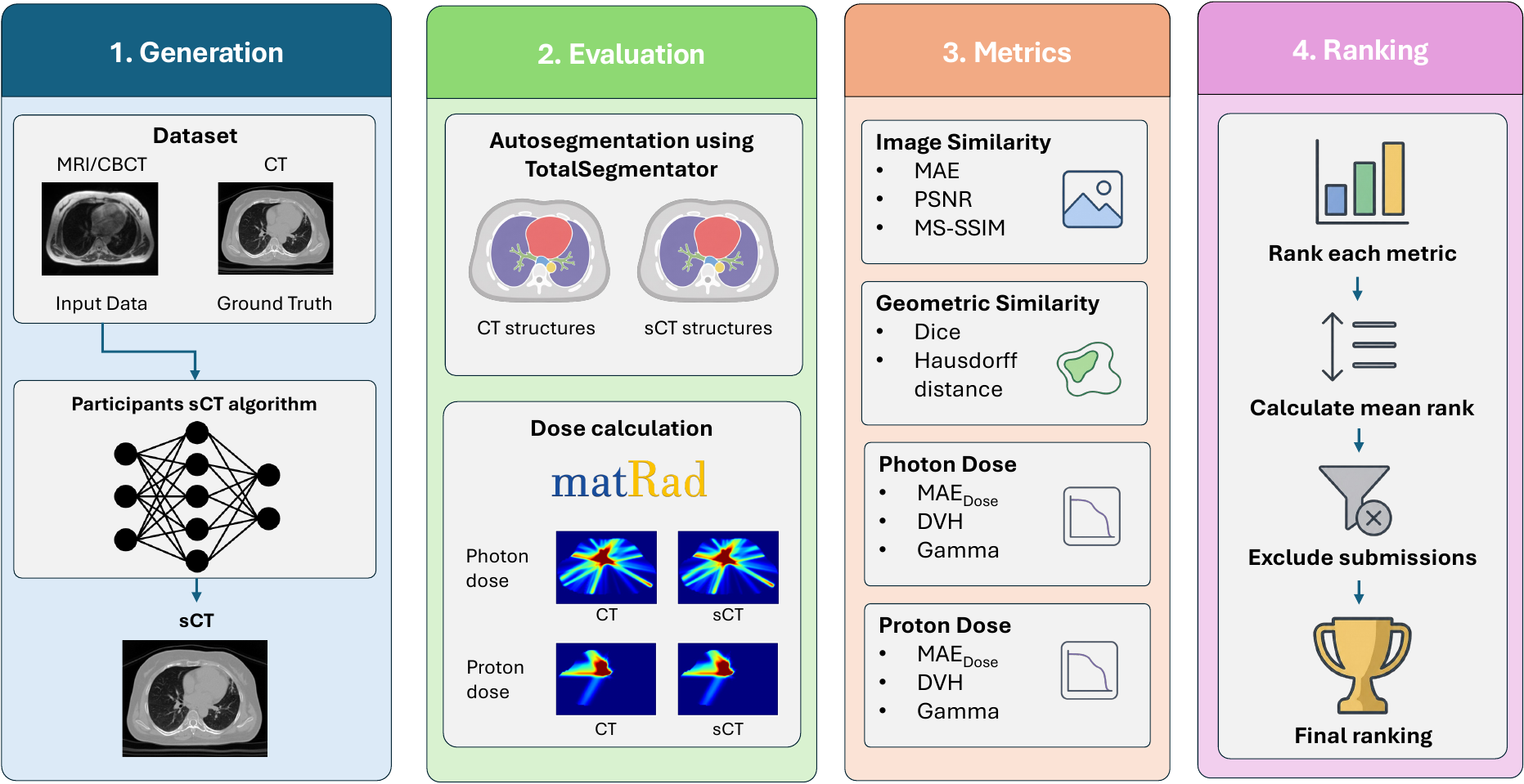}
\caption{\textbf{Overview of the SynthRAD2025 challenge pipeline}. The workflow consists of four main stages: (1) Synthetic CT Generation using submitted sCT algorithms trained with MRI/CBCT input data and ground-truth CT; (2) Evaluation pipeline, including automatic segmentation of CT and sCT structures using TotalSegmentator and photon and proton dose calculations using matRad; (3) Metrics calculations, assessing performance via image similarity, geometric similarity, and dose-based metrics for both photon and proton plans; and (4) Ranking, where metrics are ranked iniviudally, averaged, and used to determine final standings after exclusion criteria are applied.}
\label{Fig:pipeline}
\end{figure*}

The SynthRAD2025 challenge was hosted on the Grand Challenge platform (\url{https://synthrad2025.grand-challenge.org/}) and organized in conjunction with MICCAI 2025. The objective of the challenge was to develop fully automated methods to generate accurate synthetic CT (sCT) images from MRI or cone-beam CT (CBCT). An overview of the challenge design, including participants' algorithm development, the evaluation pipeline, and the ranking methodology, is presented in Figure 1. \\
Consistent with the SynthRAD2023 framework, participants could compete in two distinct tasks. Task 1 focused on MRI-to-CT synthesis to facilitate MRI-only planning and MR-guided radiotherapy workflows. Task 2 addressed CBCT-to-CT synthesis to enable high-fidelity dose calculation for CBCT-based adaptive radiotherapy.
Participants could engage in either task independently or both simultaneously. For each task, algorithms were required to generate sCTs for three anatomical regions: the head and neck, thorax, and abdomen. There were no constraints on model architecture; participants could implement either a single, unified model capable of handling all regions or separate region-specific models. \\
In a departure from the SynthRAD2023 protocol, SynthRAD2025 permitted the use of external, publicly available datasets released before March 1, 2025, provided their use was explicitly disclosed. The use of private datasets or pre-trained weights not publicly accessible by this deadline was prohibited. To foster transparency and reproducibility, the top three teams in each task were required to release their source code, weights or model publicly following the challenge.
All submissions had to be containerized using Docker for automatic evaluation within the Grand Challenge computational environment. Inference was required to run on an AWS g4dn.2xlarge instance equipped with a single 16GB VRAM GPU, 8-core CPU, and 32 GB system RAM. The maximum allowed inference time per case was 20 minutes, reflecting clinical feasibility constraints for adaptive radiotherapy workflows and accounting for Grand Challenge system overhead.
The challenge was conducted across multiple phases to support algorithm development and evaluation:
\begin{itemize}
  \item \textbf{Training Phase} (March 1, 2025 – August 18, 2025): Release of the training dataset to facilitate algorithm development.

  \item \textbf{Preliminary Test Phase} (May 1, 2025 – August 18, 2025): A period for participants to familiarize themselves with the Grand Challenge submission platform and Docker containerization requirements.
  
  \item \textbf{Validation Phase} (May 1, 2025 – August 18, 2025): Submission of sCT images for automated evaluation on validation cases. Participants were permitted up to two submissions per week; image similarity and geometric consistency metrics were displayed on a public leaderboard.
  
  \item \textbf{Test Phase} (September 16, 2025 – October 8, 2025): Submission of final containerized algorithms. To mitigate potential technical failures, a backup submission was permitted alongside the primary entry. This phase included online inference and a comprehensive evaluation across all metrics (image similarity, geometry, and dose); only the final successful submission was considered for the official ranking.

  \item \textbf{Post-Challenge Phase} (Ongoing until March 1, 2030): The platform remains open for algorithm inference and evaluation, with a persistent leaderboard maintained for continued benchmarking.
\end{itemize}

Participants achieving a top-five ranking in both tasks were invited to serve as co-authors of this challenge report. They were required to submit a detailed algorithm description in accordance with the provided checklist and format \url{https://synthrad2025.grand-challenge.org/submissions/}. The winning teams presented their methods at the SASHIMI workshop on October 23, 2025 \url{https://2025.sashimiworkshop.org/}. To ensure transparency and reproducibility, all preprocessing, evaluation, and ranking code was made publicly available \url{https://github.com/SynthRAD2025}.

\subsection{Dataset}
The SynthRAD2025 challenge dataset comprises 2362 paired imaging datasets from patients undergoing radiotherapy, collected at five European university medical centers: University Medical Center Groningen, University Medical Center Utrecht, Radboud University Medical Center, LMU University Hospital Munich, and University Hospital of Cologne. In total, the multi-institutional cohort comprises 890 MRI–CT pairs for Task 1 and 1,472 CBCT-CT pairs for Task 2, spanning three anatomical regions: head and neck (HN), thorax (TH), and abdomen (AB).
The study and data-sharing protocols were approved by the Institutional Review Board or Medical Ethics Committee at each participating institution, in accordance with local regulations. All imaging data underwent anonymization and defacing before inclusion in the dataset. Due to ethical considerations and varying data protection regulations across institutions and countries, access to detailed patient characteristics may be restricted. Consequently, demographic and clinical metadata-such as age, sex, tumor type, and disease staging-were not uniformly available across the cohort. No exclusion criteria were applied regarding patient demographics or tumor characteristics during data collection. 
The dataset was released under the Creative Commons Attribution-Non-Commercial license (CC BY-NC 4.0 International) and made publicly accessible via Zenodo, however the Center D data was released under a limited-use license that permits use only within the context of the challenge. Detailed information on dataset access, licensing, and release dates is provided in the SynthRAD 2025 dataset paper.

Data preprocessing included resampling to a uniform voxel spacing of 1 × 1 × 3 mm³, defacing all image datasets that contained the head region, and generation of a body contour mask (\(\mathcal{B}\)). More specifically, this mask corresponds to the provided binary patient-outline mask and includes a safety margin through dilation of the contour, such that evaluation is restricted to the anatomically relevant region while avoiding instability at the body boundary. For validation and test cases, deformable image registration was performed to minimize anatomical differences between CBCT/MRI and the ground-truth CT, and metric calculation was performed within this mask on the deformed reference CT. For additional details on image acquisition protocols and preprocessing, we refer readers to the previously published dataset description \citep{thummerer2025synthrad2025}.

\subsection{Baseline algorithms}
A simple bulk-density baseline was used as a reference method for evaluation. In this baseline method, voxels outside the body contour were assigned -1000 HU (air), while voxels inside the body contour were assigned 0 HU (water). This baseline provides a lower bound on performance under the assumption of homogeneous water equivalence.

\subsection{Evaluation}
\label{subsec: evaluation method}

Algorithm performance was quantified using a comprehensive evaluation framework encompassing image similarity, dose accuracy, and geometric consistency assessment. 

\subsubsection{Image similarity}
Image similarity was evaluated within the body contour masks. The following metrics quantified pixel-wise correspondence between sCT and ground-truth CT, thereby assessing how closely the generated sCT resembled the reference CT. The image similarity metrics used were mean absolute error (MAE), peak signal-to-noise ratio (PSNR), and multi-scale structural similarity index measure (MS-SSIM) \citep{wang2003multiscale}.

\textbf{MAE} was calculated to measure the average absolute difference between corresponding voxels in the sCT and CT, and was defined as
\begin{equation}
\operatorname{MAE}(\text{CT}, \text{sCT}) = \frac{1}{|\mathcal{B}|}\sum_{i\in \mathcal{B}} \left\vert\text{CT}_i - \text{sCT}_i\right\vert
\end{equation}
in which we sum over the voxels inside the body contour \(\mathcal{B}\) and normalized by the total number of masked voxels \(\,| \mathcal{B} \,|\).

\textbf{PSNR} was calculated to quantify the ratio of maximum signal intensity over the noise level in the sCT compared to the CT, defined as

\begin{equation}
\text{PSNR}(\text{CT}, \text{sCT}) = 10 \log_{10}\left(\frac{Q^2}{\frac{1}{|\mathcal{B}|}\sum_{i\in \mathcal{B}}\left(\text{CT}_i -\text{sCT}_i\right)^2}\right),
\end{equation}
where \(Q\) is the dynamic range of the voxel intensities ([\(-1024\), \(3000\)] HU). CT and sCT were clipped to this dynamic range to calculate the masked PSNR.

\textbf{ MS-SSIM} was calculated to assess structural similarity between CT and sCT. The MS-SSIM map between two images \(sCT\) and \(CT\) is computed by 
\begin{multline}
    \operatorname{MS-SSIM} = \left[\ell\left(\text{CT}_M,\text{sCT}_M\right)\right]^{\alpha_M}\\
    \prod_{j=1}^M\left[c\left(\text{CT}_j,\text{sCT}_j\right)\right]^{\beta_j}\left[s\left(\text{CT}_j,\text{sCT}_j\right)\right]^{\gamma_j},
\end{multline} computing a weighted SSIM at multiple resolution scales \(j\). At \(j=1\), the full-resolution image is used. The image at resolution scale \(j\) is obtained by applying a low-pass filter to the image at resolution scale \(j-1\), followed by a downsampling operation with a factor of 2. In total, \(M=5\) resolution scales were used with \(\beta_1=\gamma_1=0.0448\), \(\beta_2=\gamma_2=0.2856\), \(\beta_3=\gamma_3=0.3001\), \(\beta_4=\gamma_4=0.2363\), and \(\alpha_5 = \beta_5=\gamma_5=0.1333\) \citep{wang2003multiscale}.
The \(\ell\), \(c\), and \(s\) components of the MS-SSIM are defined as \begin{align*}
    \ell(x_i,y_i) &= \frac{2\mu_x^i\mu_y^i+C_1}{(\mu_x^i)^2+(\mu_y^i)^2+C_1},\\
    c(x_i,y_i) &= \frac{2\sigma_x^i\sigma_y^i+C_2}{(\sigma_x^i)^2+(\sigma_y^i)^2+C_2}\text{, and}\\
        s(x_i,y_i) &= \frac{\sigma_{xy}^i+C_3}{\sigma_x^i\sigma_y^i+C_3}
\end{align*}
where \(\mu_x^i\) and \(\sigma_x^i\) are the mean and variance, respectively, of \(x\) within an \(N\times N\times N\) window centered on voxel \(i\) and \(\sigma_{xy}^i\) is the covariance of \(x\) and \(y\) within an \(N\times N\times N\) window centered on voxel \(i\). \(N=7\) is the window size, and \(C_1 = (0.01 \cdot L)^2\), \(C_2 = (0.03 \cdot L)^2\), and \(C_3=C_2/2\) are normalization constants, where \(L = (3000 - (-1024))\) HU is the dynamic range of the volumes.

The final masked SSIM value is then obtained by computing 
\begin{equation}
    \operatorname{MS-SSIM}(\text{CT}, \text{sCT}) = \frac{1}{|\mathcal{B}|}\sum_{i\in \mathcal{B}} \operatorname{MS-SSIM}_{i}(\text{CT}, \text{sCT}),
\end{equation}
where the intensities of both the CT and sCT were clipped to [\(-1024\), \(3000\)] HU and then adjusted to be non-negative by adding 1024 HU. 


\subsubsection{Dose accuracy}
Photon and proton intensity-modulated treatment plans were optimized using the open-source matRad treatment planning system \cite{wieser2017development} based on the deformed reference CT. Dose was prescribed to the planning target volume (PTV) at the 95\% isodose level for both modalities, without robust optimization for proton plans. Specific dose prescriptions were defined for each anatomical region: 30 $\times$ 2.0~Gy for head and neck and abdomen, and 35 $\times$ 2.0~Gy for thorax cases. The center of the selected PTV was chosen as the isocenter for all plans. PTV selection was primarily performed automatically using heuristic rules based on structure naming conventions. In cases with ambiguous or inconsistent naming, manual review and correction were performed when necessary. Co-planar plans were considered, with photon plans using 9 equi-angled 6 MV beams from a generic Linac model, with the same beam angles applied across all treatment sites. Proton plans using three co-planar beams from a generic proton system available in matRad. The proton beam angles were selected per treatment site (head-and-neck, thoracic, and abdominal) and mirrored depending on isocenter position to preferentially use lower-range beams. Beam configurations were therefore standardized per site rather than optimized on an individual patient basis to comply with dose prescriptions and limit the dose to organs at risk following international guidelines for head-and-neck (\cite{deasy2010radiotherapy}, \cite{beetz2013role}, \cite{marks2010use}), lung (\cite{puckett2023consensus}), and abdomen (\cite{marks2010use}). To reduce the dose to healthy tissues and to ensure plan uniformity across patients, we used the same objective functions and constraints available in matRad for each treatment site. OAR dose limits were treated as hard constraints whenever possible and were revised on a patient-specific basis when hard constraints were not achievable. All planning goals and OAR dose limits are summarized in Table~\ref{tab:dose constraints_HeadandNeck}, ~\ref{tab:dose constraints_Lung}, and ~\ref{tab:dose constraints_Abdomen}.

\begin{table}[ht]
\centering
\footnotesize
\caption{\label{tab:dose constraints_HeadandNeck}Dose constraints and planning objectives used in matRad for the head and neck cases. Values are taken from (\cite{deasy2010radiotherapy}, \cite{beetz2013role} and \cite{marks2010use}).}
\begin{tabular}{p{3cm}p{3cm}}
\hline
\multicolumn{2}{c}{\textbf{Head and Neck}} \\
\multicolumn{2}{c}{$30 \times 2$ Gy (PTV D$_{95\%}$)} \\
\hline
\textbf{Structure} & \textbf{Constraint} \\
\hline

Spinal cord & $D_{0.03\,\mathrm{cc}} < 48$ Gy \\
Brainstem & $D_{0.03\,\mathrm{cc}} < 54$ Gy \\
Optic chiasm & $D_{0.03\,\mathrm{cc}} < 54$ Gy \\
Optic nerve & $D_{0.03\,\mathrm{cc}} < 54$ Gy \\
Retina & $D_{0.03\,\mathrm{cc}} < 45$ Gy \\
Parotid glands & $D_{\mathrm{mean}} < 25$ Gy \\
Larynx & $D_{\mathrm{mean}} < 40$ Gy \\
Submandibular glands & $D_{\mathrm{mean}} < 39$ Gy \\
Oral cavity & $D_{\mathrm{mean}} < 24$ Gy \\
Pharyngeal constrictors & $D_{\mathrm{mean}} < 50$ Gy \\
Cricopharyngeal muscles & $D_{\mathrm{mean}} < 30$ Gy \\
Carotid arteries & $D_{0.03\,\mathrm{cc}} < 40$ Gy \\
Lacrimal gland & $D_{\mathrm{mean}} < 25$ Gy \\
Lens & $D_{0.03\,\mathrm{cc}} < 5$ Gy \\
Cochlea & $D_{0.03\,\mathrm{cc}} < 40$ Gy \\
Pituitary gland & $D_{\mathrm{mean}} < 30$ Gy \\
Masseter muscle & $D_{\mathrm{mean}} < 35$ Gy \\
Esophagus & $D_{\mathrm{mean}} < 30$ Gy \\
Thyroid & $D_{\mathrm{mean}} < 40$ Gy \\

\hline
\end{tabular}
\end{table}

\begin{table}[ht]
\centering
\footnotesize
\caption{\label{tab:dose constraints_Lung}Dose constraints and planning objectives used in matRad for the lung cases. Values are taken from (\cite{puckett2023consensus}).}
\begin{tabular}{p{3cm}p{3cm}}
\hline
\multicolumn{2}{c}{\textbf{Lung}} \\
\multicolumn{2}{c}{$35 \times 2$ Gy (PTV D$_{95\%}$)} \\
\hline
\textbf{Structure} & \textbf{Constraint} \\
\hline

Spinal cord & $D_{0.03\,\mathrm{cc}} < 50.5$ Gy \\
Brainstem & $D_{0.03\,\mathrm{cc}} < 54$ Gy \\

\hline
Lungs (total–GTV) & $V_{60\,\mathrm{Gy}} < 33\%$ \\
 & $V_{5\,\mathrm{Gy}} < 33\%$ \\
 & $D_{\mathrm{mean}} < 18$ Gy \\

\hline
Brachial plexus & $D_{0.03\,\mathrm{cc}} < 66$ Gy \\

\hline
Esophagus & $D_{0.03\,\mathrm{cc}} < 73.5$ Gy \\
 & $V_{60\,\mathrm{Gy}} < 15.3\%$ \\
 & $D_{\mathrm{mean}} < 30.6$ Gy \\

\hline
\end{tabular}
\end{table}

\begin{table}[ht]
\centering
\footnotesize
\caption{\label{tab:dose constraints_Abdomen}Dose constraints and planning objectives used in matRad for the abdomen cases. Values are taken from (\cite{marks2010use}).}
\begin{tabular}{p{3cm}p{3cm}}
\hline
\multicolumn{2}{c}{\textbf{Abdomen}} \\
\multicolumn{2}{c}{$30 \times 2$ Gy (PTV D$_{95\%}$)} \\
\hline
\textbf{Structure} & \textbf{Constraint} \\
\hline

Spinal cord & $D_{0.03\,\mathrm{cc}} < 50.5$ Gy \\
Brainstem & $D_{0.03\,\mathrm{cc}} < 54$ Gy \\
Liver & $D_{\mathrm{mean}} < 30$ Gy \\
Kidneys & $D_{\mathrm{mean}} < 18$ Gy \\
Stomach & $D_{0.03\,\mathrm{cc}} < 45$ Gy \\
Bladder & $D_{0.03\,\mathrm{cc}} < 65$ Gy \\

\hline
\end{tabular}
\end{table}

To evaluate the dose calculation accuracy, the dose was recalculated on each sCT for both proton and photon treatment plans using Python re-implementation of MatRad's pencil-beam algorithms in pyRadPlan (v0.2.8) \citep{pyradplan}. This recalculation was carried out without propagating organ delineations or replanning, a deliberate measure taken to avoid potential differences arising from plan optimization. Subsequently, the differences between the planning dose distributions, originally calculated on the deformed CT, and the recalculated dose distributions on the sCT for both photon and proton plans were quantified using three specific metrics. 

\textbf{Relative mean absolute dose difference} within high dose regions $\mathcal{H} = \{i \,|\, D_{\text{CT},i} \geq 0.9 \cdot D_{\text{prescribed}}\}$ was calculated to assess the dose difference around the target and was defined as
\begin{equation}
\text{MAE}_{\text{dose}} = \frac{1}{|\mathcal{H}|}\sum_{i \in \mathcal{H}}\frac{|D_{\text{CT},i} - D_{\text{sCT},i}|}{D_{\text{prescribed}}},
\end{equation}
with $D_{(\text{s})\text{CT}}$ being the dose distribution in the (s)CT and $D_{\text{prescribed}}$ the prescribed dose.

\textbf{Dose-volume histogram (DVH)} parameters were calculated to assess the differences in the dose to the PTV and OARs. A DVH describes the relationship between dose and the volume of a given structure receiving that dose. Small differences in DVH parameters between CT and sCT indicate good agreement and thus reliable treatment planning based on the sCT. Four DVH parameters were analyzed: the near-minimum dose to the PTV, $D98_{\text{PTV}}$, defined as the dose received by at least 98\% of the target volume received; the proportion of the PTV receiving at least 95\% of the prescribed dose, $V95_{\text{PTV}}$; the near-maximum dose to a selected OAR, $D2_{\text{OAR}}$, corresponding to the dose received by 2\% of the OAR volume; and the mean dose to a selected OAR $D\text{mean}_{\text{OAR}}$. The use of the near-minimum and near-maximum dose metrics was suggested by ICRU83 \citep{Gregoire2011} (\url{https://www.fnkv.cz/soubory/216/icru-83.pdf}). For evaluation, the three OARs closest to the PTV were selected for each anatomical region. Relative absolute differences were calculated for all DVH parameters as defined by
\begin{equation}
D98_{\text{PTV}} = \frac{|D98_{\text{PTV,CT}} - D98_{\text{PTV,sCT}} + \epsilon|}{D98_{\text{PTV,CT}} + \epsilon},
\end{equation}
\begin{equation}
V95_{\text{PTV}} = \frac{|V95_{\text{PTV,CT}} - V95_{\text{PTV,sCT}} + \epsilon|}{V95_{\text{PTV,CT}} + \epsilon},
\end{equation}
\begin{equation}
D2_{\text{OARs}} = \frac{1}{n_{\text{OARs}}}\sum_{\text{OAR}} \frac{|D2_{\text{OAR,CT}} - D2_{\text{OAR,sCT}} + \epsilon|}{D2_{\text{OAR,CT}} + \epsilon},
\end{equation}
\begin{equation}
D\text{mean}_{\text{OARs}} = \frac{1}{n_{\text{OARs}}}\sum_{\text{OAR}} \frac{|D\text{mean}_{\text{OAR,CT}} - D\text{mean}_{\text{OAR,sCT}} + \epsilon|}{D\text{mean}_{\text{OAR,CT}} + \epsilon},
\end{equation}
where $\epsilon = 1\text{e}{-12}$ to avoid division by zero and $n_{\text{OARs}} = 3$ is the number of OARs. We summed the four terms to obtain one final value for the DVH metric.

\textbf{Gamma pass rates} were calculated to compare 3D dose distributions between sCT and reference CT. This calculation followed the 3D gamma pass rate approach described by \citep{low1998gamma} with a dose-difference criterion ($\Delta D$) of 2\% and a distance-to-agreement criterion ($\Delta d$) of 2 mm. The gamma index at each position vector $\mathbf{r}$ in the sCT is:
\begin{equation}
\gamma(\mathbf{r}) = \min_{\{\mathbf{r}_{\text{CT}}\}} \sqrt{\frac{|\mathbf{r} - \mathbf{r}_{\text{CT}}|^2}{\Delta d^2} + \frac{|D_{\text{sCT}}(\mathbf{r}) - D_{\text{CT}}(\mathbf{r}_{\text{CT}})|^2}{\Delta D^2}},
\end{equation}
where $\mathbf{r}_{\text{CT}}$ indicates a position vector in the CT. Gamma pass rates were evaluated within regions receiving doses $\geq 10\%$ of the prescribed dose $D_{\text{prescribed}}$, as suggested by \citep{ezzell2009imrt}.

\subsubsection{Geometric consistency}

Geometric consistency evaluation was performed to assess the preservation of anatomical structures between the reference CT and sCT. Both CT and sCT images were automatically segmented using TotalSegmentator v1.5.7 \citep{wasserthal2023totalsegmentator}, a deep learning-based tool that identifies 104 anatomical structures. 
The evaluation focused on site-specific subsets of these structures:
\begin{itemize}
    \item Head and neck: included the brain, skull, spinal cord, trachea, esophagus, thyroid gland, and vertebrae.
    \item Thorax and abdomen: included the lungs, heart, liver, stomach, kidneys, spinal cord, vertebrae, ribs, and sternum.
\end{itemize}
The segmentations were used to compute two metrics that quantify volumetric overlap and surface-distance accuracy \citep{taha2015metrics} when the structures were present in the CT.

\textbf{Multi-class Dice Similarity Coefficient (mDice)} was calculated by averaging the Dice coefficient across all successfully segmented anatomical structures, defined as
\begin{equation}
\text{mDice} = \frac{1}{N}\sum_{i=1}^{N} \frac{2|\text{Seg}_{\text{CT},i} \cap \text{Seg}_{\text{sCT},i}|}{|\text{Seg}_{\text{CT},i}|+|\text{Seg}_{\text{sCT},i}|},
\end{equation}
where $N$ represents the number of anatomical structures segmented, $\text{Seg}_{\text{CT},i}$ denotes the binary segmentation mask for structure $i$ in the reference CT, and $\text{Seg}_{\text{sCT},i}$ represents the corresponding mask in the sCT. Higher mDice values indicate better volumetric agreement between corresponding structures.

\textbf{95th percentile Hausdorff Distance (HD95)} was calculated to quantify surface boundary agreement between corresponding anatomical structures, defined as
\begin{equation}
\text{HD95} = \frac{1}{N}\sum_{i=1}^{N} \text{HD95}(\text{Seg}_{\text{CT},i}, \text{Seg}_{\text{sCT},i}),
\end{equation}
where $\text{HD95}(\text{Seg}_{\text{CT},i}, \text{Seg}_{\text{sCT},i})$ represents the 95th percentile of the bidirectional Hausdorff distance for structure $i$. The bidirectional Hausdorff distance is computed as
\begin{equation}
\text{HD}(A,B) = \max\{h(A,B), h(B,A)\},
\end{equation}
where
\begin{equation}
h(A,B) = \max_{a \in \partial A} \min_{b \in \partial B} \|a - b\|,
\end{equation}
with $\partial A$ and $\partial B$ denoting the surface boundaries of segmentations $A$ and $B$, respectively. The 95th percentile was used to provide robustness against outliers. Lower HD95 values indicate more accurate boundary preservation.

For each patient, only structures successfully segmented by TotalSegmentator in CT were included in the calculation. The patient-level mDice and HD95 were computed as the average across all valid structures.

\subsection{Eligibility and ranking}
\label{subsec: ranking method}
For each submission, evaluation metrics were calculated per case and aggregated across the test cohort, reporting the mean and standard deviation ($\mu \pm \sigma$). To ensure clinical relevance, minimum performance criteria were established for inclusion in the final ranking. Teams were excluded if their method failed to outperform the bulk-density baseline in at least one of the three image similarity metrics (MAE, PSNR, or MS-SSIM). Additionally, computational feasibility constraints were imposed, requiring that sCT generation for a single case be completed within 15 minutes on the Grand Challenge platform.

The ranking methodology differed from SynthRAD2023 and followed a rank-then-mean approach rather than a mean-then-rank strategy. For each submission, the mean metric value across all test cases was first computed for the eleven evaluation metrics: MAE, PSNR, MS-SSIM, mDice, HD95, photon MAE$_{\text{dose}}$, photon DVH, photon gamma pass rate, proton MAE$_{\text{dose}}$, proton DVH, and proton gamma pass rate. Each metric was subsequently ranked independently across all submissions, assigning rank 1 to the best-performing submission and rank $n$ to the worst-performing submission for that metric. The overall ranking of each submission was then obtained by averaging its rank across all eleven metrics.

The rank-then-mean approach was used for SynthRAD2025 based on findings from the SynthRAD2023 challenge report, which showed that aggregating metric-specific ranks by their mean provides stable rankings without requiring arbitrary normalization across metrics with different units and scales \cite{huijben2024generating}.

\subsection{Analysis}
\subsubsection{Overall sCT performance}
In addition to reporting aggregated performance metrics per submission ($\mu \pm \sigma$), we assessed whether performance differences between teams were statistically significant at the case level. Pairwise comparisons were conducted using the Wilcoxon signed-rank test \citep{wilcoxon1945}, with Holm correction applied to account for multiple testing \citep{holm1979simple}. This analysis was performed independently for each evaluation metric, enabling a detailed assessment of relative team performance. Statistical significance was defined at $\alpha = 0.05$.

\subsubsection{Model design predictors}
\label{subsec:model_design_pred}
Participants submitted a form for each task, detailing additional pre- and post-processing steps, the architecture used, the supervision approach, the spatial configuration, data augmentation techniques, training details (use of pre-trained models, loss function, optimizer), and hardware and training time. The form templates can be found in the Supplementary Material section \ref{sec:suppCD}.
We analyzed whether reported design choices were associated with sCT performance. To this end, submitted methods were grouped into six categories: (1) model and anatomical strategy, (2) backbone architecture, (3) supervision, (4) spatial configuration, (5) preprocessing, and (6) training and inference strategies.
Differences in MS-SSIM across design categories were evaluated using the Mann--Whitney U test \citep{mann1947test} ($\alpha = 0.01$), a nonparametric test suitable for comparing independent samples without assuming normality.

\subsubsection{Data influence}
To assess how dataset characteristics affected performance, we summarized each evaluation metric as \(\mu\pm\sigma\) after stratifying the test cases by task, center, and anatomical region. In particular, we compared regional and center-wise distributions of MS-SSIM, dose-based and gemoetric-based metrics to evaluate differences between MRI-to-CT and CBCT-to-CT synthesis, between institutions, and between the head-and-neck, thorax, and abdomen subsets. Differences between groups were assessed using the Mann--Whitney U test \citep{mann1947test}. To further investigate heterogeneity at the case level, we additionally performed a patient-level analysis in which metric values were inspected per test case across methods to identify consistently difficult patients, outlier failures, and recurring error patterns shared between submissions.

\subsubsection{Metric correlations}
To evaluate correlations between metrics, we first defined three metric groups: image similarity metrics (MAE, PSNR, SSIM), photon dose metrics (MAE\textsubscript{dose}, DVH\textsubscript{metric}, $\gamma$), and proton dose metrics (MAE\textsubscript{dose}, DVH\textsubscript{metric}, $\gamma$).
Correlations were then analyzed both within and across these metric groups. Intra-group relationships were first assessed through visual inspection, followed by quantitative analysis using the Spearman rank correlation coefficient 
\(\rho\) \citep{spearman}. This nonparametric measure captures monotonic relationships between ranked variables and is robust to differences in scale and distribution, enabling consistent comparison across heterogeneous evaluation metrics.

\subsubsection{Ranking stability}
We assessed the stability of the final rankings at the patient level, following the recommendation of \cite{wiesenfarth2021methods}. To do so, we used bootstrapping to evaluate variability in the ranking positions of all teams. The ranking procedure was repeated across 1,000 bootstrap samples, each comprising 120 patients randomly drawn from the test set with replacement. Teams were ranked using the rank-then-mean approach, in which each metric was individually ranked and then a mean rank was calculated for each submission.

%% file: 03_Participants.tex
\afterpage{%
\clearpage
\begin{landscape}
\makeatother
\rowcolors{1}{}{lightgray}
\setlength{\tabcolsep}{4.5pt}
\begin{table}
\centering
\caption{\label{tab:summary task1}Ranking and model details for Task 1 (MRI-to-CT synthesis). A check mark (\ding{51}) indicates that the corresponding step is applied to both MRI and CT images; otherwise, the affected modality is specified. The distinctions listed under \textit{Model \& anatomy} and \textit{Backbone architecture} are described in Section~\ref{subsec:model_design_pred}. Abbreviations: FT~--~fine-tuning; Perc.~--~perceptual loss; Adv.~--~adversarial loss; FM~--~feature matching loss; Seg.~--~segmentation-based; VLB~--~variational lower bound; AFP~--~anatomical feature-prioritized loss; def.~reg.~--~deformable registration; bias~--~bias field correction; res.~--~resampling/resizing; fg.~crop~--~foreground crop; rescale~--~intensity rescaling by 1000. All values are self reported by teams in the submission form.}
\resizebox{\linewidth}{!}{%
\begin{tabular}[]{c|c|cccc|cccc|c|cccc|cccccc|cccccc|ccc|cccc|cc}
\toprule
\textbf{Rank} & \textbf{Team} &
  \multicolumn{4}{c|}{\textbf{Model \& anatomy}} &
  \multicolumn{4}{c|}{\textbf{Backbone arch.}} &
  \multicolumn{1}{c|}{\textbf{Sup.}} &
  \multicolumn{4}{c|}{\textbf{Spatial config.}} &
  \multicolumn{6}{c|}{\textbf{Preprocessing}} &
  \multicolumn{6}{c|}{\textbf{Data augmentation}} &
  \multicolumn{3}{c|}{\textbf{Postprocess.}} &
  \multicolumn{4}{c|}{\textbf{Training}} &
  \multicolumn{2}{c}{\textbf{Hardware}} \\

\rowcolor{white}
\midrule
& &
  \rotatebox{270}{Single model} &
  \rotatebox{270}{Single model, anatomy conditional} &
  \rotatebox{270}{Identical backbones, anatomy-specific training param.} &
  \rotatebox{270}{Similar models} &
  \rotatebox{270}{CNN encoder-decoder} &
  \rotatebox{270}{GAN} &
  \rotatebox{270}{Flow matching / Diffusion model} &
  \rotatebox{270}{Ensemble of multiple models} &
  \rotatebox{270}{Supervised} &
  \rotatebox{270}{3D} &
  \rotatebox{270}{3D patch-based} &
  \rotatebox{270}{2.5D} &
  \rotatebox{270}{2D} &
  \rotatebox{270}{Iso-resampling / resizing} &
  \rotatebox{270}{Clipping} &
  \rotatebox{270}{Patient level linear normalization} &
  \rotatebox{270}{Population level normalization} &
  \rotatebox{270}{Standardization} &
  \rotatebox{270}{Other (deformable reg., bias field, foreground crop, \ldots)} &
  \rotatebox{270}{Random crop / patch} &
  \rotatebox{270}{Flipping} &
  \rotatebox{270}{Blurring} &
  \rotatebox{270}{Noise addition} &
  \rotatebox{270}{Intensity transform (bias field, contrast / histogram adj.)} &
  \rotatebox{270}{Deformation (affine, elastic, zoom)} &
  \rotatebox{270}{Average overlapping patches} &
  \rotatebox{270}{Body mask masking} &
  \rotatebox{270}{Revert normalization / padding} &
  \rotatebox{270}{Pre-trained model used} &
  \rotatebox{270}{Loss function(s)} &
  \rotatebox{270}{Optimiser} &
  \rotatebox{270}{Additional public data} &
  \rotatebox{270}{GPU hardware} &
  \rotatebox{270}{Training time (h)} \\

\midrule

1 & KoalAI &
  & & \ding{51} & &
  \ding{51}& & & \ding{51} &
  \ding{51} & &
  \ding{51} & & &
  & {\small CT} & & {\small CT} & {\small MR} & {\small CT def.~reg.} &
  & & & & & &
  \ding{51} & \ding{51} & \ding{51} &
  No & Masked MSE + Masked AFP & SGD & No &
  H100 & 24 \\

2 & ImagePasNet &
  & &  \ding{51} & &
  \ding{51} & & & &
  \ding{51} &
  & \ding{51} & & &
  & {\small CT} & {\small MR} & {\small CT} & {\small MR} & {\small CT def.~reg.} &
  & & & & & &
  \ding{51} & & \ding{51} &
  No & L1 + AFP & SGD & No &
  RTX 4070/A40 & 40 \\

3 & BreizhCT &
  & \ding{51} & & &
  \ding{51} & & & &
  \ding{51} &
  & & \ding{51} & &
  & \ding{51} & & {\small CT} & {\small MR} & {\small CT def.~reg.} &
  & \ding{51} & & & & &
  & \ding{51} & \ding{51} &
  No & MSE + Perc. (SAM~2.1) & AdamW & No &
  2$\times$RTX~6000 & 120 \\

4 & MixCT &
  \ding{51} & & & &
  & \ding{51} & & &
  \ding{51} &
  & \ding{51} & & &
  & \ding{51} & {\small MR} & {\small CT} & & {\small CT def.~reg., fg.~crop} &
  \ding{51} & \ding{51} & & & & {\small zoom} &
  & & \ding{51} &
  No & MAE + Perc. + MAE regions + Adv. + FM & AdamW & No &
  A100 & 40 \\

5 & QWER &
  & & \ding{51} & &
  & & & \ding{51} &
  \ding{51} &
  \ding{51} & & & &
  & & {\small MR} & {\small CT} & & &
  & \ding{51} & & & & &
  \ding{51} & & \ding{51} &
  No & MAE + Perc. (TotalSeg.) & SGD & No &
  A800 & 320 \\

6 & DeepSyn &
  \ding{51} & & & &
  & \ding{51} & & &
  \ding{51} &
  & & \ding{51} & &
  \ding{51} & {\small CT} & \ding{51} & {\small CT} & \ding{51} & {\small MR bias, def.~reg.} &
  \ding{51} & & & & & &
  \ding{51} & \ding{51} & \ding{51} &
  No & MAE + SSIM + PSNR & Adam & No &
  H20 & $<$1 \\

7 & Faking\_it &
  \ding{51} & & & &
  & & \ding{51} & &
  \ding{51} &
  & \ding{51} & & &
  \ding{51} & \ding{51} & {\small MR} & \ding{51} & {\small MR} & &
  \ding{51} & & \ding{51} & \ding{51} & \ding{51} & \ding{51} &
  \ding{51} & & \ding{51} &
  No & MSE + MAE + SSIM + VLB & AdamW & No &
  RTX~6000 & 240 \\

8 & MEVIS &
  & & & \ding{51} &
  \ding{51} & & & &
  \ding{51} &
  & \ding{51} & & &
  \ding{51} & {\small MR} & & {\small CT} & {\small MR} & &
  & & & & & &
  \ding{51} & \ding{51} & \ding{51} &
  No & MSE + Perc. (UMedPT) & Adam & Yes &
  A100 & 50 \\

8 & HiLab &
  \ding{51} & & & &
  \ding{51} & & & &
  \ding{51} &
  & & & \ding{51} &
  \ding{51} & {\small CT} & {\small MR} & {\small CT} & {\small MR} & &
  \ding{51} & \ding{51} & & & & &
  & & \ding{51} &
  Yes (self) & L1 & Adam & Yes &
  RTX~5090 & 60 \\

10 & sk &
  & & & \ding{51} &
  \ding{51} & & & &
  \ding{51} &
  & & \ding{51} & &
  \ding{51} & & & {\small CT} & {\small MR} & {\small MR bias} &
  & & & & & &
  & & \ding{51} &
  Yes (FT) & L1 + Perc. (VGG) & AdamW & No &
  RTX~4090 & 2 \\

11 & SEU \& Rennes &
  & & & \ding{51} &
  & & \ding{51} & &
  \ding{51} &
  \ding{51} & & & &
  \ding{51} & \ding{51} & & & {\small CT} & {\small CT def.~reg.} &
  & & & & & &
  & & \ding{51} &
  No & L1 + MSE + Bone & Adam & No &
  A6000 & 12 \\

12 & imi-graz &
  & & & \ding{51} &
  & & \ding{51} & &
  \ding{51} &
  \ding{51} & & & &
  \ding{51} & \ding{51} & & & {\small MR} & {\small CT rescale} &
  & & & & & \ding{51} &
  & & \ding{51} &
  No & L1 + MSE & AdamW & No &
  A100 & 87 \\

\bottomrule
\end{tabular}%
}
\end{table}

\end{landscape}
}
\afterpage{%
\clearpage
\begin{landscape}
\makeatother
\rowcolors{1}{}{lightgray}
\setlength{\tabcolsep}{4.5pt}
\begin{table}
\centering
\caption{\label{tab:summary task2}Ranking and model details for Task 2 (CBCT-to-CT synthesis). A check mark (\ding{51}) indicates that the corresponding step is applied to both CBCT and CT images; otherwise, the affected modality is specified (CB/CT). The distinctions listed under Model \& anatomy and Backbone architecture are described in Section~\ref{subsec:model_design_pred}. Abbreviations: FT -- fine-tuning; feat. -- pre-trained model used for feature extraction only; self -- self pre-training; Perc. -- perceptual loss; Adv. -- adversarial loss; FM -- feature matching loss; Ctx. -- contextual loss; Anat. -- anatomical loss; Grad. -- gradient loss; VLB -- variational lower bound; AFP -- anatomical feature-prioritized loss; def. reg. -- deformable registration; rescale -- intensity rescaling by 1000. All values are self-reported by teams in the submission form.}
\resizebox{\linewidth}{!}{%
\begin{tabular}[]{c|c|cccc|cccc|c|cccc|ccccccc|cccccc|cc|cccc|cc}
\toprule
\textbf{Rank} & \textbf{Team} &
  \multicolumn{4}{c|}{\textbf{Model \& anatomy}} &
  \multicolumn{4}{c|}{\textbf{Backbone arch.}} &
  \multicolumn{1}{c|}{\textbf{Sup.}} &
  \multicolumn{4}{c|}{\textbf{Spatial config.}} &
  \multicolumn{7}{c|}{\textbf{Preprocessing}} &
  \multicolumn{6}{c|}{\textbf{Data augmentation}} &
  \multicolumn{2}{c|}{\textbf{Postprocess.}} &
  \multicolumn{4}{c|}{\textbf{Training}} &
  \multicolumn{2}{c}{\textbf{Hardware}} \\

\rowcolor{white}
\midrule
& &
  \rotatebox{270}{Single model} &
  \rotatebox{270}{Single model, anatomy conditional} &
  \rotatebox{270}{Identical models, separately trained per site} &
  \rotatebox{270}{Identical backbones, anatomy-specific training param.} &
  \rotatebox{270}{CNN encoder-decoder} &
  \rotatebox{270}{GAN} &
  \rotatebox{270}{Flow matching / Diffusion model} &
  \rotatebox{270}{Ensemble of multiple models} &
  \rotatebox{270}{Supervised} &
  \rotatebox{270}{3D} &
  \rotatebox{270}{3D patch-based} &
  \rotatebox{270}{2.5D} &
  \rotatebox{270}{2D} &
  \rotatebox{270}{Iso-resampling / resizing} &
  \rotatebox{270}{Clipping} &
  \rotatebox{270}{Patient level linear normalization} &
  \rotatebox{270}{Population level normalization} &
  \rotatebox{270}{Standardization} &
  \rotatebox{270}{Other (deformable reg., overflow correction, \ldots)} &
  \rotatebox{270}{Body mask used} &
  \rotatebox{270}{Random crop / patch} &
  \rotatebox{270}{Flipping} &
  \rotatebox{270}{Blurring} &
  \rotatebox{270}{Noise addition} &
  \rotatebox{270}{Intensity transform (contrast / histogram adj.)} &
  \rotatebox{270}{Deformation (affine or elastic)} &
  \rotatebox{270}{Average overlapping patches} &
  \rotatebox{270}{Revert normalization / padding} &
  \rotatebox{270}{Pre-trained model used} &
  \rotatebox{270}{Loss function(s)} &
  \rotatebox{270}{Optimiser} &
  \rotatebox{270}{Additional public data} &
  \rotatebox{270}{GPU hardware} &
  \rotatebox{270}{Training time (h)} \\

\midrule

1 & MixCT &
  \ding{51} & & & &
  & \ding{51} & & &
  \ding{51} &
  & \ding{51} & & &
  & \ding{51} & & \ding{51} & \ding{51} & {\small CT} &
  \ding{51} &
  \ding{51} & \ding{51} & & & & &
  & \ding{51} &
  No & MAE + Perc. + Masked MAE + Adv. + FM + Dice/CE & Adam & No &
  A100 & 12 \\

2 & ImagePasNet &
  & & \ding{51} & &
  \ding{51} & & & &
  \ding{51} &
  & \ding{51} & & &
  & \ding{51} & & \ding{51} & & {\small CT} &
  No &
  & & & & & &
  \ding{51} & \ding{51} &
  No & L1 + AFP & SGD & No &
  RTX 4070/A40 & 40 \\

3 & BreizhCT &
  & \ding{51} & & &
  \ding{51} & & & &
  \ding{51} &
  & & \ding{51} & &
  & \ding{51} & {\small CB} & {\small CT} & & {\small CT} &
  \ding{51} &
  & \ding{51} & & & & &
  & \ding{51} &
  No & MSE + Perc. & AdamW & No &
  2$\times$RTX~6000 & 120 \\

4 & Et &
  \ding{51} & & & &
  \ding{51} & & & &
  \ding{51} &
  & & \ding{51} & &
  \ding{51} & \ding{51} & & & \ding{51} & \ding{51} &
  \ding{51} &
  & & & & & &
  & \ding{51} &
  Yes (FT) & L1 + Perc. & Adam & No &
  RTX~4090 & 50 \\

5 & RicardoBrioso &
  \ding{51} & & & &
  \ding{51} & & & &
  \ding{51} &
  & & \ding{51} & &
  & \ding{51} & & \ding{51} & & &
  {\small CB} &
  \ding{51} & \ding{51} & & & \ding{51} & &
  \ding{51} & \ding{51} &
  No & L1 + Perc. & AdamW & No &
  RTX~6000 & 20 \\

6 & SynthRADShu &
  & & \ding{51} & &
  & \ding{51} & & &
  \ding{51} &
  & \ding{51} & & &
  \ding{51} & \ding{51} & & & \ding{51} & &
  \ding{51} &
  \ding{51} & \ding{51} & & & & &
  \ding{51} & \ding{51} &
  No & GAN + Perc. & Adam & No &
  A100 & 48 \\

7 & HiLab &
  & & \ding{51} & &
  \ding{51} & & & &
  \ding{51} &
  & & & \ding{51} &
  \ding{51} & \ding{51} & & \ding{51} & & &
  No &
  \ding{51} & \ding{51} & & & & &
  & \ding{51} &
  Yes (self) & L1 & Adam & Yes &
  RTX~5090 & 44 \\

8 & Faking\_it &
  \ding{51} & & & &
  & & \ding{51} & &
  \ding{51} &
  & \ding{51} & & &
  \ding{51} & \ding{51} & & \ding{51} & & &
  \ding{51} &
  \ding{51} & & \ding{51} & \ding{51} & \ding{51} & \ding{51} &
  \ding{51} & \ding{51} &
  No & MSE + MAE + SSIM + VLB & AdamW & No &
  RTX~6000 & 240 \\

9 & KoalAI &
  & & \ding{51}  & &
  \ding{51} & & &\ding{51} &
  \ding{51} &
  & \ding{51} & & &
  & \ding{51} & & \ding{51} & & {\small CT} &
  {\small CT} &
  & & & & & &
  \ding{51} & \ding{51} &
  No & Masked MSE + Masked AFP & SGD & No &
  H100 & 24 \\

10 & abdullahshazli18 &
  & & \ding{51} & &
  & \ding{51} & & &
  \ding{51} &
  & \ding{51} & & &
  & \ding{51} & \ding{51} & & & &
  \ding{51} &
  & & & & & &
  & \ding{51} &
  Yes (feat.) & MAE + Ctx. + Anat. + SSIM + Grad. + Adv. & Adam & No &
  RTX~4070~Ti & 6 \\

11 & Sirus &
  & & \ding{51} & &
  & & \ding{51} & &
  \ding{51} &
  & & & \ding{51} &
  \ding{51} & & & & & &
  & & & & & &
  \ding{51} & &
  No & No & LPIPS & Adam & -
  & - & - \\

12 & imi-graz &
  & & \ding{51} & &
  & & \ding{51} & &
  \ding{51} &
  \ding{51} & & & &
  \ding{51} & \ding{51} & & & & \ding{51} &
  No &
  & & & & & \ding{51} &
  & \ding{51} &
  No & L1 + MSE & AdamW & No &
  A100 & 57 \\

13 & sk &
  & & \ding{51} & &
  \ding{51} & & & &
  \ding{51} &
  & & \ding{51} & &
  \ding{51} & & & \ding{51} & & &
  \ding{51} &
  & & & & & &
  & \ding{51} &
  Yes (FT) & L1 + Perc. & AdamW & No &
  RTX~4090 & 2 \\

\bottomrule
\end{tabular}%
}
\end{table}

\end{landscape}
}

\afterpage{%
\clearpage
\begin{landscape}
\makeatother
\rowcolors{1}{}{lightgray}
\setlength{\tabcolsep}{4.5pt}

\begin{table}
\centering
\caption{\label{tab:metric_results}All quantitative metrics \(\left(\mu\pm\sigma\right)\) produced by every participant in task 1 (MRI-to-CT) and task 2 (CBCT-to-CT). There were three image-based and six dose-based metrics: three for photon treatment and three for proton treatment. The best results for each task and metric are marked in boldface. }
\subcaption*{Metric table 1: The quantitative metrics for task 1 (MRI-to-CT). Participants who scored worse than the water baseline on one image metric were excluded from the final ranking. }

\begin{adjustbox}{max width=1.25\textwidth}
\footnotesize
\begin{tabular}{c|c|ccc|cc|ccc|ccc}
\toprule
\rowcolor{white}
\textbf{Rank} & \textbf{Submission}
  & \multicolumn{3}{c|}{\textbf{Image Similarity}}
  & \multicolumn{2}{c|}{\textbf{Segmentation}}
  & \multicolumn{6}{c}{\textbf{Dose Calculation}} \\
\cmidrule(lr){1-13}
\rowcolor{white}
  & & & & & & &
  \multicolumn{3}{c|}{\textbf{Photon}}
  & \multicolumn{3}{c}{\textbf{Proton}} \\
\cmidrule(lr){8-10} \cmidrule(lr){11-13}
\rowcolor{white}
 & & MAE (HU, $\downarrow$) & PSNR (dB, $\uparrow$) & MS-SSIM ($\uparrow$)
 & HD95 (mm, $\downarrow$) & DICE ($\uparrow$)
 & $\gamma_\text{2\%/2mm}$ ($\uparrow$) & DVH ($\downarrow$) & MAE (Gy, $\downarrow$)
 & $\gamma_\text{2\%/2mm}$ ($\uparrow$) & DVH ($\downarrow$) & MAE (Gy, $\downarrow$) \\
\midrule
1 & KoalAI & $\mathbf{64.8\pm21.3}$ & $\mathbf{30.0\pm2.8}$ & $\mathbf{0.936\pm0.050}$ & $6.01\pm3.25$ & $0.78\pm0.11$ & $98.33\pm5.43$ & $\mathbf{0.011 \pm0.005}$& $0.006\pm0.009$ & $84.04\pm10.56$ & $0.064\pm0.026$ & $0.024\pm0.068$ \\
2 & ImagePasNet & $65.5\pm21.9$ & $29.6\pm2.7$ & $0.933\pm0.053$ & $6.32\pm3.61$ & $0.77\pm0.12$ & $98.50\pm5.29$ & $\mathbf{0.011\pm0.005}$ & $\mathbf{0.006\pm0.010}$ & $\mathbf{84.56\pm10.13}$ & $\mathbf{0.060\pm0.026}$ & $0.024\pm0.067$ \\
3 & BreizhCT & $67.2\pm22.9$ & $30.0\pm2.7$ & $0.935\pm0.046$ & $7.51\pm4.08$ & $0.74\pm0.12$ & $\mathbf{98.88\pm4.57}$ & $0.013\pm0.007$ & $0.006\pm0.009$ & $82.19\pm10.19$ & $0.067\pm0.031$ & $0.027\pm0.068$ \\
4 & MixCT & $67.9\pm22.0$ & $29.6\pm2.8$ & $0.931\pm0.053$ & $\mathbf{5.77\pm3.31}$ & $\mathbf{0.79\pm0.11}$ & $98.22\pm5.78$ & $0.016\pm0.007$ & $0.007\pm0.012$ & $81.92\pm11.57$ & $0.075\pm0.026$ & $\mathbf{0.023\pm0.019}$ \\
5 & QWER & $75.7\pm22.6$ & $28.8\pm2.4$ & $0.922\pm0.053$ & $7.69\pm4.24$ & $0.72\pm0.12$ & $98.29\pm5.74$ & 
$0.014\pm0.006$ & $0.007\pm0.012$ & $81.91\pm10.73$ & $0.073\pm0.029$ & $0.027\pm0.068$ \\
6 & DeepSyn & $83.3\pm25.0$ & $28.0\pm2.2$ & $0.913\pm0.052$ & $8.17\pm4.16$ & $0.70\pm0.12$ & $98.14\pm5.30$ & $0.015\pm0.006$ & $0.007\pm0.011$ & $79.15\pm11.16$ & $0.087\pm0.037$ & $0.026\pm0.019$ \\
7 & Faking\_it & $96.9\pm26.8$ & $26.8\pm2.0$ & $0.881\pm0.067$ & $7.60\pm4.30$ & $0.69\pm0.13$ & $96.58\pm7.60$ & $0.017\pm0.008$ & $0.009\pm0.013$ & $77.04\pm12.44$ & $0.115\pm0.043$ & $0.034\pm0.069$ \\
8 & MEVIS & $105.2\pm22.0$ & $26.9\pm1.8$ & $0.882\pm0.052$ & $10.63\pm5.31$ & $0.63\pm0.12$ & $97.91\pm5.06$ & $0.018\pm0.008$ & $0.008\pm0.009$ & $75.32\pm12.68$ & $0.104\pm0.041$ & $0.038\pm0.094$ \\
8 & HiLab & $89.8\pm26.0$ & $27.1\pm2.1$ & $0.882\pm0.071$ & $22.65\pm15.41$ & $0.54\pm0.19$ & $96.67\pm8.55$ & $0.019\pm0.008$ & $0.009\pm0.012$ & $77.49\pm12.81$ & $0.112\pm0.038$ & $0.038\pm0.095$ \\
10 & sk & $99.1\pm22.1$ & $26.6\pm1.6$ & $0.871\pm0.055$ & $24.06\pm15.11$ & $0.50\pm0.18$ & $96.29\pm7.02$ & 
$0.026\pm0.011$ & $0.010\pm0.011$ & $74.29\pm13.03$ & $0.149\pm0.044$ & $0.036\pm0.068$ \\
11 & SEU \& Rennes & $134.1\pm36.1$ & $24.7\pm2.0$ & $0.788\pm0.065$ & $35.36\pm20.82$ & $0.28\pm0.12$ & $93.83\pm11.28$ & $0.029\pm0.014$ & $0.014\pm0.018$ & $68.82\pm14.46$ & $0.234\pm0.083$ & $0.049\pm0.113$ \\
12 & imi-graz & $145.8\pm28.1$ & $24.7\pm1.5$ & $0.825\pm0.061$ & $17.91\pm9.98$ & $0.45\pm0.15$ & $82.52\pm12.22$ & $0.086\pm0.048$ & $0.033\pm0.020$ & $55.17\pm9.96$ & $0.464\pm0.226$ & $0.083\pm0.075$ \\
- & \textit{baseline} & $\mathit{309.3\pm56.2}$ & $\mathit{18.6\pm1.0}$ & $\mathit{0.466\pm0.089}$ & $\mathit{136.34\pm64.61}$ & $\mathit{0.01\pm0.01}$ & $\mathit{78.08\pm27.98}$ & $\mathit{0.093\pm0.025}$ & $\mathit{0.056\pm0.150}$ & $\mathit{59.59\pm18.85}$ & $\mathit{0.451\pm0.131}$ & $\mathit{0.152\pm0.310}$ \\
\bottomrule
\end{tabular}%
\end{adjustbox}

\vspace{0.5cm}
\subcaption*{Metric table 2: The quantitative metrics for task 2 (CBCT-to-CT). Participants who scored worse than the water baseline on one image metric were excluded from the final ranking. }

\begin{adjustbox}{max width=1.25\textwidth}
\footnotesize
\begin{tabular}{c|c|ccc|cc|ccc|ccc}
\toprule
\rowcolor{white}
\textbf{Rank} & \textbf{Team}
  & \multicolumn{3}{c|}{\textbf{Image Similarity}}
  & \multicolumn{2}{c|}{\textbf{Segmentation}}
  & \multicolumn{6}{c}{\textbf{Dose Calculation}} \\
\cmidrule(lr){1-13}
\rowcolor{white}
 & & & & & & &
  \multicolumn{3}{c|}{\textbf{Photon}}
  & \multicolumn{3}{c}{\textbf{Proton}} \\
\cmidrule(lr){8-10} \cmidrule(lr){11-13}
\rowcolor{white}
 & & MAE (HU, $\downarrow$) & PSNR (dB, $\uparrow$) & MS-SSIM ($\uparrow$)
 & HD95 (mm, $\downarrow$) & DICE ($\uparrow$)
 & $\gamma_\text{2\%/2mm}$ ($\uparrow$) & DVH ($\downarrow$) & MAE (Gy, $\downarrow$)
 & $\gamma_\text{2\%/2mm}$ ($\uparrow$) & DVH ($\downarrow$) & MAE (Gy, $\downarrow$) \\
\midrule
1 & MixCT& $\mathbf{48.3\pm13.4}$ & $\mathbf{32.6\pm2.3}$ & $\mathbf{0.968\pm0.025}$ & $\mathbf{4.53\pm3.03}$ & $\mathbf{0.86\pm0.07}$ & $99.30\pm1.18$ & $\mathbf{0.013\pm0.005}$ & $\mathbf{0.004\pm0.003}$ & $\mathbf{88.64\pm7.87}$ & $\mathbf{0.034\pm0.014}$ & $\mathbf{0.017\pm0.012}$ \\
2 &ImagePasNet & $52.5\pm15.1$ & $31.9\pm2.3$ & $0.964\pm0.026$ & $4.87\pm3.19$ & $0.85\pm0.08$ & $\mathbf{99.31\pm1.08}$ & $0.015\pm0.006$ & $0.005\pm0.003$ & $87.77\pm8.16$ & $0.036\pm0.016$ & $0.018\pm0.013$ \\
3 &BreizhCT & $53.1\pm17.4$ & $32.5\pm2.3$ & $0.966\pm0.025$ & $5.08\pm3.36$ & $0.84\pm0.08$ & $99.31\pm1.10$ & $0.015\pm0.007$ & $0.005\pm0.004$ & $86.41\pm8.43$ & $0.036\pm0.019$ & $0.020\pm0.014$ \\
4 &Et & $53.6\pm15.3$ & $31.9\pm2.2$ & $0.963\pm0.026$ & $4.99\pm3.32$ & $0.84\pm0.08$ & $99.23\pm1.14$ & $0.015\pm0.006$ & $0.005\pm0.003$ & $87.41\pm8.02$ & $0.039\pm0.016$ & $0.018\pm0.013$ \\
5 &RicardoBrioso & $62.7\pm18.1$ & $31.0\pm2.3$ & $0.952\pm0.030$ & $6.65\pm4.45$ & $0.80\pm0.09$ & $98.95\pm1.52$ & $0.017\pm0.007$ & $0.006\pm0.004$ & $84.99\pm8.59$ & $0.046\pm0.021$ & $0.021\pm0.015$ \\
6 &SynthRADShu & $66.7\pm17.4$ & $30.0\pm2.1$ & $0.945\pm0.032$ & $5.46\pm3.41$ & $0.82\pm0.08$ & $98.70\pm1.99$ & $0.018\pm0.007$ & $0.006\pm0.004$ & $84.23\pm9.17$ & $0.050\pm0.027$ & $0.022\pm0.015$ \\
7 &HiLab & $65.1\pm17.7$ & $30.3\pm2.2$ & $0.944\pm0.033$ & $7.48\pm5.60$ & $0.78\pm0.11$ & $98.59\pm2.30$ & $0.022\pm0.007$ & $0.006\pm0.005$ & $83.71\pm9.17$ & $0.054\pm0.026$ & $0.023\pm0.015$ \\
8 &Faking\_it & $78.2\pm19.9$ & $29.0\pm2.0$ & $0.935\pm0.029$ & $6.04\pm3.61$ & $0.76\pm0.08$ & $98.57\pm1.62$ & $0.023\pm0.010$ & $0.006\pm0.004$ & $81.97\pm8.57$ & $0.059\pm0.029$ & $0.024\pm0.013$ \\
9 &KoalAI & $88.2\pm43.7$ & $28.7\pm2.7$ & $0.934\pm0.044$ & $5.81\pm3.91$ & $0.81\pm0.10$ & $96.37\pm7.13$ & $0.026\pm0.012$ & $0.009\pm0.007$ & $78.82\pm11.71$ & $0.076\pm0.033$ & $0.030\pm0.019$ \\
10 &abdullahshazli18 & $84.8\pm22.7$ & $29.4\pm2.3$ & $0.934\pm0.036$ & $8.35\pm5.47$ & $0.75\pm0.12$ & $97.45\pm4.60$ & $0.033\pm0.013$ & $0.009\pm0.007$ & $78.83\pm10.21$ & $0.086\pm0.039$ & $0.030\pm0.019$ \\
11 &Sirus & $88.8\pm20.6$ & $28.1\pm1.6$ & $0.920\pm0.036$ & $7.13\pm4.63$ & $0.75\pm0.09$ & $97.33\pm3.82$ & $0.038\pm0.018$ & $0.011\pm0.006$ & $74.45\pm10.08$ & $0.106\pm0.062$ & $0.036\pm0.019$ \\
12 &imi-graz & $115.1\pm24.2$ & $26.4\pm1.6$ & $0.881\pm0.048$ & $13.08\pm8.23$ & $0.58\pm0.14$ & $96.88\pm3.49$ & $0.046\pm0.021$ & $0.011\pm0.005$ & $71.24\pm8.58$ & $0.141\pm0.073$ & $0.040\pm0.017$ \\
13 &sk & $153.9\pm107.1$ & $25.8\pm4.1$ & $0.815\pm0.168$ & $34.67\pm42.05$ & $0.51\pm0.33$ & $76.14\pm32.17$ & $0.073\pm0.015$ & $0.051\pm0.089$ & $64.84\pm25.08$ & $0.254\pm0.048$ & $0.066\pm0.069$ \\
- &\textit{baseline} & $\mathit{309.0\pm73.3}$ & $\mathit{18.8\pm1.4}$ & $\mathit{0.505\pm0.107}$ & $\mathit{133.48\pm59.39}$ & $\mathit{0.00\pm0.01}$ & $\mathit{76.39\pm25.15}$ & $\mathit{0.184\pm0.038}$ & $\mathit{0.036\pm0.093}$ & $\mathit{59.14\pm15.90}$ & $\mathit{0.482\pm0.176}$ & $\mathit{0.094\pm0.207}$ \\
\bottomrule
\end{tabular}
\end{adjustbox}

\end{table}
\end{landscape}

\begin{figure*}[htbp]
\centering
\includegraphics[width=0.99\linewidth]{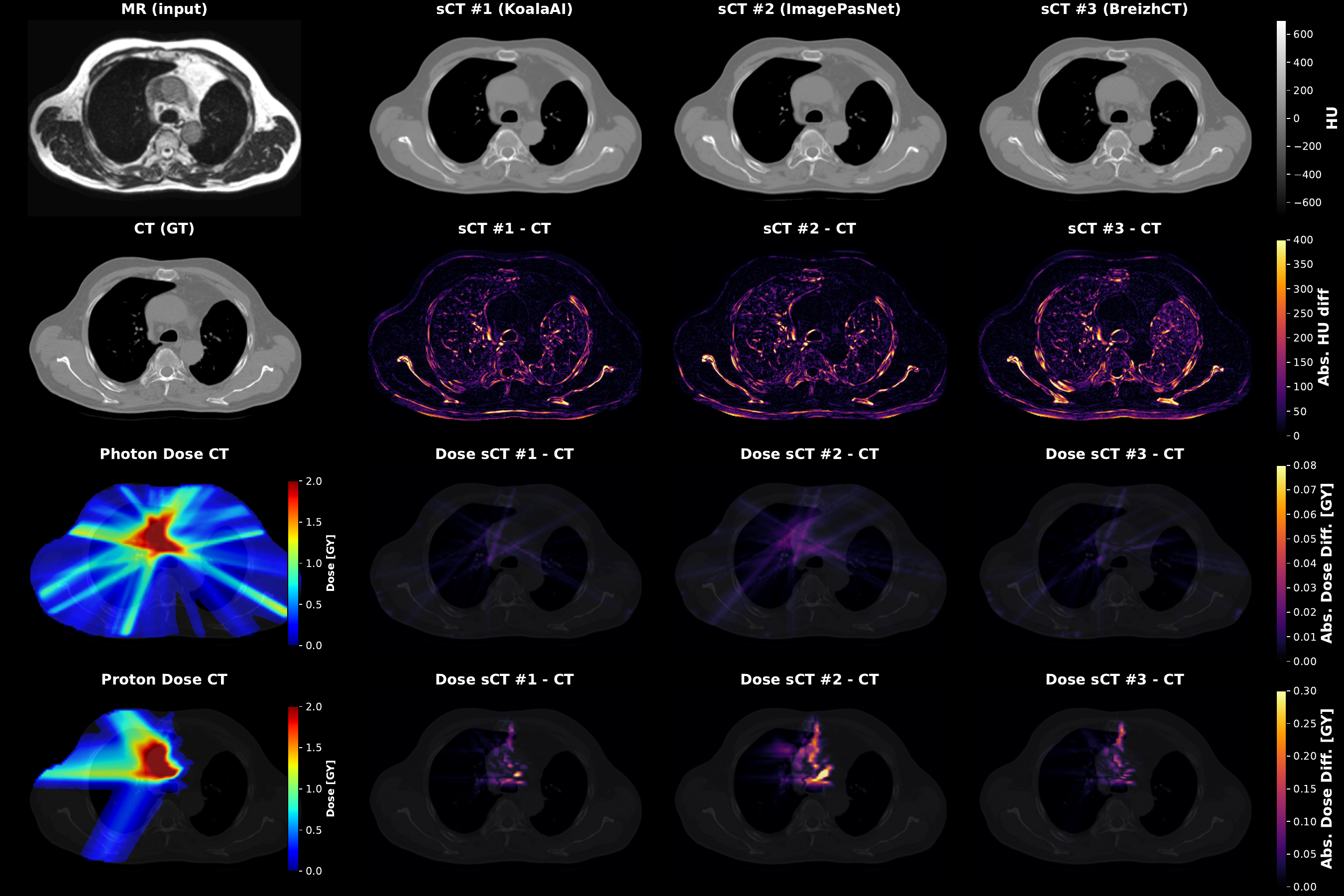}
\caption{Representative example of synthetic CT (sCT) generation for Task 1 (MRI-to-CT, thorax) by the three top-performing methods (KoalAI, ImagePasNet, BreizhCT). The first column shows the input MRI and reference CT (GT) together with the planned photon and proton dose distributions on the reference CT. The first row displays the corresponding sCT predictions, the second row the absolute HU difference maps relative to the reference CT, and the third and fourth rows the absolute dose differences (|CT dose − sCT dose|) for photon and proton plans, respectively. Note the different colorbar ranges for photon (0–0.08 Gy) and proton (0–0.30 Gy) dose differences.}
\label{fig:example_task1}
\end{figure*}

\begin{figure*}[htbp]
\centering
\includegraphics[width=0.7\linewidth]{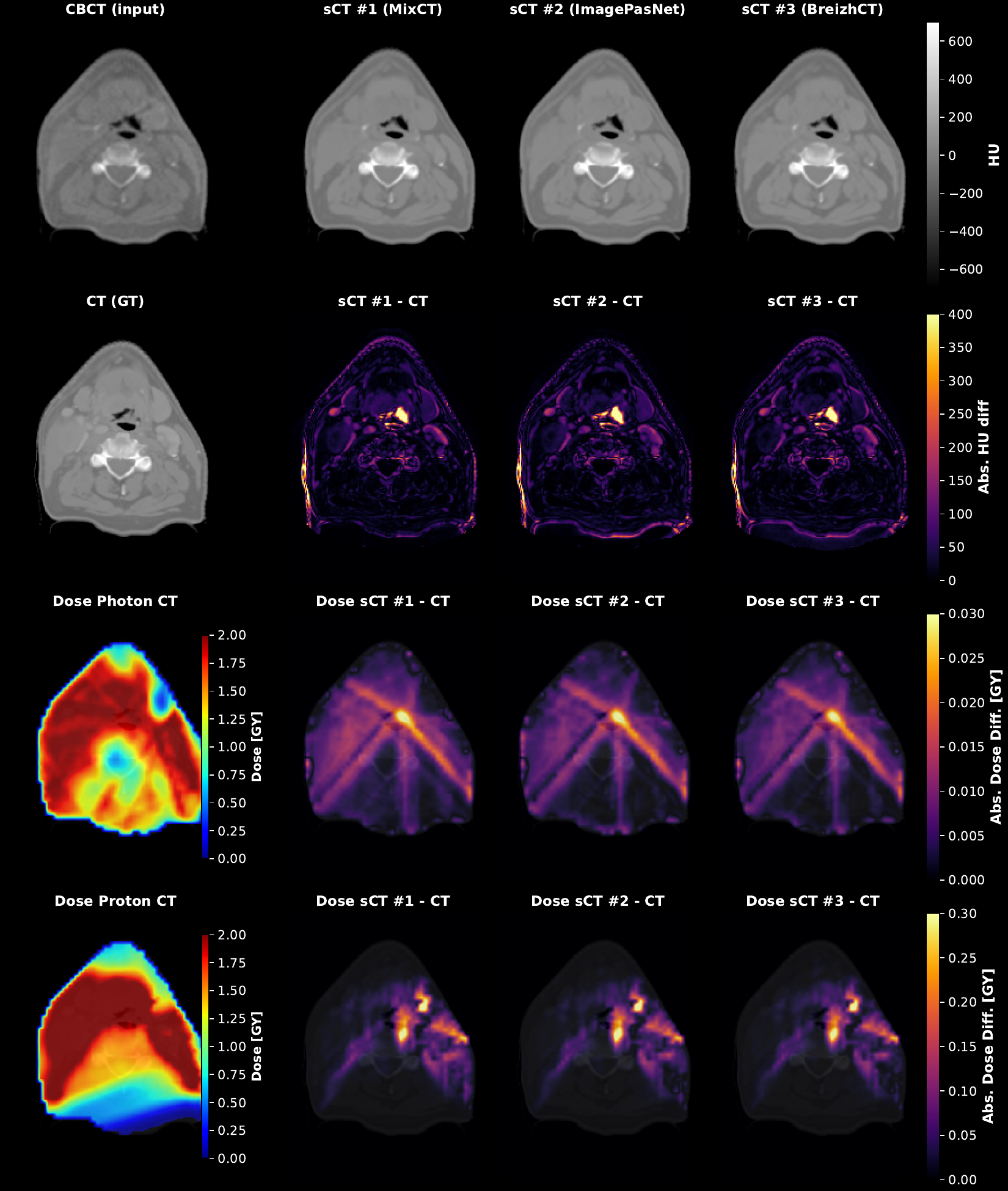}
\caption{Representative example of synthetic CT (sCT) generation for Task 2 (CBCT-to-CT, head-and-neck) by the three top-performing methods (MixCT, ImagePasNet, BreizhCT). The first column shows the input CBCT and reference CT (GT) together with the planned photon and proton dose distributions on the reference CT. The first row displays the corresponding sCT predictions, the second row the absolute HU difference maps relative to the reference CT, and the third and fourth rows the absolute dose differences (|CT dose − sCT dose|) for photon and proton plans, respectively. Note the different colorbar ranges for photon (0–0.03 Gy) and proton (0–0.30 Gy) dose differences.}
\label{fig:example_task2}
\end{figure*}
}

\section{Participation}
\begin{table}
\small
\centering
\caption{Details on the challenge participation. Participants without a team are displayed as a one-person team.}
\label{tab:partipants}
\begin{tabular}{llll}
\toprule
& \textbf{Validation} & \textbf{Preliminary test} & \textbf{Test} \\
\midrule
\multirow{3}{*}{\textbf{Task 1}} & 228 valid & 108 valid & 12 included,\\
& submissions & submissions & 1 excluded team\\
& (40 teams) &  (21 teams) &  \\
\midrule
\multirow{3}{*}{\textbf{Task 2}} & 124 valid & 44 valid & \multirow{2}{*}{13 included and}\\
 & submissions & submissions & \multirow{2}{*}{2 excluded teams}\\
 & (25 teams) &  (15 teams)& \\
\bottomrule
\end{tabular}
\end{table}
The SynthRAD2025 Grand Challenge attracted broad international participation, with research teams exploring a wide range of deep learning approaches for synthetic CT (sCT) generation from MRI and CBCT. Teams participated in one or both tasks, MRI-to-CT (Task 1) and CBCT-to-CT (Task 2), employing either unified or task-specific pipelines.
A total of 674 participants registered for the challenge on grand-challenge.org up to 15 August 2025, forming 78 teams. Throughout the different challenge phases (training, validation, and test), participation gradually decreased, with 14 teams submitting results in the final test phase for both task 1 and task 2. Based on the inclusion criteria described in Section \ref{subsec: ranking method}, 12 submissions were retained for the final Task 1 analysis and 13 submissions were retained for the final Task 2 analysis.
Across both tasks, this corresponded to 17 unique teams, of which 8 participated in both tasks, often leveraging similar architectures with minor task-specific adaptations.
Based on the inclusion criteria described in Section \ref{subsec: ranking method}, 12 teams were included in the final analysis for Task 1 and 13 for task 2. Among these, several teams participated in both tasks, often leveraging similar architectures with minor task-specific adaptations.

\subsection{MixCT (task 1 \& 2)}
\label{method mixCT}
MixCT employed a 3D patch-based generative adversarial network based on ConvNeXt-inspired \citep{Liu2022ConvNeXt} convolutional blocks for both tasks. The generator (GeNeXt) followed a U-shaped encoder–decoder architecture, in which each stage consisted of stacked ConvNeXt blocks that combined depthwise separable convolutions, pointwise convolutions, layer normalization, and GELU activations. Residual connections were applied within each block to facilitate gradient flow and stabilize training.
A conditional PatchGAN discriminator was used to operate on local 3D patches and enforce high-frequency realism. In addition, a multi-head discriminator design was introduced, incorporating an auxiliary segmentation branch to guide the generator towards anatomically consistent outputs.
Preprocessing involved deformable registration, foreground cropping using the body mask, and percentile-based normalization for MRI and CBCT. CT intensities were clipped to modality-specific ranges (e.g., [−1024, 1000] HU for MRI-based synthesis and [−1024, 1500] HU for CBCT) before linear scaling. Training was performed on 3D patches with sizes of 32×160×192 (MRI) and 32×128×128 (CBCT).
The loss function combined MAE loss, perceptual loss based on ConvNeXt features, adversarial loss, feature matching loss, and segmentation-based losses. The model was trained using AdamW optimizers with separate learning rates for the generator and the discriminator.
During inference, overlapping patches were extracted using a sliding-window approach with 0.8 overlap, and the resulting patches were averaged to reconstruct the final sCT volume.
The implementation is publicly available at: \url{https://github.com/yaxiaa924/GANeXt-SynthRAD2025}

\subsection{ImagePasNet (task 1 \& 2)}
\label{method ImagePasNet}
ImagePasNet employed an adaptation of residual U-Net (nnResU-Net) architecture for image-to-image translation\citep{isensee2021nnu,longuefosse2024adapted}. MRI inputs were per-case z-normalized, while CBCT and CT volumes were clipped to [-1024, 3071] HU and dataset-level z-normalized. Training was performed on 3D patches with region-specific sizes, e.g. 56×192×192 for head-and-neck and 48×192×224 for thorax/abdomen.\\
A key contribution was the anatomical feature-prioritized (AFP) loss, which computes feature-space differences using a pretrained segmentation network on each task separately, encouraging alignment of anatomically relevant structures \citep{longuefosse2025anatomical}. Training was performed on all the anatomical regions using segmentation labels from TotalSegmentator \citep{wasserthal2023totalsegmentator}.
Models were trained using a stochastic gradient descent (SGD) optimizer with momentum (0.99) and polynomial learning-rate decay for up to 1500 epochs, followed by 500 epochs of fine-tuning with a combined L1 and AFP loss. At inference, dense sliding-window reconstruction with overlap of 0.3 was used to reduce boundary artifacts, and predictions were rescaled back to HUs. 
The implementation is publicly available at:  \url{https://github.com/Phyrise/nnUNet_translation.git}

\subsection{KoalAI (task 1 \& 2)}
KoalAI used an ensemble of 3D ResUNet-L architectures derived from nnU-Net \citep{isensee2021nnu}, and were configured within the nnU-Net v2 framework \citep{longuefosse2024adapted}, which automatically adapted patch sizes and related training parameters according to memory constraints and anatomical resolution.
The model was trained using a dedicated Masked Anatomical Perception (MAP) loss that jointly optimizes voxel-wise intensity accuracy and anatomical consistency within constrained volumes of interest. This was achieved via 1) a masked MSE loss for enforcing global intensity agreement between the predicted and reference CT within a body mask, and 2) a masked variant of AFP loss \cite{longuefosse2025anatomical} to align multi-scale anatomical features, extracted from predicted and reference CT within a body mask, using a pretrained student TotalSegmentator network \citep{wasserthal2023totalsegmentator}. The masking mechanism in MAP loss allows a more focused optimisation on anatomically relevant regions, reduces gradient dilution caused by background voxels, and encourages faster convergence. Preprocessing included deformable registration to generate paired MRI--CT and CBCT--CT datasets. MRI inputs were normalized using individual-level z-score normalization, while CT and CBCT were normalized using masked population-level z-score normalization after intensity clipping to approximately [−1024, 3000] HU. The use of masked normalization ensures that statistics are computed only within anatomically relevant regions. KoalAI trained separate models for each anatomical region. Training was performed using SGD with an initial learning rate of 0.01 and polynomial decay over 1000 epochs. A five-fold cross-validation strategy was employed to improve generalization, and final predictions were obtained by averaging the outputs across folds. During inference, predicted volumes were transformed back to HUs by inverting the normalization process, then masked with the body contour to suppress background artifacts, setting voxels outside the body to −1024 HU.
The implementation is publicly available at: \url{https://github.com/aehrc/nnsyn}

\subsection{BreizhCT (task 1 \& 2)}
\label{method BreizhCT}
BreizhCT employed a 2.5D U-Net++ \citep{zhou2018unet++} architecture with a ResNet-34 encoder, leveraging nested and dense skip connections to enhance multi-scale feature fusion. The U-Net++ design introduces intermediate convolutional pathways between the encoder and decoder stages. The model operated on stacked adjacent slices (2.5D input). For the CBCT-CT task, the model input consisted of three consecutive slices (target slice with one superior and one inferior slice).
For MR-CT, the model input consisted of five consecutive slices (the target slice and two superior and two inferior slices). The loss function combined L1 loss with multiple perceptual losses derived from pretrained networks (VGG-19, SAM, and TotalSegmentator), enforcing both intensity fidelity and structural consistency. Training was performed using AdamW with stepwise learning-rate decay, random flip augmentation, and early stopping. In addition, the team investigated two registration strategies: standard Elastix-based alignment and a feature-based IMPACT registration method. Final predictions were obtained using five-fold cross-validation ensembling combined with test-time augmentation.
The implementation is publicly available at: \url{https://github.com/vboussot/Synthrad2025_Task_1}

\subsection{RicardoBriso (task 2)}
\label{method RicardoBriso}
RicardoBriso proposed a lightweight 2.5D U-Net architecture based on the Tiny U-Net framework \citep{chen2024tinyu}, enhanced with cascaded multi-receptive field (CMRF) blocks and attention-gated skip connections. The encoder–decoder structure followed the standard U-Net topology. Still, each convolutional block was replaced with a CMRF module, which splits feature channels and processes them via cascaded depthwise convolutions to efficiently capture multi-scale receptive fields. Attention gates were inserted in the skip connections to suppress irrelevant features and highlight anatomically relevant regions. The model input consisted of three consecutive slices (the target slice and one superior and one inferior slice), forming a 2.5D representation. Training was performed on 2D patches of size 256×256 extracted from full-resolution slices. Preprocessing included clipping CBCT and CT intensities to [-1000, 1000] HU and [−1024, 3071] HU, respectively, and then normalizing to [0, 1]. The loss function combined foreground- and background-weighted L1 loss, perceptual loss (VGG-19), and SSIM loss, with weights based on the body mask. Training was conducted in two stages by adjusting the loss-component weights: an initial training phase ($\sim$50 epochs) emphasized perceptual quality by assigning greater weight to the perceptual loss, followed by a fine-tuning phase (5--10 epochs) that prioritized pixel-level accuracy by increasing the relative weight of the voxel-wise loss. Final predictions were obtained through five-fold ensembling and test-time augmentation using flipped inputs.

\subsection{Et (task 2)}
\label{method et}
Team Et implemented a  2.5D UNet++ architecture with a ResNeXt101 backbone \citep{zhou2018unet++, xie2017aggregated}. The model extended the U-Net++ framework by incorporating grouped convolutions and increasing the encoder cardinality. Multiple adjacent CBCT slices were stacked as input channels, enabling the network to leverage limited 3D contextual information while operating within a 2D framework. The encoder utilized pre-trained ResNeXt blocks. Preprocessing involved deformable registration, resizing to 512×512 resolution, and intensity clipping followed by normalization to [−1, 1]. CT images were clipped to [−1024, 3071] HU, while CBCT intensities were limited to the 99.5th percentile. The model was trained with a combination of MAE and perceptual loss, optimized with AdamW for 40 epochs under a stepwise learning-rate schedule. During inference, predictions were generated slice-wise and rescaled back to HUs.

\subsection{Qwer (task 1)}
\label{method Qwer}
Qwer proposed an nnU-Net-based framework for MRI-to-CT synthesis. The architecture followed the standard nnU-Net encoder--decoder design, with convolutional blocks, normalization layers, and skip connections. Two training strategies were explored: a unified model trained across all anatomical regions and region-specific models trained separately. Preprocessing included deformable registration using GroupMorph \citep{tan2024groupmorph}, MRI normalization based on tissue-specific statistics derived from segmented fat regions, and CT intensity clipping to [−1000, 1000] HU followed by scaling to [0, 2]. Input volumes were resized to 192$\times$192$\times$128. The loss function combined MAE with a perceptual loss computed from features extracted by a U-Net pretrained from task 291 in TotalSegmentator \citep{d2024totalsegmentator}. Training followed standard nnU-Net settings, including a batch size of 2, 1000 epochs, and polynomial learning rate decay. At inference, patch-based prediction with 0.5 overlap was used, and final outputs were obtained by ensembling predictions from both unified and region-specific models, combined with test-time augmentation.

%% file: 04_Results.tex
\section{Results}
\label{sec:results}

\subsection{Overall sCT generation performance }
\label{subsec:results overall performance}

Quantitative results for all participating teams across Task 1 (MRI-to-CT) and Task 2 (CBCT-to-CT) are presented in \autoref{tab:metric_results}. Across both tasks, a clear performance gap was observed between top-performing and lower-ranked methods. In Task 1, MAE ranged from $64.8 \pm 21.3$ HU (KoalAI) to $145.8 \pm 28.1$ HU (imi-graz), a 2.3-fold gap between best- and worst-ranked submission. Dice scores spanned from $0.79 \pm 0.11$ (MixCT) down to $0.28 \pm 0.12$ (SEU \& Rennes), indicating substantial variability across submissions. A similar trend was evident in Task 2, where MAE values ranged from $48.3 \pm 13.4$ HU (MixCT) to $153.9 \pm 107.1$ HU (sk), a 3.2-fold gap in MAE. HD95 values varied from $4.53 \pm 3.03$ mm (MixCT) to $34.67 \pm 42.05$ mm (sk). All ranked submissions in both tasks substantially outperformed the water-equivalent baseline on image-similarity and segmentation metrics, whereas for dose-calculation metrics the gap between the worst-performing methods and the baseline was considerably smaller.

\subsection{Submitted method characteristics and design factors}
In Task 1, the submitted approaches were distributed almost equally among single-model, similar-but-separate model, and anatomy-specific modelling strategies, whereas only one submission used an anatomy-conditional model. In Task 2, site-specific training was more common: 7/13 submissions used separate models for each anatomical site, compared with 4/13 using a single model, 1/13 using an anatomy-conditional model, and 1/13 using anatomy-specific parameterisation.
CNN encoder--decoder architectures were the most frequently used backbone, accounting for 7/12 submissions in Task 1 and 7/13 submissions in Task 2 (Figure~\ref{fig:backbone_performance}). GAN-based methods were less common, representing 2/12 and 3/13 submissions, respectively, while flow-matching or diffusion-based approaches accounted for 3/12 submissions in Task 1 and 3/13 submissions in Task 2. Explicit multi-model ensembles were uncommon and were reported only in Task 1. In the backbone-wise analysis, CNN encoder--decoder and GAN-based methods generally achieved the highest MS-SSIM values, with substantially overlapping distributions across tasks and anatomical regions. GAN-based approaches reached performance comparable to CNN encoder--decoder models in several subtasks, whereas flow-matching and diffusion-based approaches generally showed lower median MS-SSIM values. Statistically significant differences between backbone groups were observed within several subtasks.
All 25 analysed submissions used supervised learning based on paired, deformably registered data. Preprocessing commonly included iso-resampling or resizing (7/12 in Task 1 and 7/13 in Task 2) and intensity clipping (5/12 and 11/13, respectively), while body masking was particularly frequent in Task 2 (9/13). Pretrained models were used by 3/12 submissions in task 1 and 5/13 submissions in task 2, and the use of additional public data was uncommon (2/12 and 1/13, respectively).
Spatial configuration varied across tasks. In Task 1, submissions comprised 5/12 3D patch-based approaches, 3/12 full-3D approaches, 3/12 2.5D approaches, and 1/12 2D approach. In Task 2, 3D patch-based models formed the largest group, with 5/13 submissions, followed by 4/13 2.5D approaches, 2/13 full-3D approaches, and 2/13 2D approaches. The spatial-configuration analysis showed that performance differences were task- and region-dependent (Figure~\ref{fig:dimension_per_task}). In Task 1, MS-SSIM distributions overlapped substantially across spatial configurations, suggesting limited separation by dimensionality despite some statistically significant pairwise differences. In Task 2, 3D patch-based and 2.5D approaches generally achieved high MS-SSIM values, whereas full-3D and 2D approaches showed more variable performance across anatomical regions. Several statistically significant differences were observed, particularly in Task 2, but no spatial configuration was uniformly superior across all subtasks.
Training and inference strategies were broadly similar across teams. Overlapping patch aggregation was reported in 6/12 Task 1 submissions and 6/13 Task 2 submissions, while flipping-based augmentation was used in 4/12 and 6/13 submissions, respectively. Overall, the strongest-performing submissions combined conventional supervised training pipelines with heterogeneous choices of backbone architecture and spatial configuration, indicating that performance was shaped by multiple interacting design factors rather than by a single architectural choice.
Training and inference strategies were comparatively similar across teams. Overlapping patch aggregation was reported in 6 of 12 Task 1 submissions and 6 of 13 Task 2 submissions, while flipping-based augmentation was used in 4 of 12 and 6 of 13 submissions, respectively. Overall, the strongest-performing submissions combined conventional supervised training pipelines with heterogeneous choices in backbone architecture and spatial configuration.

\subsection{Task 1: MRI-to-CT synthesis}

In Task 1, the four top-performing methods (KoalAI, ImagePasNet, BreizhCT, and MixCT)  formed a tight cluster in terms of image similarity metrics with MAE, PSNR and MS-SSIM within $3.1$ HU, $0.4$ dB and 0.005, respectively. Team KoalAI achieved the most balanced performance across all metric groups (image similarity, segmentation, and dose), being ranked first or second in 10 out of the 11 metrics.

Segmentation metrics behaved differently from image similarity and dose metrics: the highest-ranked team in DSC ($0.79 \pm 0.11$) and HD95 ($5.77 \pm 3.31$ mm), MixCT, was only the fourth-ranked submission overall. This divergence indicates that geometric fidelity of anatomical structures and voxel-wise HU accuracy capture distinct aspects of synthesis quality, and that strong performance on one does not guarantee strong performance on the other (see also \autoref{sec:metric_cor}).

Photon dose metrics showed limited separation among top methods, with $\gamma_{\text{photon}}$ pass rates between $98.1\%$ and $98.9\%$ for the top 6 teams. In contrast,  proton $\gamma$ pass rates were significantly lower and exhibited greater variability, ranging from $79.2\%$ to $84.6\%$ among the top 6 methods and even falling as low as $55.2\%$ for the lowest-ranked submission. This disparity between photon and proton pass rates highlights the well-known increased sensitivity of proton dose calculations to HU inaccuracies.

\autoref{fig:example_task1} presents sCTs, HU difference maps and dose distributions of the top 3 teams (KoalAI, ImagePasNet, BreizhCT) for an exemplary thorax case. The top-performing methods all produce visually realistic sCTs with accurate global anatomy. Differences between the top-ranked methods were minimal and all methods showed similar trends. HU difference maps revealed systematic errors in anatomically challenging regions. In thoracic cases, errors were most pronounced in low-density lung tissue and bones, where limited MRI signal or misalignment with the reference CT caused all three top-ranked submissions to show higher errors than in surrounding tissues. Photon dose error maps showed very low errors for the top-performing methods, with deviations below $3 \pm 3$ cGy. The spatial distribution of these errors was also similar across methods, reflecting their comparable HU patterns. 

Pairwise significance maps for Task 1 (Supplementary Fig. 1) revealed different discriminative power across metric categories. Image similarity (MAE, MS-SSIM, PSNR) and segmentation-based metrics (Dice, HD95) showed the strongest statistical separation, with most pairwise comparisons reaching significance. However, among the four top-ranked methods (KoalAI, ImagePasNet, BreizhCT, MixCT) some image similarity differences were statistically insignificant (e.g.MS-SSIM between BreizhCT and ImagePasNet/MixCT), indicating narrow margins at the top of the leaderboard. Photon and proton dose metrics showed noticeably fewer significant differences than image similarity metrics, with the DVH-based metric showing the weakest separation.

The two lowest-ranked submissions, SEU \& Rennes and imi-graz, still substantially outperformed the water-equivalent baseline on image similarity (MAE of $134.1 \pm 36.1$ and $145.8 \pm 28.1$ HU vs. $309.3 \pm 56.2$ HU) but approached baseline performance on several dose metrics, with imi-graz showing a proton DVH error ($0.464 \pm 0.226$ Gy) comparable to the baseline ($0.451 \pm 0.131$ Gy).

\subsection{Task 2: CBCT-to-CT synthesis}

In Task 2, overall performance improved across all metric categories compared to Task 1. The four top-ranked methods (MixCT, ImagePasNet, BreizhCT, and Et) also formed a tight cluster in terms of image similarity metrics, with MAE, PSNR, and MS-SSIM within $5.3$ HU, $0.7$ dB, and $0.005$, respectively. Team MixCT achieved the a strong performance across all metric groups, ranking first in 10 out of the 11 metrics, including MAE ($48.3\pm 13.4$ HU), Dice ($0.86\pm 0.07$), and $\gamma_{\text{proton}}$ ($88.64 \pm 7.87 \%$). 
In contrast to Task 1, segmentation rankings closely tracked image similarity rankings, with the top four submissions achieving Dice scores between $0.84$ and $0.86$ and HD95 values between $4.53$ and $5.08$mm.
Photon dose metrics showed near-ceiling performance, with $\gamma_{\text{photon}}$ pass rates consistently above 99\% for the top four submissions. As in Task 1, $\gamma_{\text{proton}}$ pass rates were lower and exhibited greater variability, ranging from 86.4 to 88.6 \% among the top four methods and falling to 64.8\% for the lowest-ranked submission (sk). 

\autoref{fig:example_task2} presents sCTs, HU difference maps, and dose distributions of the top 3 teams (MixCT, ImagePasNet, BreizhCT) for an exemplary head-and-neck case. The top-performing methods all produced visually realistic sCTs with accurate global anatomy, and soft tissue and bone structures accurately reconstructed. The exemplary patient showed a mismatch in the air cavities between CBCT/sCT and the reference CT, leading to locally increased errors in that region. Differences between the top-ranked methods were minimal, and all top submissions showed comparable visual quality. Photon dose error maps confirmed the high dose calculation performance, with deviations spatially uniform across methods and consistent with the high $\gamma_{\text{photon}}$ pass rates. Proton dose errors remained localized along beam paths. 
Across anatomical regions, thoracic and abdominal cases still exhibited greater variability and larger local errors than head-and-neck cases, likely due to motion artifacts, scatter, and the incomplete field of view in CBCT acquisitions.

Pairwise significance maps for Task 2 (\autoref{fig:significance_maps_appendix_task1}) showed a similar trend as for Task 1, with image similarity and segmentation-based metrics showing statistically significant differences between most of the closely ranked teams. For dose-based metrics statistically significant separation in Task 2 was stronger than in Task 1, however, dose based metrics still showed less statistical significance than other metric categories.

Compared with Task 1, the spread between top- and mid-ranked methods was smaller in Task 2. However, the lowest-ranked methods still showed notable degradation, with MAE exceeding $100$ HU and MS-SSIM dropping below $0.90$, accompanied by reduced segmentation and dose performance. The two lowest-ranked submissions, imi-graz and sk, substantially outperformed the water-equivalent baseline on image similarity (MAE of $115.1 \pm 24.2$ and $153.9 \pm 107.1$ HU vs. $309.0 \pm 73.3$ HU) but approached baseline performance on several dose metrics. sk in particular showed a $\gamma_{\text{photon}}$ pass rate ($76.1 \pm 32.2$\%) comparable to the baseline ($76.4 \pm 25.2\%$) and a photon $MAE_{dose}$ error ($0.051 \pm 0.089$ Gy) that exceeded the baseline ($0.036 \pm 0.093$ Gy).

\subsection{Metric correlations}
\label{sec:metric_cor}
\begin{figure*}[t]
\centering
\subfloat[Task 1\label{figa:correlation_image}]{%
  \includegraphics[clip,height=7.1cm]{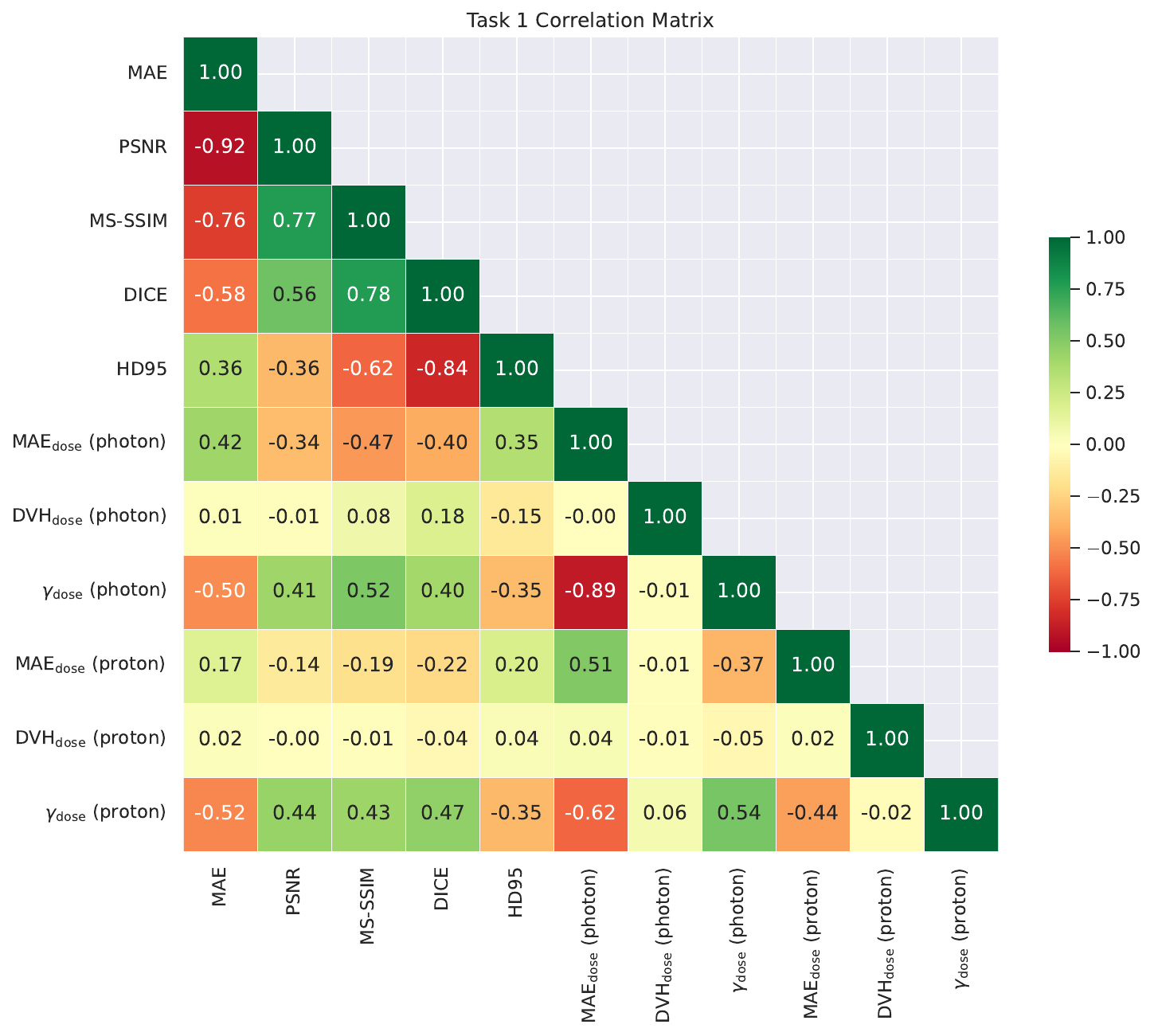}%
}
\hfill
\subfloat[Task 2\label{figb:correlation_photon}]{%
  \includegraphics[clip,height=7.1cm]{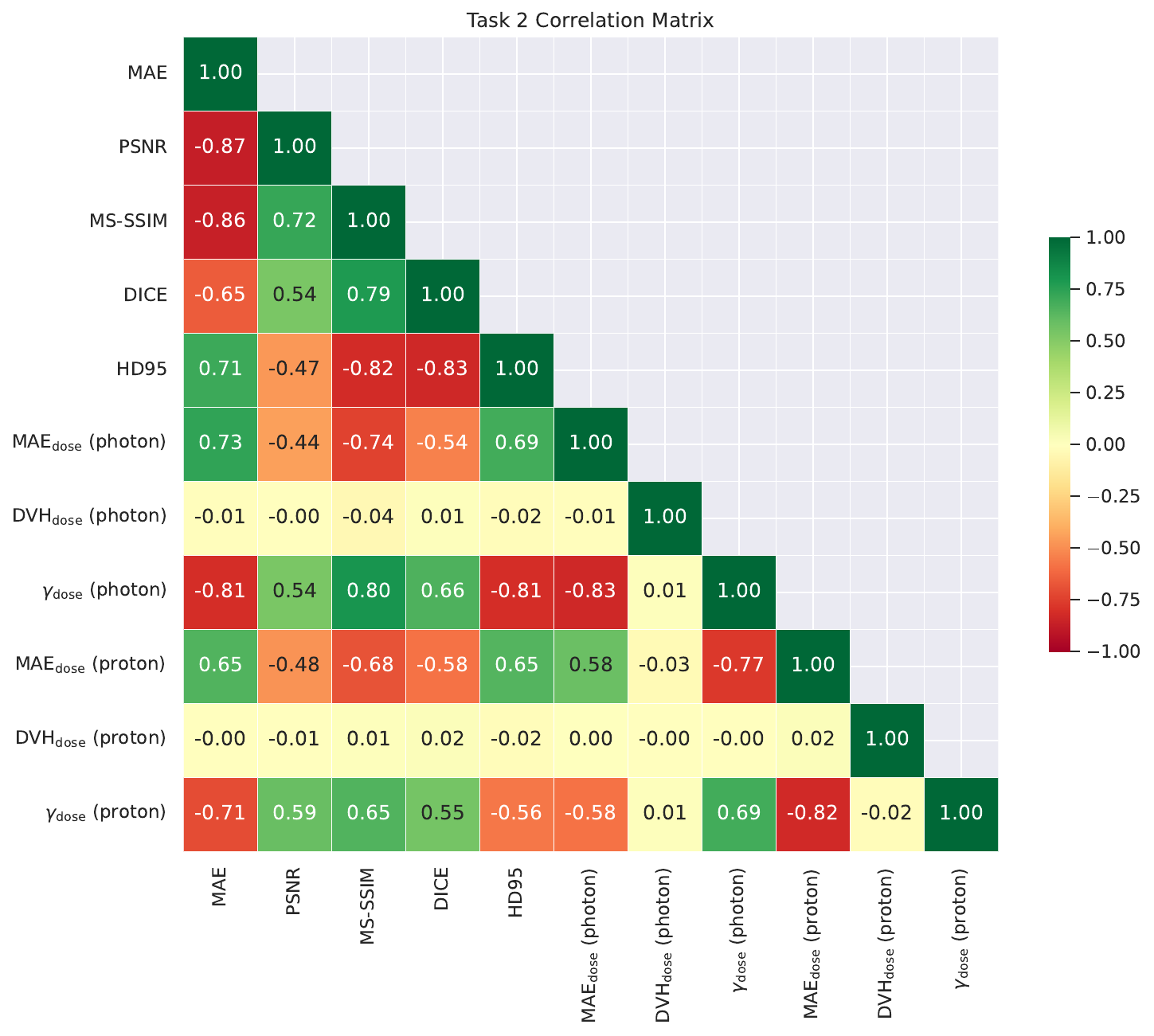}%
}
\caption{Spearman rank correlation coefficients ($\rho$) between evaluation metrics for (a) Task 1 (MRI-to-CT) and (b) Task 2 (CBCT-to-CT). Each matrix includes image similarity metrics (MAE, PSNR, MS-SSIM), segmentation metrics (Dice, HD95), and photon and proton dose metrics ($MAE_{dose}$, $DVH_{dose}$, $\gamma_{dose}$). Green denotes positive correlation and red denotes negative correlation.}
\label{fig:correlation_metrics}
\end{figure*}

\begin{figure*}[htbp]
\centering
\includegraphics[clip,height=5.5cm]{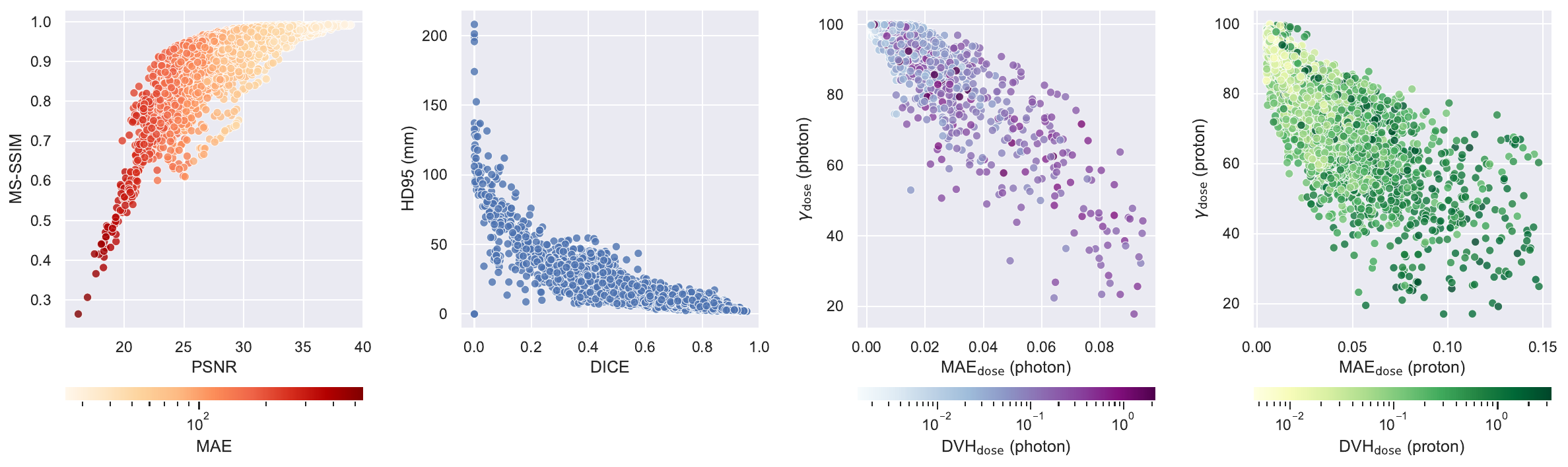}
\caption{Scatter plots illustrating pairwise relationships between the evaluation metrics across all submissions and test cases, complementing the Spearman correlation analysis in Figure 3. From left to right: MS-SSIM versus PSNR, with points colored by MAE (HU); $\gamma_{photon}$ pass rate versus photon $MAE_{dose}$ (Gy), colored by photon DVH error; $\gamma_{proton}$ versus proton $MAE_{dose}$ (Gy), colored by proton DVH error; and HD95 (mm) versus Dice.}
\label{fig:correlation_scatter}
\end{figure*}

Spearman rank correlations between all evaluation metrics are summarised in \autoref{fig:correlation_metrics} for both tasks. Inter-group correlations are visualized in scatter plots shown in \autoref{fig:correlation_scatter}. 

Image similarity metrics were strongly correlated with each other. MAE was strongly anti-correlated with PSNR ($\rho = -0.92$ in Task 1, $\rho = -0.87$ in Task 2) and with MS-SSIM (Task 1 $\rho = -0.76$ and Task 2 $\rho = -0.86$), while PSNR and MS-SSIM were positively correlated ($\rho = 0.77$ in Task 1 and $\rho = 0.72$ in Task 2). These coefficients indicate that the three image similarity metrics captured largely overlapping aspects of voxel-wise similarity.

Segmentation metrics were strongly linked to image similarity, particularly through MS-SSIM. Dice correlated with MS-SSIM at $\rho = 0.78$ in Task 1 and $\rho = 0.79$ in Task 2, and with MAE at $\rho = -0.58$ and $\rho = -0.65$, respectively. HD95 mirrored Dice closely, with correlation coefficients of $-0.84$ in Task 1 and $-0.83$ in Task 2 between the two segmentation metrics, and was more strongly tied to image similarity in Task 2 (HD95 vs. MAE: $\rho = 0.71$) than in Task 1 ($\rho = 0.36$). This pattern indicates that, in CBCT-to-CT synthesis, methods producing higher voxel-wise fidelity also tended to preserve organ boundaries more accurately, whereas in MRI-to-CT synthesis a stronger decoupling between intensity accuracy and geometric consistency was observed. 

The relationship between image similarity and dose metrics differed noticeably between the two tasks. In Task 1, photon and proton dose metrics correlated only moderately with image similarity (e.g. $\gamma_{\text{photon}}$ vs. MAE: $\rho = -0.50$ and $\gamma_{\text{proton}}$ vs. MAE: $\rho = -0.52$). In Task 2, these associations were substantially stronger. $\gamma_{\text{photon}}$ correlated with MS-SSIM at $\rho = 0.80$ and with MAE at $\rho = -0.81$, and the corresponding proton correlation coefficients reached $\rho = 0.65$ and $\rho = -0.71$. Within the dose group, the dose metrics were internally consistent: photon $MAE_{dose}$ and $\gamma_{\text{photon}}$ were strongly anti-correlated in both tasks ($\rho = -0.89$ in Task 1 and $\rho = -0.83$ in Task 2), similar for proton $MAE_{dose}$ and $\gamma_{\text{proton}}$ which resulted in correlation coefficients of $−0.44$ and $−0.82$, respectively. Photon and proton $\gamma$ pass rates were moderately to strongly correlated ($\rho = 0.54$ in Task 1, $\rho = 0.69$ in Task 2), confirming that methods performing well for photon plans tended also to perform well for proton plans. DVH-based metrics, in contrast, showed near-zero correlations with all other evaluation criteria in both tasks ($|\rho| \leq 0.18$ across all pairs). Photon and proton DVH metrics were also uncorrelated with each other and with their respective $\gamma$ and $MAE_{dose}$ counterparts. This finding suggests that in this challenge setup DVH-based summaries capture clinically relevant aspects of plan quality, such as target coverage and OAR dose, that are not reflected in voxel-wise image similarity, segmentation overlap, or even spatially aggregated dose-difference measures.

\subsection{Regional and center-wise analysis}
\begin{figure*}
    \centering
    \includegraphics[width=\linewidth]{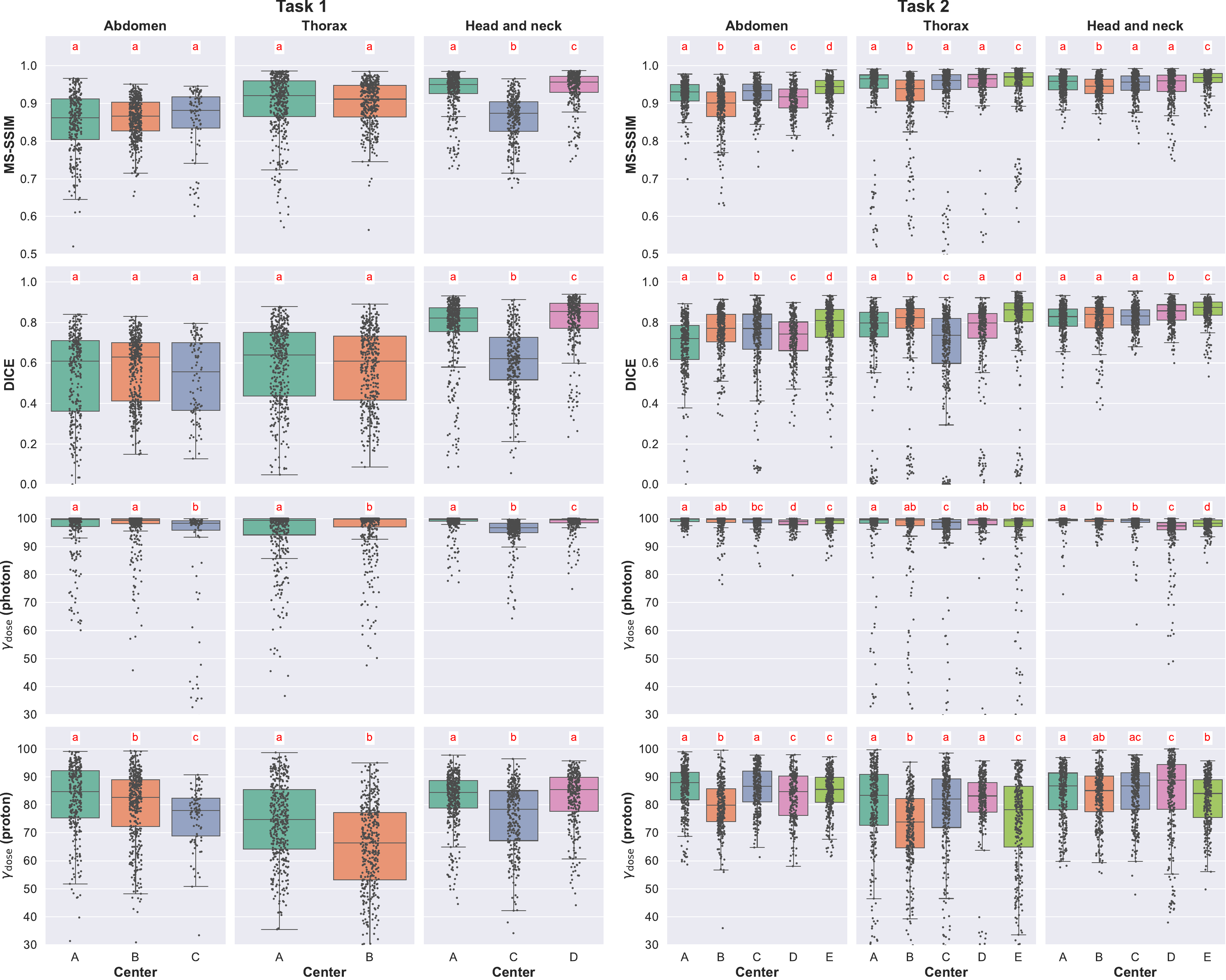}
    \caption{Boxplots of MS-SSIM, Dice, and $\gamma_{photon}$ and $\gamma_{proton}$ pass rates, stratified by anatomical region (abdomen, thorax, head and neck) and contributing center for Task 1 (MRI-to-CT, left) and Task 2 (CBCT-to-CT, right). Letters above each box denote statistically homogeneous groups based on pairwise Mann–Whitney U tests with Holm correction ($\alpha$ = 0.05): centers sharing a letter within a panel are not significantly different, whereas centers labeled with disjoint letters differ significantly. In the interest of better readability, some outliers might be cut-off on the y-axis.}
    \label{fig:data_invest}
\end{figure*}

The regional analysis presented in Figure~\ref{fig:data_invest} revealed substantial variability across anatomical sites, datasets and tasks. In Task 1, head-and-neck cases achieved the highest image and geometric similarity. Noticeably, in the head-and-neck region, data from center C, resulted in significantly worse results than data from centers A and D. Abdomen and Thorax did not show any significant inter-center differences. Dose-based metrics exhibited a different pattern: $\gamma_{\text{photon}}$ was uniformly high across all regions and centers, with modest but statistically significant differences emerging in the abdomen (center C lower than A and B), thorax (center A lower than center B) and head-and-neck (center C noticeably lower than A and D). Proton dose agreement was systematically lower and more variable than photon dose agreement, particularly in the thorax, where center B showed the lowest $\gamma_{\text{proton}}$ values. In the abdomen, all three centers differed significantly from one another for proton dose, and in head-and-neck, centers A and D performed comparably while center C was significantly lower.

In Task 2, which involved five contributing centers, \autoref{fig:data_invest} highlights the previous finding of overall better performance and less variability. However, inter-center differences were present across all metrics and regions. MS-SSIM and DICE distributions showed multiple statistically distinct groupings per region, with data from center E resulting in the best results across all three anatomical regions. Photon dose metrics remained generally high across all regions and datasets, but still exhibited small but statistically significant differences between groups. Proton dose metrics again showed a larger inter-center spread, with median $\gamma_{proton}$ values ranging from approximately 80\% (abdomen center B) up to 90\% (head-and-neck center D).

\subsection{Model design predictors}

\begin{figure*}[htbp]
    \centering
    \includegraphics[width=\linewidth]{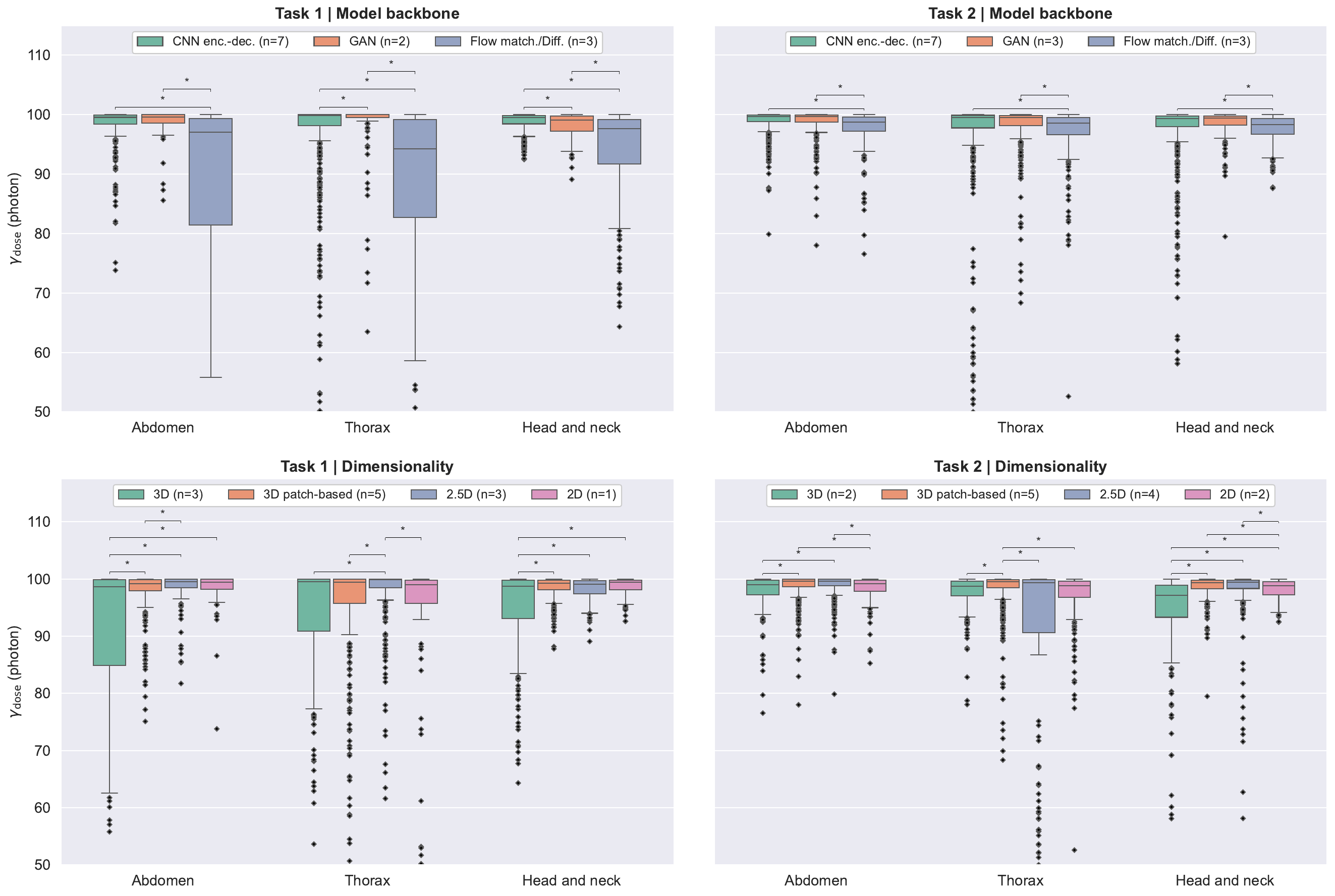}
    \caption{Boxplots of photon gamma pass rates ($\gamma_{photon}$ 2\%/mm criterion) for all patients in each subtask across the abdomen, thorax, and head-and-neck regions, grouped by model design choice. The top row groups submissions by backbone architecture (CNN encoder–decoder, GAN, flow matching/diffusion), and the bottom row groups submissions by spatial configuration (3D, 3D patch-based, 2.5D, 2D). The n in the legend indicates the number of teams represented in each group. Asterisks denote statistically significant differences between groups within a given task–region combination (Mann–Whitney U test, $\alpha$ = 0.01). Outliers below 50\% were omitted in the plot for better readability.}
    \label{fig:backbone_performance}
\end{figure*}

To assess how methodological design choices influenced synthesis quality, we stratified submissions by model backbone and network dimensionality and compared their $\gamma_{\text{photon}}$-pass rates across anatomical regions for both tasks (\autoref{fig:backbone_performance}). 

For the model backbone analysis, submissions were grouped into convolutional encoder–decoder networks, generative adversarial networks (GANs), and flow-matching/diffusion models. In Task 1, CNN encoder–decoders and GANs achieved consistently high $\gamma_{\text{photon}}$-pass rates with tight distributions across all three anatomical regions, whereas flow-matching/diffusion models exhibited wider distributions and lower medians, particularly in the abdomen and thorax. Pairwise comparisons showed that flow-matching/diffusion models differed significantly from both CNN-based and GAN-based methods in all three regions. In Task 2, distributions across the three categories were more tightly clustered with significant differences persisting only between flow-matching/diffusion models and the other two groups.

To evaluate the impact of the spatial dimension, submissions were grouped into full-volume 3D, 3D patch-based, 2.5D, and 2D approaches. In both tasks, full-volume 3D models exhibited the widest distributions and lowest medians, most notably in the abdomen for Task 1, where the lower whisker extended below $\gamma_{\text{photon}}$=65\%. Patch-based 3D, 2.5D, and 2D approaches produced higher and more tightly distributed $\gamma_{\text{photon}}$-pass rates across nearly all region–task combinations, with multiple statistically significant pairwise differences favouring these strategies over full-volume 3D processing in both tasks. In Task 2, 2.5D-based trained models seem to be an outlier in the thorax region, with significantly worse performance than in other approaches and regions.

\label{subsec: results data influence}

\subsection{Ranking stability}

\begin{figure}[htbp]
    \subfloat[Task 1]{%
    \includegraphics[width=\linewidth]
    {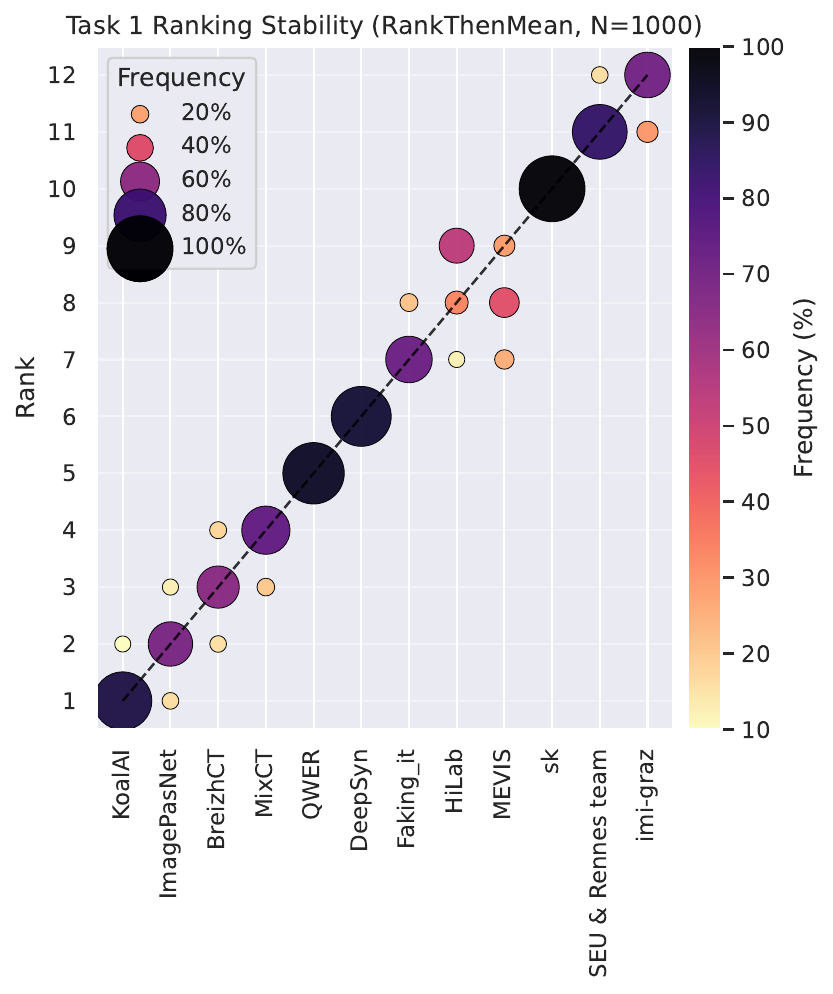}{}}
    \hfill
    \subfloat[Task 2]{%
    \includegraphics[width=\linewidth]
    {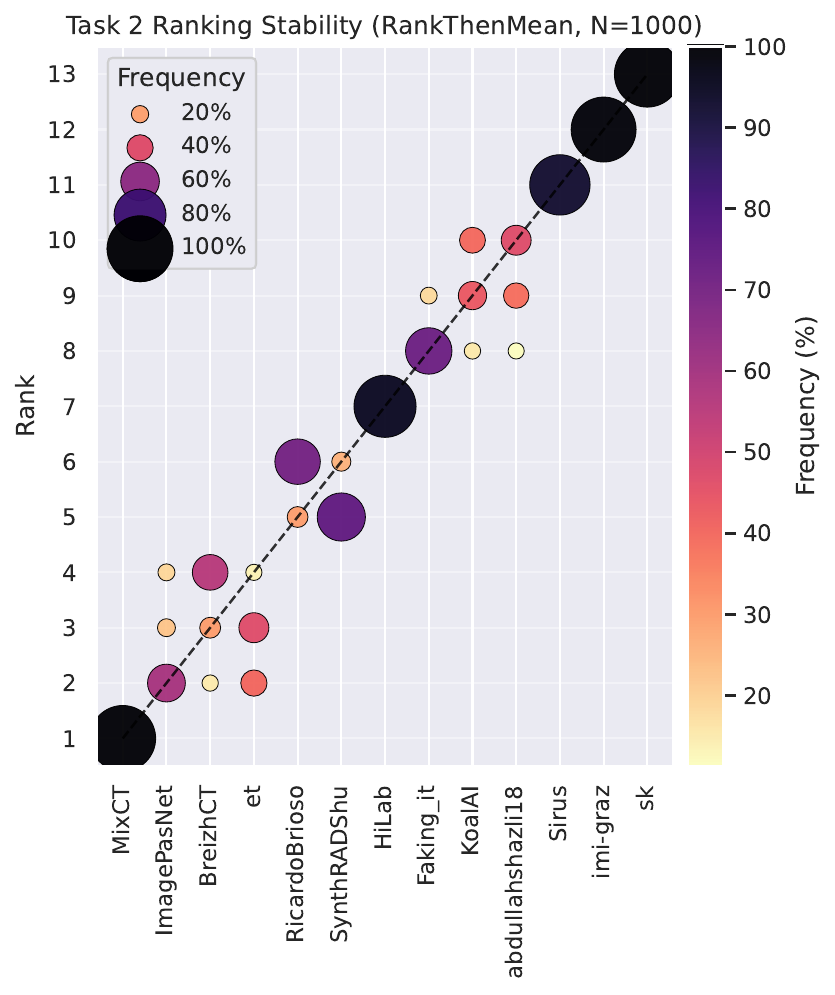}{}}
    \caption{Visualization of ranking stability. Blob size is proportional to the frequency of the rank achieved on bootstrapping (N = 1000).}
    \label{fig:ranking-stability}
\end{figure}

\autoref{fig:ranking-stability} shows that the final Rank-Then-Mean rankings were generally stable for both tasks. In Task 1, the top-ranked teams were consistently assigned high ranks, with KoalAI most frequently ranked first and ImagePassNet second. The largest uncertainty occurred among neighbouring teams in the middle of the leaderboard, particularly around ranks 7--9. The lowest-ranked teams showed only minor variation, mainly between adjacent ranks. In Task 2, stability was similarly high for both the best- and worst-performing teams. However, reduced stability was observed not only among mid-ranked teams around ranks 8--10, but also among higher-ranked teams occupying ranks 2--4. Overall, the analysis indicates that the Rank-Then-Mean rankings were robust, with rank fluctuations largely restricted to adjacent teams with similar performance.

%% file: 05_Discussion.tex
\section{Discussion}
\label{sec:discussion}
\subsection{Summary of the main findings}
SynthRAD2025 provided a multi-center, multi-region benchmark for evaluating deep learning-based sCT generation from MRI and CBCT. Compared with SynthRAD2023, SynthRAD2025 expanded the anatomical scope beyond the brain and pelvis, thereby providing a more demanding benchmark for generalizable sCT generation \citep{huijben2024generating}. Across both tasks, the leading submissions achieved high image similarity, good geometric agreement, and high photon dose agreement, but Task 2 consistently outperformed Task 1. The lowest MAE was $64.8 \pm 21.3$ HU in Task 1 and $48.3 \pm 13.4$ HU in Task 2, while photon $\gamma$ pass rates reached $98.1\%$--$98.9\%$ among the top six Task 1 methods and exceeded $99\%$ among the top four Task 2 methods. In both tasks, proton $\gamma$ pass rates remained lower and more variable than photon $\gamma$ pass rates.
The results further showed that sCT performance was not captured by any single metric. In Task 1, the top four teams were separated by only $3.1$ HU in MAE, $0.4$ dB in PSNR, and 0.005 in MS-SSIM, yet the strongest geometric performance was achieved by MixCT (rank 4) rather than by the overall top 3-ranked teams. In task 2, image similarity and geometric rankings were more concordant, with the top four teams achieving Dice values of $0.84$--$0.86$ and HD95 values of $4.53$--$5.08$ mm. All ranked submissions outperformed the water-equivalent baseline on image-similarity and segmentation metrics, whereas the lowest-ranked methods in both tasks approached baseline performance on several dose metrics.

\subsection{Interpretation in relation to the study aim}
The SynthRAD2025 challenge aimed to benchmark fully automated sCT generation methods across institutions, anatomical regions, and two clinically relevant source modalities. The results showed that current supervised deep learning pipelines could perform strongly across this heterogeneous benchmark, but also that performance remained contingent on task, anatomy, and endpoint. Thus, the present data supported substantial progress toward generalizable sCT generation, while arguing against the assumption that good performance on one task or one metric would translate uniformly across settings.
The consistent advantage of Task 2 suggested that CBCT-to-CT synthesis was less challenging than MRI-to-CT synthesis within the present framework. Although the challenge was not designed to establish mechanism, this pattern was compatible with a smaller domain shift between CBCT and CT than between MRI and CT. The closer alignment of image similarity, segmentation, and dose rankings in Task 2 supported this interpretation. By contrast, Task 1 showed a clearer dissociation between voxel-wise similarity and structural fidelity: MixCT achieved the best Dice ($0.79 \pm 0.11$) and HD95 ($5.77 \pm 3.31$ mm) but ranked fourth overall, and proton dose metrics showed greater dispersion than photon dose metrics.
The correlation analysis refined this interpretation. Image similarity metrics were strongly interrelated in both tasks, with MAE and PSNR showing strong inverse correlations ($\rho=-0.92$ in Task 1 and $\rho=-0.87$ in Task 2). MS-SSIM was also strongly associated with Dice ($\rho=0.78$ in Task 1 and $\rho=0.79$ in Task 2). However, the relationship between image similarity and dose was task dependent: it was only moderate in Task 1 but stronger in Task 2, whereas DVH-based metrics remained weakly correlated with almost all other endpoints in both tasks. These results indicated that methods with similar voxel-wise fidelity could nevertheless differ in clinically relevant downstream performance, particularly for beam-path-dependent dose calculation.

\subsection{Comparison with prior work}
These observations were consistent with the rationale established by SynthRAD2023, which showed that image-based metrics alone did not capture radiotherapy relevance \citep{huijben2024generating}. Compared with SynthRAD2023, which focused on brain and pelvic anatomies, SynthRAD2025 examined head-and-neck, thoracic, and abdominal imaging and therefore encompassed a broader range of anatomical complexity and motion-related variability. Despite this increased heterogeneity, the central message was similar: image fidelity alone did not provide an adequate surrogate for radiotherapy relevance. At the same time, the present results added distinction. Whereas SynthRAD2023 highlighted a negligible association between image similarity and dose accuracy, SynthRAD2025 suggested that this relationship was task dependent. In the current challenge, DVH metrics remained weakly coupled to the rest of the evaluation framework, but image similarity and dose metrics were more closely aligned in Task 2 than in Task 1. Thus, the link between image fidelity and dosimetric fidelity appeared to depend on modality and benchmark context, rather than being uniformly weak \citep{huijben2024generating, chourak2022qa, reinke2024metricpitfalls}.
The dosimetric findings were also consistent with prior studies showing that sCT-based dose calculation could reach clinically acceptable accuracy in MRI-only and CBCT-guided workflows \citep{dinkla2018brain_sct, wyatt2023pelvic_sct, ohara2022cbct_sct}. However, the present multi-center evaluation showed that near-ceiling photon dose agreement could coexist with greater variability in proton dose agreement and with residual differences across anatomical sites and centers. The current data therefore extended prior single-center experience by emphasizing robustness across heterogeneous data, rather than average performance alone.

\subsection{Clinical and methodological implications}
From a radiotherapy perspective, the challenge results are encouraging. For photon therapy, the leading methods achieved very high agreement with the reference CT, particularly in Task 2, where the top submissions exceeded $99\%$ photon $\gamma$ pass rate. Within this benchmark, these results suggested that sCT generation could support accurate photon dose recalculation for many cases. Proton results were less permissive: proton $\gamma$ pass rates remained lower ($79.2\%$--$84.6\%$ among the top six methods in Task 1 and $86.4\%$--$88.6\%$ among the top four methods in Task 2), consistent with greater sensitivity to residual density and boundary errors. This likely reflected not only the stronger dependence of proton dose on accurate path length and stopping-power estimation, but also the benchmark setup itself: dose was recalculated on deformably matched reference CTs, and proton plans were not robustly optimized. Under these conditions, residual registration errors or local boundary mismatches may have affected proton dose agreement more strongly than photon dose agreement. Head-and-neck cases generally yielded the most stable performance, whereas thoracic and abdominal cases showed broader distributions and stronger center effects. These patterns suggested that clinical translation would require site-specific validation, particularly for motion-prone anatomies and proton workflows.
Methodologically, the results did not identify a single dominant architectural strategy. CNN encoder--decoder and GAN-based methods generally showed tighter and higher-performing distributions than flow-matching/diffusion methods, whereas full-volume 3D approaches showed the widest variability in photon $\gamma$ performance. The absence of a clear advantage for flow-matching/diffusion methods should, however, have been interpreted cautiously. These approaches were represented by only three submissions per task, and they were generally more complex than conventional CNN-based pipelines in both optimization and inference. Within the present challenge, that added complexity did not translate into a consistent performance advantage. More broadly, the strongest submissions were characterized less by one specific model class than by well-executed supervised pipelines built on paired, deformably registered data. These findings argued for benchmark designs that integrate image similarity, geometric consistency, and dose recalculation, and for model-development strategies that optimize clinically relevant downstream performance rather than image fidelity alone.
As in SynthRAD2023, the challenge results should not be interpreted as immediate evidence for universal clinical deployment. Before adopting a given method in routine practice, site-specific validation remains essential, including assessments of local scanners, imaging protocols, immobilization workflows, and treatment planning systems. Nevertheless, the overall quality of the best-performing submissions indicates that deep learning-based sCT generation is maturing into a clinically relevant technology for both MRI-guided and CBCT-guided adaptive radiotherapy workflows \citep{Cusumano2026}. 

\subsection{Limitations of the SynthRAD2025 dataset and setup}
Several limitations qualified the interpretation of these results. First, the benchmark relied on paired data generated through deformable registration, so residual mismatch between source images and reference CT may have contributed to apparent errors, particularly in MRI-to-CT synthesis and in motion-prone thoracic and abdominal cases \citep{brock2017tg132}. Second, the heterogeneity that strengthened the benchmark also introduced confounding through acquisition protocol, image quality, anatomical coverage, and center-specific case mix; the present analyses identified robust patterns, but did not disentangle their causes. Third, the model-design comparisons were based on small subgroups and pooled complete pipelines that differed in many respects beyond backbone or dimensionality. Finally, although dose metrics contributed to the final ranking, participants primarily optimized against image-based validation signals, which may have favored methods tuned to image similarity rather than explicit dosimetric objectives \citep{huijben2024generating, maierhein2018rankings}.

\subsection{Future directions}
SynthRAD2025 demonstrates that high-quality multi-region sCT generation is feasible, but the task is not yet solved. Several directions for future work emerge from the present challenge.
First, improved handling of anatomically complex and motion-prone regions remains necessary, particularly for the thorax and abdomen.
Second, the moderate coupling between image and dose metrics suggests that future methods should optimize directly for clinically relevant objectives. This may include dose-aware loss functions, organ-specific structural supervision, or hybrid evaluation pipelines that combine image similarity, segmentation consistency, and dose recalculation during development \citep{spadea2021sctreview}, \citep{chourak2022qa}.
Third, registration quality remains a major bottleneck, especially for MRI-to-CT synthesis. Future datasets and challenges may benefit from improved pairing strategies, repeated imaging with closer temporal spacing, or benchmark setups that explicitly quantify the effect of residual registration error \citep{brock2017tg132}.

%% file: 06_Conclusion.tex
\section{Conclusion}
\label{sec:conclusion}

SynthRAD2025 provides a comprehensive benchmark for synthetic CT generation from MRI and CBCT across head-and-neck, thoracic, and abdominal anatomy. The challenge results indicate that current deep learning methods can generate high-quality sCTs with strong image similarity, good geometric consistency, and highly accurate photon dose calculations, particularly for CBCT-to-CT synthesis.
Across both tasks, the best-performing methods achieved tightly clustered performance, with Task 2 consistently outperforming Task 1. CNN-based encoder--decoder models and GAN-based approaches showed the most robust overall performance. In contrast, diffusion- and flow-based approaches have not yet demonstrated a consistent advantage in this challenge setting. The analysis further showed that image similarity metrics are strongly correlated with one another and with segmentation quality, but only moderately associated with dose metrics, emphasizing that image quality alone is insufficient to assess clinical utility.
Overall, the SynthRAD2025 challenge highlights the strong potential of deep learning for sCT generation in adaptive radiotherapy, while also underscoring the remaining importance of robust evaluation, dose-based validation, and generalization across anatomically and clinically heterogeneous data.

%% file: 07_suppA.tex
\section*{Supplementary Document A: Participant methods}
\label{sec:suppA}
Each subsection briefly describes the methods used by the participating teams. The top five methods were presented in the main paper. The team names correspond to the submission reported on the leaderboard at \url{https://synthrad2025.grand-challenge.org/evaluation/test-task-1-mri/leaderboard/} and \url{https://synthrad2025.grand-challenge.org/evaluation/test-task-1-cbct/leaderboard/}

\subsection{SEU \& Rennes (task 1)}
SEU \& Rennesn proposed a diffusion-based MRI-to-CT synthesis framework using a 3D Swin-VNet architecture. The method learns to iteratively denoise CT images from Gaussian noise, conditioned on MRI input, via a forward–reverse diffusion process.
SEU \& Rennes used inputs consisting of concatenated noisy CT volumes and MRI images. The architecture combines convolutional layers with Swin transformer blocks to capture both local and global contextual information. Temporal conditioning was incorporated via sinusoidal timestep embeddings injected into all network layers.
SEU \& Rennes trained the model using AdamW with weight decay and a learning rate of approximately 1e−4 for up to 1000 epochs. Residual and skip connections were used extensively to stabilize training and preserve spatial information. Although patch sizes were not explicitly specified, the model operates on full 3D volumes with normalized inputs.

\subsection{HiLab (task 1 \& 2)} \label{method HiLab}
HiLab employed a 2D UNet++ architecture trained using a two-stage strategy consisting of self-supervised pre-training followed by supervised fine-tuning. The network operated slice-wise and reconstructed full 3D volumes after inference. During pre-training, CBCT-like degradations were simulated from CT images, including motion blur, Fourier-domain low-frequency perturbations, and Radon-domain occlusions, applied individually or in combination to improve robustness to CBCT artifacts.

The backbone network followed a nested UNet++ design with dense skip connections and intermediate convolutional pathways, enabling enhanced multi-scale feature fusion. MRI inputs were normalized using mean–standard deviation normalization. CT and CBCT volumes were clipped to standard Hounsfield Unit ranges (e.g., [−1024, 3071]) and linearly scaled to a normalized range before training. A single model was trained jointly across anatomical regions for Task 1, whereas Task 2 benefited from the artifact-simulation strategy.

Training was performed using slice-wise reconstruction losses (L1/MAE), and inference consisted of slice-wise prediction followed by stacking into 3D volumes. This design enabled efficient training while maintaining robustness across anatomical sites and imaging conditions. 

\subsection{IMI-Graz (task 1 \& 2)} \label{method imi-graz}
IMI-Graz proposed a fully 3D flow-matching framework, in which synthetic CT generation is formulated as a continuous transformation from Gaussian noise to a target CT image via a learned velocity field. The model used a 3D U-Net architecture to predict velocity fields, conditioned on the input modality via a lightweight encoder that extracts contextual features.

All input volumes were resampled to 128×128×128 voxels with 1 mm isotropic spacing. MRI inputs were z-score normalized and clipped to [−3, 3], while CBCT and CT volumes were clipped to [−1024, 3071] HU and scaled by a factor of 1/1000. The model input consisted of a noisy CT sample concatenated with modality-specific features extracted from the conditioning image.

The loss function combined L1 and MSE losses computed on the predicted velocity fields. Training was performed using AdamW with a learning rate of $1\times10^{-4}$ and weight decay of $1\times10^{-5}$, with data augmentation including random rotations and translations. Separate models were trained for each anatomical region and task, and inference involved reconstructing the CT volume, followed by resampling to the original resolution and intensity scale. 

\subsection{SEU \& Rennes (task 1)} \label{method SEU_Rennes}
The SEU \& Rennes team proposed a diffusion-based framework for MRI-to-CT synthesis using a Swin-VNet architecture within a denoising diffusion probabilistic model (DDPM) formulation. The method models CT generation as a reverse diffusion process, in which Gaussian noise is progressively denoised to produce a synthetic CT conditioned on the MRI input.

The architecture consisted of a 3D encoder–decoder structure combining convolutional layers with Swin Transformer blocks. The encoder included an initial residual convolutional block followed by multiple Swin-based stages, enabling both local feature extraction and global context modeling. The decoder mirrored this structure to reconstruct the CT image progressively. Inputs consisted of concatenating noisy CT volumes at time step \(t\) with their corresponding MRI volumes.

Although explicit clipping ranges were not specified, the framework operates within the standard diffusion paradigm, in which intensities are normalized before training. Training was performed across multiple diffusion time steps to optimize reconstruction of denoised CT images. At inference, the reverse diffusion process was applied to generate full 3D CT volumes conditioned on MRI inputs.

\subsection{SK (task 1 \& 2)} \label{method sk}
Team SK employed a U-Net–based architecture with a pretrained ConvNeXt-L encoder and a convolutional decoder. The model operated on 2.5D inputs by stacking adjacent slices as input channels, thereby incorporating limited 3D contextual information while maintaining computational efficiency.

The encoder used fixed, pretrained ConvNeXt-L features, while the decoder reconstructed CT intensities using convolutional layers. Separate models were trained for each task and anatomical region. CT and CBCT images were clipped to clinically relevant Hounsfield Unit ranges (e.g., [−1024, 3071]) and normalized before training, while MRI inputs were normalized using modality-specific statistics.

The loss function combined L1 loss and VGG-based perceptual loss to balance pixel-wise accuracy and structural fidelity. Training was performed using AdamW with cosine-annealed learning-rate scheduling. Inference was conducted slice-wise using multi-slice inputs, followed by reconstruction of full 3D volumes. 

\subsection{MEVIS (task 1)} \label{method MEVIS}
MEVIS proposed a 3D patch-based hybrid U-Net architecture that integrates Residual Dilated Swin Transformer (RDSformer) blocks into the U-Net backbone. The model processed 3D patches of size 96×96×96 and reconstructed full volumes using a sliding-window inference strategy.

The architecture incorporated channel-spatial attention mechanisms and learnable 3D positional encodings to enhance both local and global feature representation. Separate models were trained for the head-and-neck and abdomen/thorax regions to reflect anatomical variability. MRI inputs were used directly without explicit clipping, while CT targets were implicitly handled through regression-based losses.

Training was performed using a combination of mean squared error and perceptual loss, optimized with AdamW and polynomial learning rate decay. During inference, overlapping patches were aggregated to reconstruct the final synthetic CT volume, ensuring spatial consistency across patch boundaries.

\subsection{Faking\_it (task 1 \& 2)} \label{method fakingit}
Faking\_it proposed a 3D Variable-Step Denoising Diffusion Probabilistic Model (VS-DDPM) with a Swin-ViT backbone. The model employed a variable number of diffusion steps, allowing adaptation between computational efficiency and synthesis quality.

A unified preprocessing pipeline was applied across both tasks, including rigid registration, intensity normalization, and cropping to the body mask. Volumes were processed as patches of size 128×128×32. Although explicit clipping ranges were not specified, normalization was applied consistently across modalities before training.

The loss function combined MSE, MAE, SSIM loss, and a variational lower bound objective. Training was performed using AdamW with cosine learning-rate scheduling, and variable diffusion step schedules were applied during training. Final predictions were reconstructed from overlapping patches to form full 3D volumes. 

\subsection{DeepSyn (task 1)} \label{method DeepSyn}
DeepSyn employed a conditional generative adversarial network (cGAN) consisting of a ResNet-based generator and a patch-based discriminator. The generator incorporated residual blocks to improve gradient flow and capture long-range dependencies, while the discriminator enforced local image realism.

The model operated on 2D slices with input size 384×384×3, where adjacent slices were stacked as channels. MRI inputs were normalized to [0, 1] using patient-specific statistics, while CT images were clipped to [−1000, 2600] HU and normalized accordingly before training.

The loss function combined L1 loss, SSIM loss, PSNR-based loss, and adversarial loss. Training was performed using Adam with an initial learning rate of 2×$10^{-4}$ and linear decay scheduling. Inference was performed slice-wise, followed by reconstruction into full 3D volumes. 

\subsection{SynthRADShu (task 2)} \label{method SynthRADShu}
SynthRADShu proposed a 3D GAN-based framework using a Hypergraph-based Multi-resolution Residual Attention U-Net (HMRAUNet) as the generator and a multi-resolution U-Net discriminator. The architecture incorporated hypergraph-based feature modeling and multi-resolution attention mechanisms.

Training was performed on 3D patches of size 96×96×96 using a two-stage full-parameter fine-tuning strategy, with learning rates of 1×$10^{-4}$ for pre-training and 1×$10^{-5}$ for fine-tuning. Preprocessing included intensity scaling and spatial normalization, although explicit clipping ranges were not specified.

The loss function combined voxel-wise adversarial loss and a 2.5D perceptual loss. Data augmentation included random flipping and padding. Inference was performed using patch-based prediction followed by reconstruction of full volumes.

\subsection{Sirus (task 2)} \label{method Sirus}
Sirus proposed a flow-based generative framework based on Bi-directional Discrete Process Matching (Bi-DPM), modeling the transformation between CBCT and CT as a bidirectional matching of intermediate states in forward and backward ODE flows.

The model processes 3D volumes as 2D slices to improve computational efficiency. A unified normalization strategy was applied across inputs, although explicit clipping ranges were not specified. The framework leverages paired data to enforce consistency between modalities.

Training focuses on matching intermediate states across forward and backward trajectories, enabling accurate reconstruction without predefined transport paths. Inference generates CT volumes through learned bidirectional mappings.

\subsection{AbdullahShazli18 (task 2)} \label{method AbdullahShazli18}
AbdullahShazli18 employed a fully 3D DynUNet architecture from MONAI for CBCT-to-CT synthesis. The model was trained on sub-volumes of size 128×128×64 with batch size 1, using mask-guided cropping.

CBCT and CT intensities were clipped to [−1000, 1000] HU and normalized to [0, 1]. The architecture utilized anisotropic kernels, residual connections, and flexible receptive fields to capture multi-scale volumetric features.

The loss function combined MAE and MSE reconstruction losses, 3D SSIM, contextual loss based on MedicalNet features, anatomical feature-consistency loss, and optional adversarial and gradient-consistency terms. Training was implemented in PyTorch Lightning with MONAI, and inference used patch-based reconstruction of full volumes.

%% file: 08_suppB.tex
\section*{Supplementary Document B: Supplementary analyses and results}
\label{sec:suppB}
This document provides supplementary analyses that support the interpretation of the SynthRAD2025 challenge results for both tasks: MRI-to-CT synthesis (Task 1) and CBCT-to-CT synthesis (Task 2). Specifically, it presents pairwise statistical comparisons of submission performance for each evaluation metric and qualitative examples of two challenging test cases.

\begin{figure}[h]
    \centering
    \includegraphics[width=0.85\linewidth]{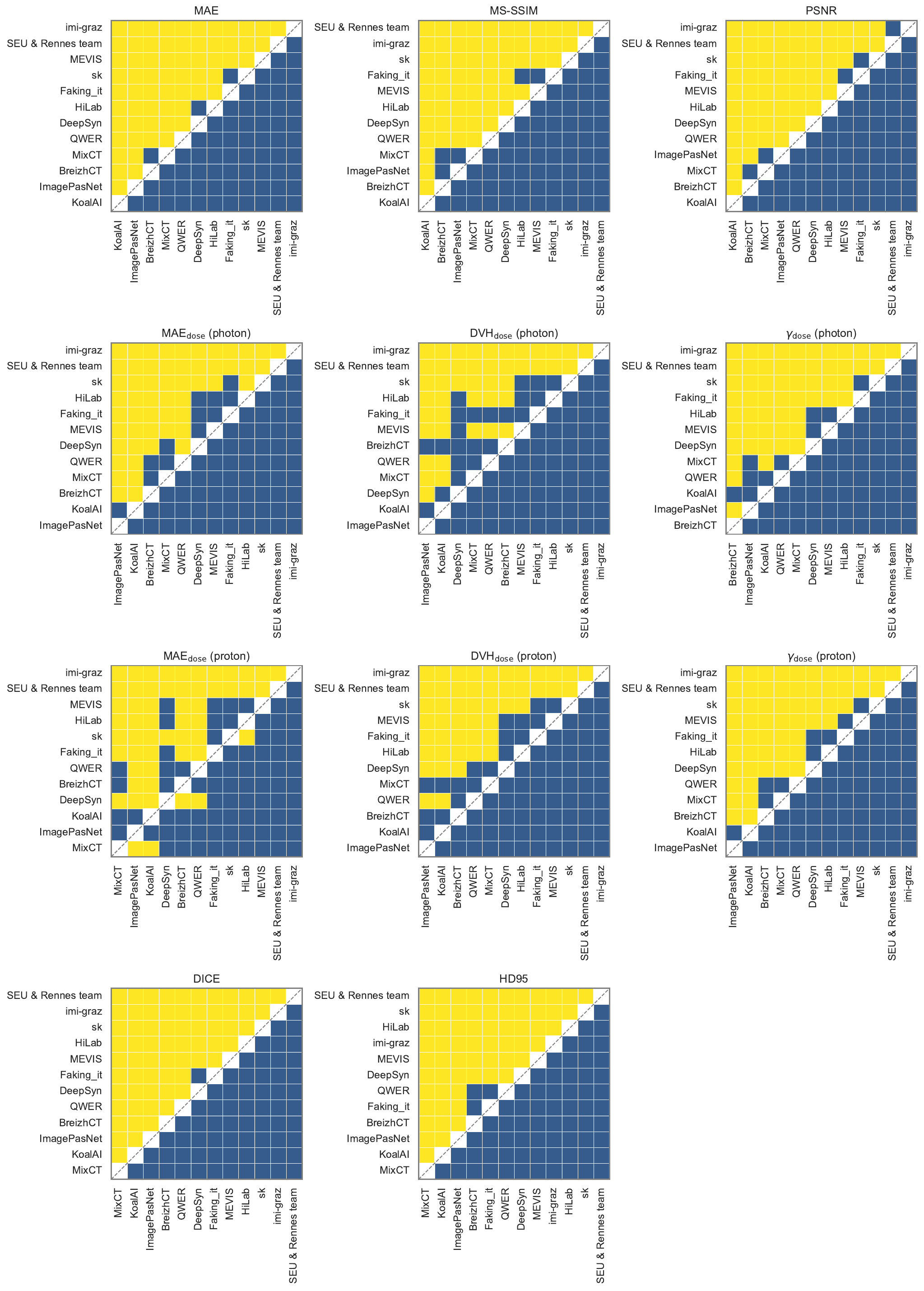}
    \caption{Significance maps for task 1 (MRI-to-CT). Pairwise comparisons of team performance for each evaluation metric are shown, with teams ordered by their ranking on that metric. Yellow regions in the upper triangle indicate that the method on the x-axis significantly outperforms the method on the y-axis ($p < 0.05$, Wilcoxon signed-rank test with Holm correction), while blue indicates no statistically significant difference. In the lower triangle, blue regions indicate that the method on the x-axis performs significantly worse than the method on the y-axis.}
    \label{fig:significance_maps_appendix_task1}
\end{figure}
\begin{figure}[h]
    \centering
    \includegraphics[width=0.85\linewidth]{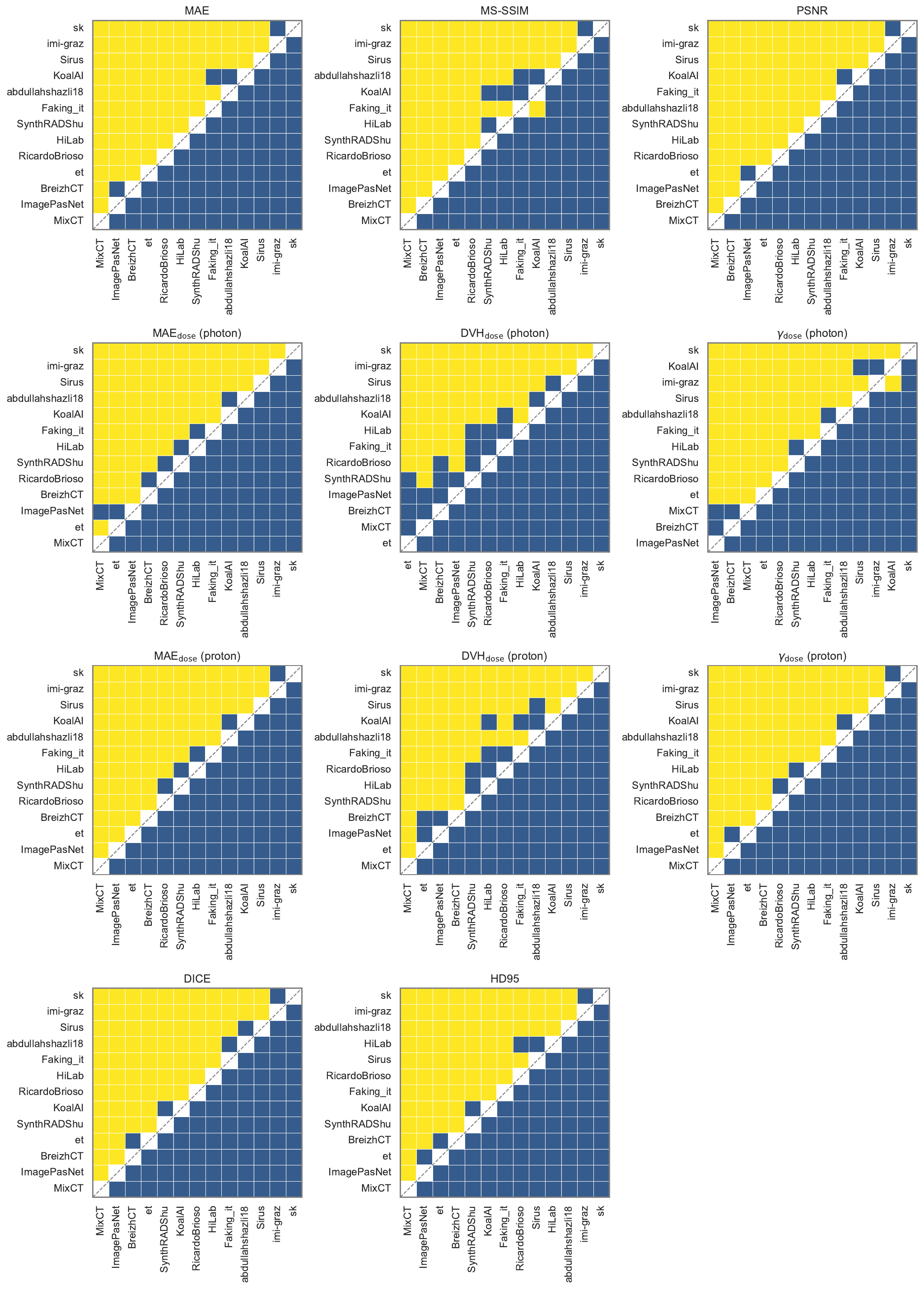}
    \caption{Significance maps for task 2 (CBCT-to-CT). Pairwise comparisons of team performance for each evaluation metric are shown, with teams ordered by their ranking on that metric. Yellow regions in the upper triangle indicate that the method on the x-axis significantly outperforms the method on the y-axis ($p < 0.05$, Wilcoxon signed-rank test with Holm correction), while blue indicates no statistically significant difference. In the lower triangle, blue regions indicate that the method on the x-axis performs significantly worse than the method on the y-axis.}
    \label{fig:significance_maps_appendix_task2}
\end{figure}

\subsection{Statistical comparison across submissions}
\label{app:sct_perf}
To assess whether differences in performance between submissions were statistically significant, we performed pairwise Wilcoxon signed-rank tests for each evaluation metric and adjusted the resulting $p$ values using Holm's procedure. The significance level was set to $\alpha=0.05$ (\Cref{fig:significance_maps_appendix_task1,fig:significance_maps_appendix_task2}). In each significance map, submissions were ordered from best to worst for the respective metric. Accordingly, extended yellow regions in the upper triangle indicated that higher-ranked submissions significantly outperformed lower-ranked submissions, whereas blue regions indicated no statistically significant difference.

For the image-similarity metrics (MAE, PSNR, and MS-SSIM), both tasks showed a clear hierarchical structure. Higher-ranked submissions significantly outperformed most lower-ranked submissions, producing large contiguous regions of significance in the upper triangle of the matrices. This pattern was especially pronounced when top-ranked methods were compared with low-ranked methods, indicating that gains in voxel-wise accuracy and structural similarity were robust across cases.

The geometric metrics (Dice and HD95) also showed substantial discriminatory power in both tasks. Dice, in particular, displayed a pronounced hierarchical structure that closely resembled MAE and PSNR, with higher-ranked submissions significantly outperforming most lower-ranked submissions. HD95 showed the same overall trend, although with more localized regions of non-significant differences, indicating that boundary agreement was separated somewhat less uniformly than volumetric overlap. Overall, the geometric metrics broadly supported the ranking structure observed for the image-similarity metrics, while still showing reduced discrimination for some closely ranked submissions.

Dose-based metrics showed a more heterogeneous pattern. Photon and proton $\gamma$ pass rates retained a clear hierarchical structure in both tasks, and $MAE_{\text{dose}}$ also separated submissions, particularly in Task 2. By contrast, the DVH-based metrics showed many non-significant pairwise comparisons and were the least discriminative endpoints in both tasks. Thus, the weaker statistical separation in dose space was driven primarily by the DVH summaries rather than uniformly by all dose-based metrics.

Among the dose-based endpoints, proton metrics appeared somewhat more variable than their photon counterparts, especially for the DVH-based comparisons. Nevertheless, the main impression from both tasks was that dose-based evaluation did not behave as a single homogeneous group: $\gamma$ pass rates and, in part, $MAE_{\text{dose}}$ preserved substantial ranking information, whereas DVH-based summaries showed markedly weaker separation.

Overall, these analyses supported the main manuscript by showing that image-similarity and geometric metrics, and to some extent also $\gamma$ pass rates, detected systematic performance differences between submissions. In contrast, DVH-based metrics were less sensitive to those differences, particularly among the leading methods. This pattern suggested that the different metric families captured complementary aspects of performance and should not be interpreted as interchangeable.

\begin{figure}[h]

 \subfloat[Example sCTs for outlier patient 1ABA066.]{%
   \includegraphics[clip,width=0.99\linewidth]{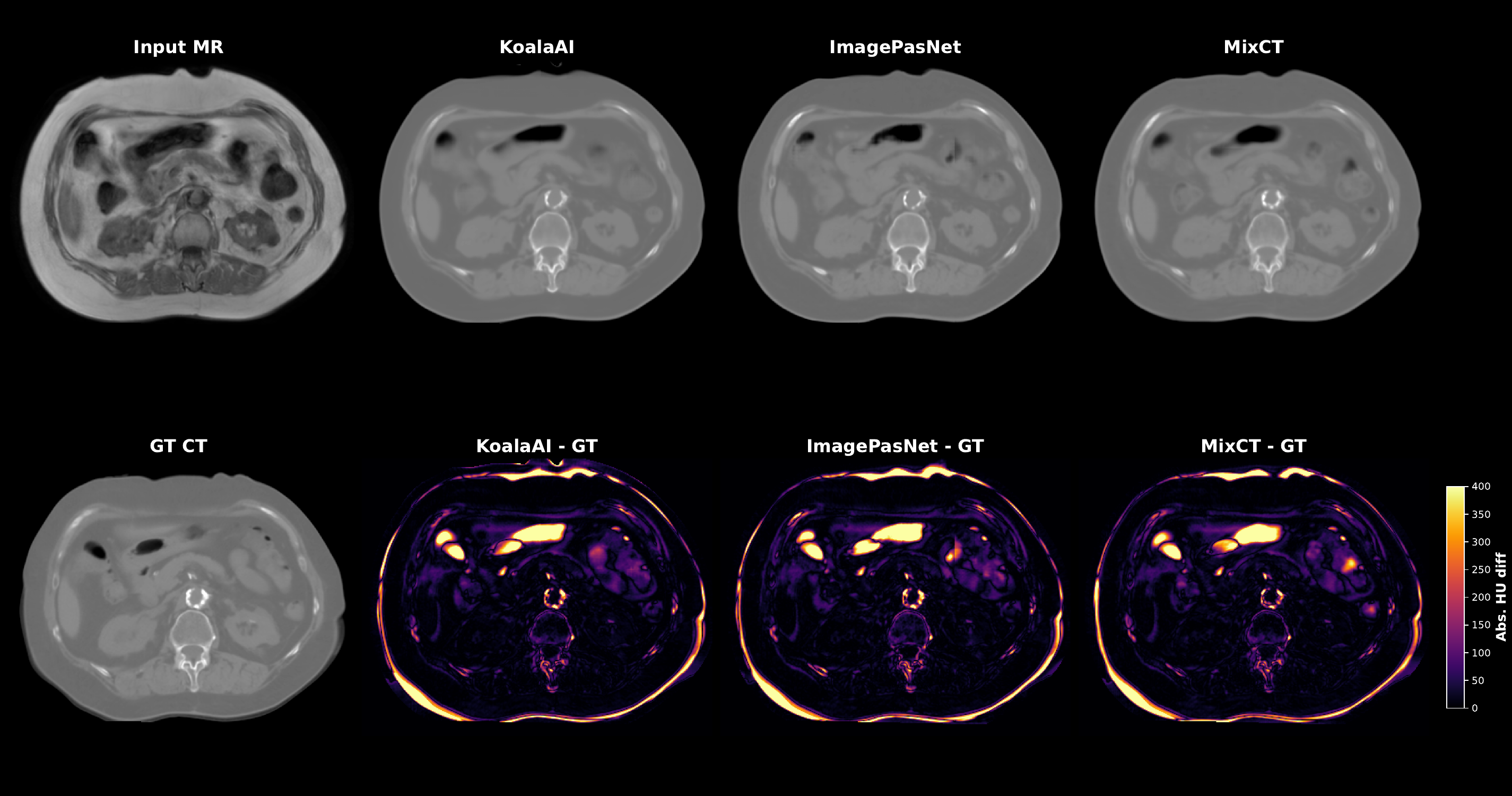}%
 }

 \vspace{1cm}
 \subfloat[Example sCTs for outlier patient 2ABB074.]{%
   \includegraphics[clip,width=0.99\linewidth]{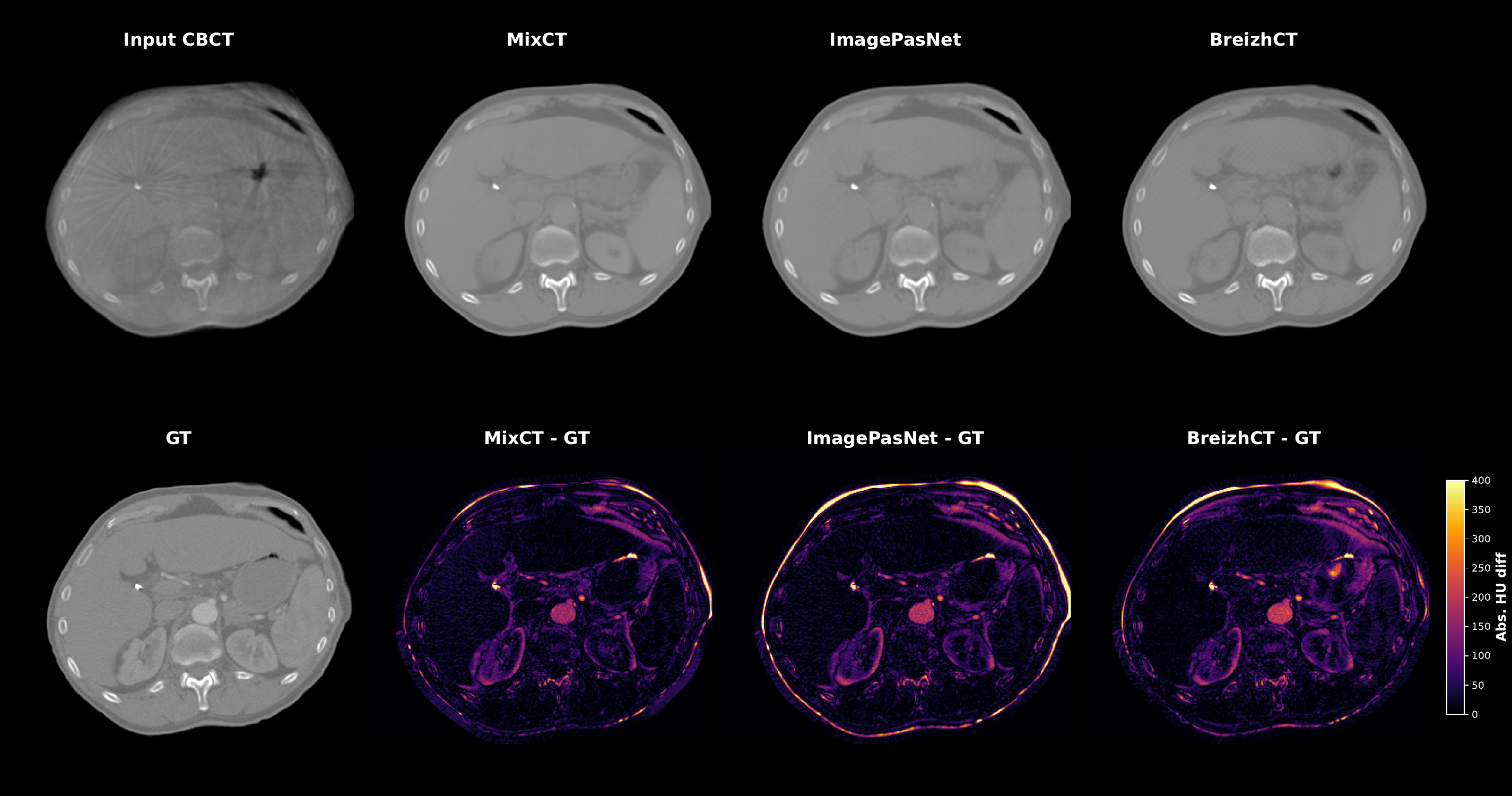}%
 }

 \caption{Examples of underperforming patients: patient 1ABA066 for task 1 (MRI-to-CT; a) and patient 2ABB074 for task 2 (CBCT-to-CT: b). Theese cases demostrated the lowest MS-SSIM values across all participants and may therefore represent particularly challenging cases for sCT generation.
 The model input is shown in the upper left, and the ground truth is in the center-left. The sCT of the top three participants for tasks 1 and 2 is shown in the top row. The absolute difference between the ground-truth CT and the result after masking with the provided mask is shown in the bottom row.}
 \label{fig:qualitative_evaluation_underperformance}
 \end{figure}

\subsection{Qualitative analysis of challenging cases}

To complement the aggregate statistical analyses, \Cref{fig:qualitative_evaluation_underperformance} shows one difficult test case from each task. These examples illustrate that low-performing cases can arise for different reasons. 

%% file: 09_suppC.tex
\section*{Supplementary Documents C: Algorithm details submission forms}
\label{sec:suppCD}
Files SynthRAD2025\_SubmissionForm\_Task1.pdf and SynthRAD2025\_SubmissionForm\_Task2.pdf contain the forms that participants submitted to provide details of their algorithms for tasks 1 and 2, respectively.

\includepdf[pages=-]{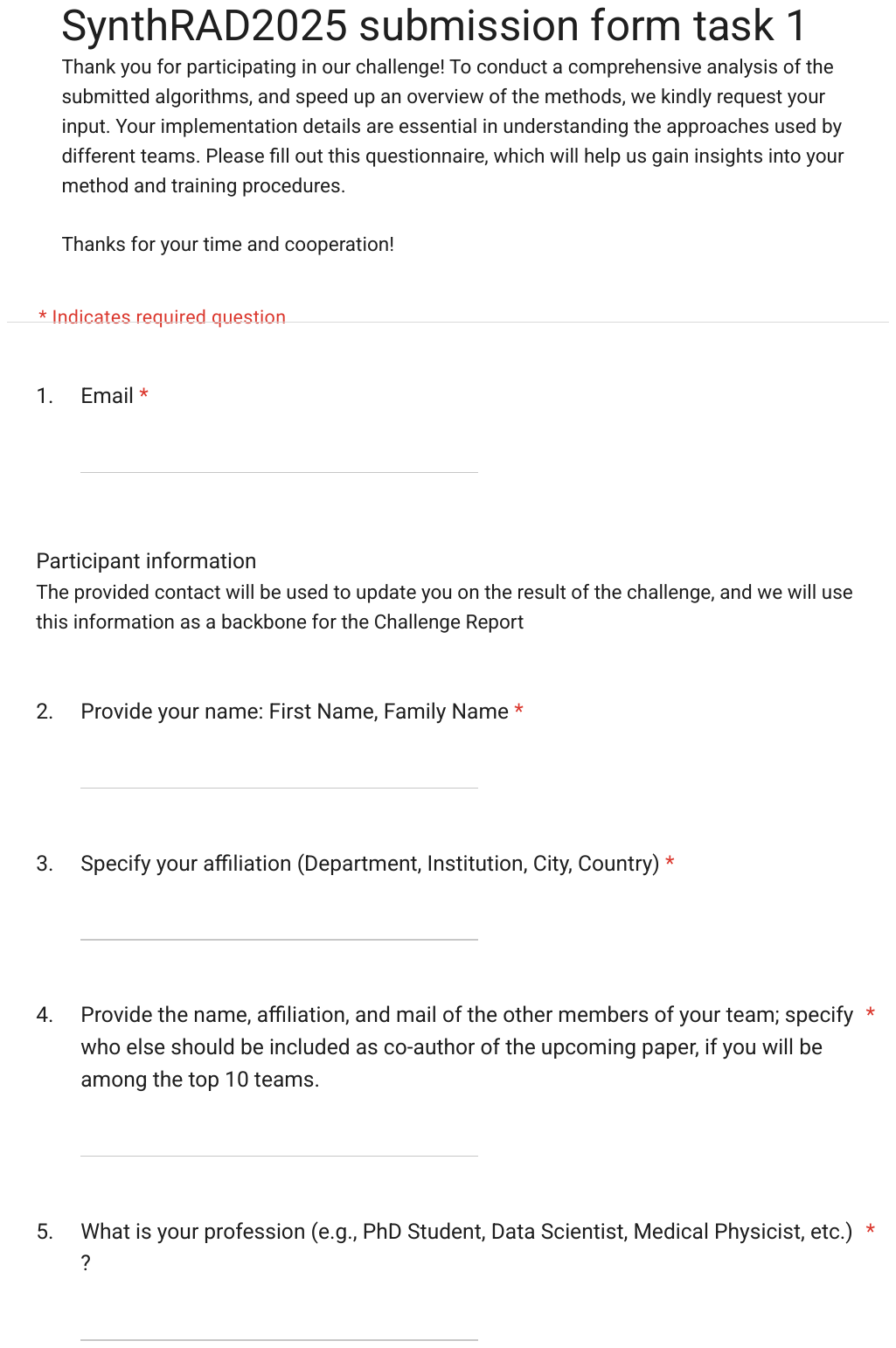}
\includepdf[pages=-]{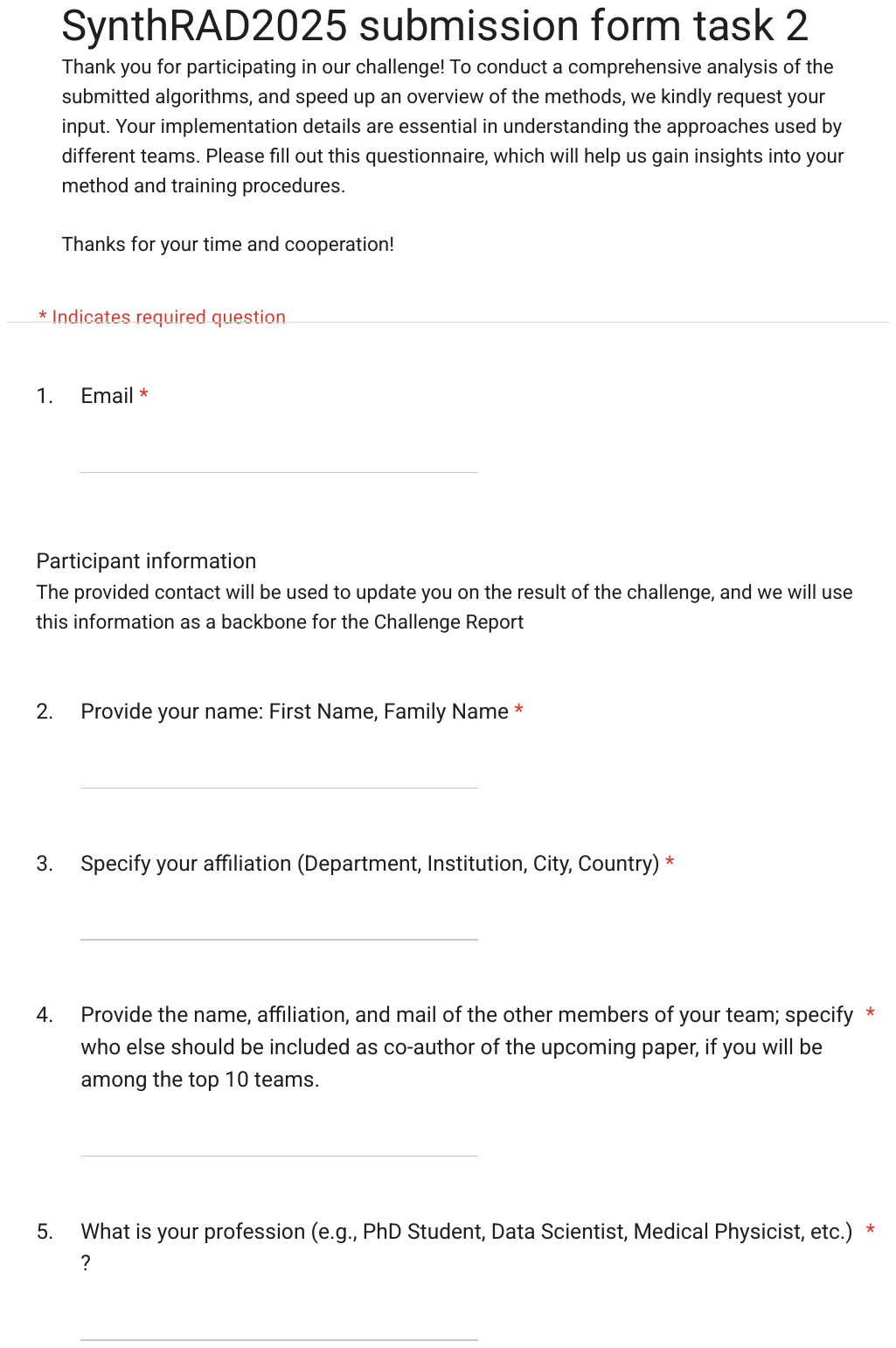}

%% file: refs.bib
@article{huijben2024generating,
  title={{Generating synthetic computed tomography for radiotherapy: SynthRAD2023 challenge report}},
  author={Huijben, Evi MC and Terpstra, Maarten L and Pai, Suraj and Thummerer, Adrian and Koopmans, Peter and Afonso, Manya and Van Eijnatten, Maureen and Gurney-Champion, Oliver and Chen, Zeli and Zhang, Yiwen and others},
  journal={Medical image analysis},
  volume={97},
  pages={103276},
  year={2024},
  publisher={Elsevier}
}

@article{thummerer2025synthrad2025,
  title={{SynthRAD2025 Grand Challenge dataset: Generating synthetic CTs for radiotherapy from head to abdomen}},
  author={Thummerer, Adrian and van der Bijl, Erik and Galapon, Arthur Jr and Kamp, Florian and Savenije, Mark and Muijs, Christina and Aluwini, Shafak and Steenbakkers, Roel JHM and Beuel, Stephanie and Intven, Martijn PW and others},
  journal={Medical physics},
  volume={52},
  number={7},
  pages={e17981},
  year={2025},
  publisher={Wiley Online Library}
}

@article{Thummerer2023synth,
  title={{SynthRAD2023 Grand Challenge dataset: Generating synthetic CT for radiotherapy}},
  author={Thummerer, Adrian and van der Bijl, Erik and Galapon Jr, Arthur and Verhoeff, Joost J C and Langendijk, Johannes A and Both, Stefan and van den Berg, Cornelis A T and Maspero, Matteo},
  journal={Medical Physics},
  year={2023},
  volume = {50},
  issue ={7},
  pages = {4664-4674},
  doi = {https://doi.org/10.1002/mp.16529},
  publisher={Wiley}
}

@article{deasy2010radiotherapy,
  title={{Radiotherapy dose--volume effects on salivary gland function}},
  author={Deasy, Joseph O and Moiseenko, Vitali and Marks, Lawrence and Chao, KS Clifford and Nam, Jiho and Eisbruch, Avraham},
  journal={International Journal of Radiation Oncology* Biology* Physics},
  volume={76},
  number={3},
  pages={S58--S63},
  year={2010},
  publisher={Elsevier}
}

@article{beetz2013role,
  title={Role of minor salivary glands in developing patient-rated xerostomia and sticky saliva during day and night},
  author={Beetz, Ivo and Schilstra, Cornelis and Visink, Arjan and van der Schaaf, Arjen and Bijl, Henk P and van der Laan, Bernard FAM and Steenbakkers, Roel JHM and Langendijk, Johannes A},
  journal={Radiotherapy and Oncology},
  volume={109},
  number={2},
  pages={311--316},
  year={2013},
  publisher={Elsevier}
}

@article{marks2010use,
  title={Use of normal tissue complication probability models in the clinic},
  author={Marks, Lawrence B and Yorke, Ellen D and Jackson, Andrew and Ten Haken, Randall K and Constine, Louis S and Eisbruch, Avraham and Bentzen, S{\o}ren M and Nam, Jiho and Deasy, Joseph O},
  journal={International Journal of Radiation Oncology* Biology* Physics},
  volume={76},
  number={3},
  pages={S10--S19},
  year={2010},
  publisher={Elsevier}
}

@article{puckett2023consensus,
  title={Consensus quality measures and dose constraints for lung cancer from the veterans affairs radiation oncology quality surveillance program and ASTRO expert panel},
  author={Puckett, Lindsay L and Titi, Mohammad and Kujundzic, Ksenija and Dawes, Samantha L and Gore, Elizabeth M and Katsoulakis, Evangelia and Park, John H and Solanki, Abhishek A and Kapoor, Rishabh and Kelly, Maria and others},
  journal={Practical radiation oncology},
  volume={13},
  number={5},
  pages={413--428},
  year={2023},
  publisher={Elsevier}
}

@article{wieser2017development,
  title={{Development of the open-source dose calculation and optimization toolkit matRad}},
  author={Wieser, Hans-Peter and Cisternas, Eduardo and Wahl, Niklas and Ulrich, Silke and Stadler, Alexander and Mescher, Henning and M{\"u}ller, Lucas-Raphael and Klinge, Thomas and Gabrys, Hubert and Burigo, Lucas and others},
  journal={Medical physics},
  volume={44},
  number={6},
  pages={2556--2568},
  year={2017},
  publisher={Wiley Online Library}
}

@article{ezzell2009imrt,
  title={{IMRT commissioning: multiple institution planning and dosimetry comparisons, a report from AAPM Task Group 119}},
  author={Ezzell, Gary A and Burmeister, Jay W and Dogan, Nesrin and LoSasso, Thomas J and Mechalakos, James G and Mihailidis, Dimitris and Molineu, Andrea and Palta, Jatinder R and Ramsey, Chester R and Salter, Bill J and others},
  journal={Medical physics},
  volume={36},
  number={11},
  pages={5359--5373},
  year={2009},
  publisher={Wiley Online Library}
}

@article{wasserthal2023totalsegmentator,
  title={TotalSegmentator: robust segmentation of 104 anatomic structures in CT images},
  author={Wasserthal, Jakob and Breit, Hanns-Christian and Meyer, Manfred T and Pradella, Maurice and Hinck, Daniel and Sauter, Alexander W and Heye, Tobias and Boll, Daniel T and Cyriac, Joshy and Yang, Shan and others},
  journal={Radiology: Artificial Intelligence},
  volume={5},
  number={5},
  pages={e230024},
  year={2023},
  publisher={Radiological Society of North America}
}

@article{Edmund2017,
  title = {A review of substitute CT generation for MRI-only radiation therapy},
  volume = {12},
  ISSN = {1748-717X},
  url = {http://dx.doi.org/10.1186/s13014-016-0747-y},
  DOI = {10.1186/s13014-016-0747-y},
  number = {1},
  journal = {Radiation Oncology},
  publisher = {Springer Science and Business Media LLC},
  author = {Edmund,  Jens M. and Nyholm,  Tufve},
  year = {2017},
}

@article{glide2021adaptive,
  title={Adaptive radiation therapy (ART) strategies and technical considerations: a state of the ART review from NRG oncology},
  author={Glide-Hurst, Carri K and Lee, Percy and Yock, Adam D and Olsen, Jeffrey R and Cao, Minsong and Siddiqui, Farzan and Parker, William and Doemer, Anthony and Rong, Yi and Kishan, Amar U and others},
  journal={International Journal of Radiation Oncology* Biology* Physics},
  volume={109},
  number={4},
  pages={1054--1075},
  year={2021},
  publisher={Elsevier}
}

@inproceedings{sonke2019adaptive,
  title={Adaptive radiotherapy for anatomical changes},
  author={Sonke, Jan-Jakob and Aznar, Marianne and Rasch, Coen},
  booktitle={Seminars in radiation oncology},
  volume={29},
  number={3},
  pages={245--257},
  year={2019},
  organization={Elsevier}
}

@article{dobbs1983use,
  title={The use of CT in radiotherapy treatment planning},
  author={Dobbs, H Jane and Parker, RP and Hodson, NJ and Hobday, Pauline and Husband, Janet E},
  journal={Radiotherapy and oncology},
  volume={1},
  number={2},
  pages={133--141},
  year={1983},
  publisher={Elsevier}
}

@article{schmidt2015radiotherapy,
  title={Radiotherapy planning using MRI},
  author={Schmidt, Maria A and Payne, Geoffrey S},
  journal={Physics in Medicine \& Biology},
  volume={60},
  number={22},
  pages={R323},
  year={2015},
  publisher={IOP Publishing}
}

@InProceedings{unet,
author={Ronneberger, Olaf
and Fischer, Philipp
and Brox, Thomas},
editor={Navab, Nassir
and Hornegger, Joachim
and Wells, William M.
and Frangi, Alejandro F.},
title={{U-Net: Convolutional Networks for Biomedical Image Segmentation}},
booktitle={Medical Image Computing and Computer-Assisted Intervention -- MICCAI 2015},
year={2015},
publisher={Springer International Publishing},
address={Cham},
pages={234--241},
isbn={978-3-319-24574-4}
}

@inproceedings{gan,
 author = {Goodfellow, Ian and Pouget-Abadie, Jean and Mirza, Mehdi and Xu, Bing and Warde-Farley, David and Ozair, Sherjil and Courville, Aaron and Bengio, Yoshua},
 booktitle = {Advances in Neural Information Processing Systems},
 editor = {Z. Ghahramani and M. Welling and C. Cortes and N. Lawrence and K.Q. Weinberger},
 publisher = {Curran Associates, Inc.},
 title = {{Generative Adversarial Nets}},
 url = {https://proceedings.neurips.cc/paper_files/paper/2014/file/5ca3e9b122f61f8f06494c97b1afccf3-Paper.pdf},
 volume = {27},
 year = {2014}
}

@inproceedings{CycleGAN2017,
  title={{Unpaired Image-to-Image Translation using Cycle-Consistent Adversarial Networks}},
  author={Zhu, Jun-Yan and Park, Taesung and Isola, Phillip and Efros, Alexei A},
  booktitle = {Proceedings of the IEEE International Conference on Computer Vision (ICCV)},
  year={2017}
}

@article{pix2pix2017,
  title={{Image-to-Image Translation with Conditional Adversarial Networks}},
  author={Isola, Phillip and Zhu, Jun-Yan and Zhou, Tinghui and Efros, Alexei A},
  journal={Proceedings of the IEEE Conference on Computer Vision and Pattern Recognition (CVPR)},
  pages={1125--1134},
  year={2017}
}

@inproceedings{transformer,
 author = {Vaswani, Ashish and Shazeer, Noam and Parmar, Niki and Uszkoreit, Jakob and Jones, Llion and Gomez, Aidan N and Kaiser, \L ukasz and Polosukhin, Illia},
 booktitle = {Advances in Neural Information Processing Systems},
 editor = {I. Guyon and U. Von Luxburg and S. Bengio and H. Wallach and R. Fergus and S. Vishwanathan and R. Garnett},
 publisher = {Curran Associates, Inc.},
 title = {{Attention is All you Need}},
 url = {https://proceedings.neurips.cc/paper_files/paper/2017/file/3f5ee243547dee91fbd053c1c4a845aa-Paper.pdf},
 volume = {30},
 year = {2017}
}

@article{diffusion,
  title={{Denoising Diffusion Probabilistic Models}},
  author={Jonathan Ho and Ajay Jain and Pieter Abbeel},
  year={2020},
  journal={arXiv preprint arxiv:2006.11239}
}

@inproceedings{wang2003multiscale,
  title={Multiscale structural similarity for image quality assessment},
  author={Wang, Zhou and Simoncelli, Eero P and Bovik, Alan C},
  booktitle={The thrity-seventh asilomar conference on signals, systems \& computers, 2003},
  volume={2},
  pages={1398--1402},
  year={2003},
  organization={Ieee}
}

@article{wilcoxon1945,
  author    = {Wilcoxon, Frank},
  title     = {Individual Comparisons by Ranking Methods},
  journal   = {Biometrics Bulletin},
  year      = {1945},
  volume    = {1},
  number    = {6},
  pages     = {80--83},
  doi       = {10.2307/3001968},
}

@article{spearman,
 ISSN = {00029556},
 author = {C. Spearman},
 journal = {The American Journal of Psychology},
 number = {1},
 pages = {72--101},
 publisher = {University of Illinois Press},
 title = {The Proof and Measurement of Association between Two Things},
 volume = {15},
 year = {1904},
doi = {10.2307/1412159}
}

@article{mann1947test,
  title={On a test of whether one of two random variables is stochastically larger than the other},
  author={Mann, Henry B and Whitney, Donald R},
  journal={The annals of mathematical statistics},
  pages={50--60},
  year={1947},
  volume={18},
  number={1},     
  publisher={JSTOR},
  doi={10.1214/aoms/1177730491},
}

@article{holm1979simple,
  title={A simple sequentially rejective multiple test procedure},
  author={Holm, Sture},
  journal={Scandinavian journal of statistics},
  pages={65--70},
  year={1979},
  publisher={JSTOR}
}

@article{spadea2021sctreview,
  title={Deep learning based synthetic-CT generation in radiotherapy and PET: A review},
  author={Spadea, Maria Francesca and Maspero, Matteo and Zaffino, Paolo and Seco, Joao},
  journal={Medical Physics},
  volume={48},
  number={11},
  pages={6537--6566},
  year={2021},
  doi={10.1002/mp.15150}
}

@article{brock2017tg132,
  title={Use of image registration and fusion algorithms and techniques in radiotherapy: Report of the AAPM Radiation Therapy Committee Task Group No. 132},
  author={Brock, Kristy K and Mutic, Sasa and McNutt, Todd R and Li, Hua and Kessler, Marc L},
  journal={Medical Physics},
  volume={44},
  number={7},
  pages={e43--e76},
  year={2017},
  doi={10.1002/mp.12256}
}

@article{low1998gamma,
  title={A technique for the quantitative evaluation of dose distributions},
  author={Low, Daniel A and Harms, W B and Mutic, S and Purdy, J A},
  journal={Medical Physics},
  volume={25},
  number={5},
  pages={656--661},
  year={1998},
  doi={10.1118/1.598248}
}

@article{taha2015metrics,
  title={Metrics for evaluating 3D medical image segmentation: analysis, selection, and tool},
  author={Taha, Abdel Aziz and Hanbury, Allan},
  journal={BMC Medical Imaging},
  volume={15},
  pages={29},
  year={2015},
  doi={10.1186/s12880-015-0068-x}
}

@article{maierhein2018rankings,
    title={Why rankings of biomedical image analysis competitions should be interpreted with care},
    author={Maier-Hein, Lena and Eisenmann, Matthias and Reinke, Annika and others},
    journal={Nature Communications},
    volume={9},
    number={1},
    pages={5217},
    year={2018},
    doi={10.1038/s41467-018-07619-7}
}

@article{chourak2022qa,
  title={Quality assurance for MRI-only radiation therapy: A voxel-wise population-based methodology for image and dose assessment of synthetic CT generation methods},
  author={Chourak, Hilda and Barateau, Ana{\"i}s and Tahri, Safaa and Cadin, Capucine and Lafond, Caroline and Nunes, Jean-Claude and Boue-Rafle, Adrien and Perazzi, Mathias and Greer, Peter B and Dowling, Jason and de Crevoisier, Renaud and Acosta, Oscar},
  journal={Frontiers in Oncology},
  volume={12},
  pages={968689},
  year={2022},
  doi={10.3389/fonc.2022.968689}
}

@article{rusanov2022cbctreview,
  title={Deep learning methods for enhancing cone-beam CT image quality toward adaptive radiation therapy: A systematic review},
  author={Rusanov, Branimir and Hassan, Ghulam Mubashar and Reynolds, Mark and Sabet, Mahsheed and Kendrick, Jake and Rowshanfarzad, Pejman and Ebert, Martin},
  journal={Medical Physics},
  volume={49},
  number={9},
  pages={6019--6054},
  year={2022},
  doi={10.1002/mp.15840}
}

@article{dinkla2018brain_sct,
  title={MR-Only Brain Radiation Therapy: Dosimetric Evaluation of Synthetic CTs Generated by a Dilated Convolutional Neural Network},
  author={Dinkla, Anna M and Wolterink, Jelmer M and Maspero, Matteo and Savenije, Mark H F S and Verhoeff, Joost J C and Seravalli, Enrica and I{\v{s}}gum, Ivana and Seevinck, Peter R and van den Berg, Cornelis A T},
  journal={International Journal of Radiation Oncology* Biology* Physics},
  volume={102},
  number={4},
  pages={801--812},
  year={2018},
  doi={10.1016/j.ijrobp.2018.05.058}
}

@article{wyatt2023pelvic_sct,
  title={Comprehensive dose evaluation of a Deep Learning based synthetic Computed Tomography algorithm for pelvic Magnetic Resonance-only radiotherapy},
  author={Wyatt, Jonathan J and Kaushik, Sandeep and Cozzini, Cristina and Pearson, Rachel A and Petit, Steven and Capala, Marta and Hernandez-Tamames, Juan A and Hidegh{\'e}ty, Katalin and Maxwell, Ross J and Wiesinger, Florian and McCallum, Hazel M},
  journal={Radiotherapy and Oncology},
  volume={184},
  pages={109692},
  year={2023},
  doi={10.1016/j.radonc.2023.109692}
}

@article{ohara2022cbct_sct,
  title={Assessment of CBCT-based synthetic CT generation accuracy for adaptive radiotherapy planning},
  author={O'Hara, Christopher J and Bird, David and Al-Qaisieh, Bashar and Speight, Richard},
  journal={Journal of Applied Clinical Medical Physics},
  volume={23},
  number={11},
  pages={e13737},
  year={2022},
  doi={10.1002/acm2.13737}
}

@article{reinke2024metricpitfalls,
  title={Understanding metric-related pitfalls in image analysis validation},
  author={Reinke, Annika and Tizabi, Minu D and Baumgartner, Michael and others},
  journal={Nature Methods},
  year={2024},
  doi={10.1038/s41592-023-02150-0}
}

@article{Jaffray2012,
  title={Image-guided radiotherapy: from current concept to future perspectives},
  author={Jaffray, David A},
  journal={Nature reviews Clinical oncology},
  volume={9},
  number={12},
  pages={688--699},
  year={2012},
  publisher={Nature Publishing Group UK London}
}

@article{Gregoire2011,
  title={State of the art on dose prescription, reporting and recording in Intensity-Modulated Radiation Therapy (ICRU report No. 83)},
  author={Gr{\'e}goire, Vincent and Mackie, Thomas R},
  journal={Cancer/radioth{\'e}rapie},
  volume={15},
  number={6-7},
  pages={555--559},
  year={2011},
  publisher={Elsevier}
}

@article{Cusumano2026,
  title={{Standardizing MRI-only radiotherapy commissioning: Benchmark dataset and acceptance levels from the MESCAL initiative}},
  author={Cusumano, Davide and Maspero, Matteo and Vellini, Luca and Alvarez-Michael, Emilie and Barateau, Ana{\"\i}s and Bessieres, Igor and Bohoudi, Omar and Dal Bello, Riccardo and Dufreneix, St{\'e}phane and Kurz, Christopher and others},
  journal={Radiotherapy and Oncology},
  pages={111530},
  year={2026},
  publisher={Elsevier}
}

@misc{Thummerer_2024_14051075,
  author       = {Thummerer, Adrian and
                  Galapon, Arhur Jr. and
                  Kurz, Christopher and
                  van der Bijl, Erik and
                  Kamp, Florian and
                  Landry, Guillaume and
                  Terpstra, Maarten and
                  Maspero, Matteo and
                  Wahl, Niklas and
                  Rogowski, Viktor},
  title        = {Synthesizing computed tomography for radiotherapy
                   challenge (SynthRAD2025)
                  },
  month        = nov,
  year         = 2024,
  publisher    = {Zenodo},
  doi          = {10.5281/zenodo.14051075},
  url          = {https://doi.org/10.5281/zenodo.14051075},
}

@article{Liu2022ConvNeXt,
  title={A ConvNet for the 2020s},
  author={Liu, Zhuang and Mao, Hanzi and Wu, Chao-Yuan and Feichtenhofer, Christoph and Darrell, Trevor and Xie, Saining},
  journal={arXiv preprint arXiv:2201.03545},
  year={2022}
}

@article{isensee2021nnu,
  title={nnU-Net: a self-configuring method for deep learning-based biomedical image segmentation},
  author={Isensee, Fabian and Jaeger, Paul F and Kohl, Simon AA and Petersen, Jens and Maier-Hein, Klaus H},
  journal={Nature methods},
  volume={18},
  number={2},
  pages={203--211},
  year={2021},
  publisher={Nature Publishing Group US New York}
}

@inproceedings{zhou2018unet++,
  title={Unet++: A nested u-net architecture for medical image segmentation},
  author={Zhou, Zongwei and Rahman Siddiquee, Md Mahfuzur and Tajbakhsh, Nima and Liang, Jianming},
  booktitle={International workshop on deep learning in medical image analysis},
  pages={3--11},
  year={2018},
  organization={Springer}
}

@inproceedings{xie2017aggregated,
  title={Aggregated Residual Transformations for Deep Neural Networks},
  author={Xie, Saining and Girshick, Ross and Dollár, Piotr and Tu, Zhuowen and He, Kaiming},
  booktitle={Proceedings of the IEEE Conference on Computer Vision and Pattern Recognition},
  year={2017}
}

@article{tan2024groupmorph,
  title={GroupMorph: medical image registration via grouping network with contextual fusion},
  author={Tan, Zuopeng and Zhang, Lihe and Lv, Yanan and Ma, Yili and Lu, Huchuan},
  journal={IEEE Transactions on Medical Imaging},
  volume={43},
  number={11},
  pages={3807--3819},
  year={2024},
  publisher={IEEE}
}

@article{d2024totalsegmentator,
  title={TotalSegmentator MRI: robust sequence-independent segmentation of multiple anatomic structures in MRI},
  author={D'Antonoli, Tugba Akinci and Berger, Lucas K and Indrakanti, Ashraya K ancd Vishwanathan, Nathan and Wei{\ss}, Jakob and Jung, Matthias and Berkarda, Zeynep and Rau, Alexander and Reisert, Marco and K{\"u}stner, Thomas and others},
  journal={arXiv preprint arXiv:2405.19492},
  year={2024}
}

@inproceedings{chen2024tinyu,
  title={Tinyu-net: Lighter yet better u-net with cascaded multi-receptive fields},
  author={Chen, Junren and Chen, Rui and Wang, Wei and Cheng, Junlong and Zhang, Lei and Chen, Liangyin},
  booktitle={International conference on medical image computing and computer-assisted intervention},
  pages={626--635},
  year={2024},
  organization={Springer}
}

@article{wiesenfarth2021methods,
  title={Methods and open-source toolkit for analyzing and visualizing challenge results},
  author={Wiesenfarth, Manuel and Reinke, Annika and Landman, Bennett A and Eisenmann, Matthias and Saiz, Laura Aguilera and Cardoso, M Jorge and Maier-Hein, Lena and Kopp-Schneider, Annette},
  journal={Scientific reports},
  volume={11},
  number={1},
  pages={2369},
  year={2021},
  publisher={Nature Publishing Group UK London}
}

@inproceedings{longuefosse2024adapted,
  title={Adapted nnU-Net: a robust baseline for cross-modality synthesis and medical image inpainting},
  author={Longuefosse, Arthur and Bot, Edern Le and De Senneville, Baudouin Denis and Giraud, R{\'e}mi and Mansencal, Boris and Coup{\'e}, Pierrick and Desbarats, Pascal and Baldacci, Fabien},
  booktitle={International Workshop on Simulation and Synthesis in Medical Imaging},
  pages={24--33},
  year={2024},
  organization={Springer}
}

@article{longuefosse2025anatomical,
  title={Anatomical feature-prioritized loss for enhanced MR to CT translation},
  author={Longuefosse, Arthur and Denis de Senneville, Baudouin and Dournes, Ga{\"e}l and Benlala, Ilyes and Baldacci, Fabien and Desbarats, Pascal},
  journal={Physics in Medicine \& Biology},
  volume={70},
  number={14},
  pages={145012},
  year={2025},
  publisher={IOP Publishing}
}

@software{pyradplan,
  title        = {pyRadPlan},
  author       = {
    Becher, Tobias and
    Bucher, Lina and
    Ermeneux, Louis and
    Hardt, Jennifer Josephine and
    Kontopoulos, Antonios and
    Laub, Julius and
    Leininger, Florian and
    Ortkamp, Tim and
    Santiago Aguilar, Victoria and
    Stanic, Goran and
    Schulz, Samir and
    Wangon Zekou, Wilfrid and
    Wahl, Niklas
  },
  year         = {2025},
  doi          = {10.5281/zenodo.15006909},
  url          = {https://pyradplan.readthedocs.io/},
  repository   = {https://github.com/e0404/pyRadPlan},
  license      = {BSD-3-Clause},
  type         = {software}
}
